\documentclass[fleqn,usenatbib]{mnras}

\usepackage{newtxtext,newtxmath}
\usepackage[flushleft]{threeparttable}
\usepackage{lscape}

\usepackage{float,xcolor}
\usepackage{rotating}
\usepackage{tablefootnote}

\usepackage[T1]{fontenc}

\DeclareRobustCommand{\VAN}[3]{#2}
\let\VANthebibliography\thebibliography
\def\thebibliography{\DeclareRobustCommand{\VAN}[3]{##3}\VANthebibliography}

\usepackage{graphicx}	
\usepackage{amsmath}	
\usepackage{graphicx}
\usepackage{float}

\title[SPHERE/IRDIS view on evolved binary discs]{Second-generation protoplanetary discs around evolved binaries: A\,high-resolution polarimetric view with SPHERE/IRDIS}

\author[K. Andrych et al.]{
Kateryna Andrych$^{1,2}$\thanks{E-mail: kateryna.andrych@students.mq.edu.au (KA);\newline devika.kamath@mq.edu.au (DK)},
Devika Kamath$^{1,2,3}$\footnotemark[1],
Jacques Kluska$^{4}$,
Hans Van Winckel$^{4}$,
Steve Ertel$^{5,6}$, \newauthor
Akke Corporaal$^{4}$
\\
$^{1}$School of Mathematical and Physical Sciences, Macquarie University, Balaclava Road, Sydney, NSW 2109, Australia\\
$^{2}$Research Centre for Astronomy, Astrophysics and Astrophotonics, Macquarie University, Balaclava Road, Sydney, NSW 2109, Australia\\
$^{3}$INAF, Osservatorio Astronomico di Roma, Via Frascati 33, 00077, Monte Porzio Catone, Italy \\
$^{4}$Institute of Astronomy, KU Leuven, Celestijnenlaan 200D, 3001 Leuven, Belgium\\
$^{5}$Department of Astronomy and Steward Observatory, University of Arizona, 933 N. Cherry Avenue, Tucson, AZ 85721-0065, USA\\
$^{6}$Large Binocular Telescope Observatory, University of Arizona, 933 N. Cherry Avenue, Tucson, AZ 85721-0065, USA
}

\date{Accepted 2023 June 27. Received 2023 June 27; in original form 2023 May 9}

\pubyear{2023}

\begin{document}
\label{firstpage}
\pagerange{\pageref{firstpage}--\pageref{lastpage}}
\maketitle

\begin{abstract}

Binary post-asymptotic giant branch (post-AGB) stars are products of a poorly understood binary interaction process that occurs during the AGB phase. These systems comprise of a post-AGB primary star, a main-sequence secondary companion and a stable circumbinary disc. Studying the structure and properties of these circumbinary discs is crucial for gaining insight into the binary interaction process that governs post-AGB binaries as well as comprehending the disc's creation, evolution, and its interaction with the post-AGB binary system. We aim to use near-infrared polarimetric imaging to investigate the morphology and potential substructures of circumbinary discs around eight representative post-AGB binary stars. To achieve this, we performed polarimetric differential imaging in $H$ and $Y$ bands using the high-angular resolution capabilities of the European Southern Observatory-Very Large Telescope/SPHERE-Infra-Red Dual-beam Imaging and Spectroscopy instrument. We resolved the extended circumbinary disc structure for a diverse sample of eight post-AGB binary systems. Our analysis provided the first estimates of the disc scale-height for two of the systems: IW\,Car and IRAS\,15469-5311. We also investigated the morphological differences between the full discs (with the inner rim at the dust sublimation radius) and transition discs (which are expected to have larger inner cavities), as well as similarities to protoplanetary disks around young stellar objects. We found that the transition discs displayed a more intricate and asymmetric configuration. Surprisingly, no correlation was found between the over-resolved flux in near-IR interferometric data and the polarimetric observations, suggesting that scattering of light on the disc surface may not be the primary cause of the observed over-resolved flux component.

\end{abstract}




\section{Introduction}

\label{sec:intro}

Towards the end of the asymptotic giant branch (AGB) phase, low and intermediate-mass single stars (0.8 -- 8 M$_\odot$) lose the majority of their envelope through mass loss driven by stellar winds. In the case of binary systems, the mass loss is induced by binary interaction, which in some cases, produces a dusty circumbinary disc in the post-AGB phase \citep[e.g.,][]{VanWinckel2003ARA&A..41..391V, VanWinckel2018arXiv180900871V}. 

While binary post-AGB stars were detected serendipitously, they show a characteristic spectral energy distribution (SED) wherein the dust excess peaks in the near-infrared \citep[near-IR, e.g.][]{DeRuyter2006A&A...448..641D, VanWinckel2009A&A...505.1221V, Kamath2014MNRAS.439.2211K, Kamath2015MNRAS.454.1468K, Gezer2015MNRAS.453..133G, Kluska2022}. Such a near-IR excess is commonly attributed to circumstellar dust and gas residing in a stable compact disc with an inner rim near the sublimation temperature. Observational studies have shown that post-AGB circumbinary discs exhibit Keplerian rotation, as deduced from $^{12}{\rm CO}$ position-velocity maps \citep[e.g.,][]{Bujarrabal2015A&A...575L...7B}. In addition, these discs display signatures of crystallisation \citep{Gielen2011A&A...533A..99G} and dust grain growth \citep{Scicluna2020MNRAS.494.2925S}. Therefore, post-AGB binaries are complex systems composed of the post-AGB star (the primary star), a secondary companion \citep[likely to be a main-sequence star,][]{Oomen2018}, and a circumbinary disc. Furthermore, recent time-series spectroscopic studies \citep[e.g.,][]{Bollen2017A&A...607A..60B, Bollen2022arXiv220808752B} have revealed that many post-AGB binaries also show the presence of high-velocity outflows (referred to as jets). Although the jet-launching mechanisms are not firmly constrained, the authors conclude that jets are launched by the companion, either as a stellar jet or from the accretion disc around the companion as a disc wind.

The circumbinary disc plays a key role in the evolution of the binary post-AGB star. For example, post-AGB binaries display a peculiar chemical composition known as photospheric chemical depletion, where refractory elements are depleted from the stellar photosphere   \citep[e.g.,][]{VanWinckel2003ARA&A..41..391V, Maas2005, Giridhar2010MNRAS.406..290G, Kamath2019MNRAS.486.3524K, Kluska2022}. The exact mechanism behind this depletion is not yet fully understood, but it is believed to result from the chemical fractionation of gas and dust in the circumbinary disc, followed by the re-accretion of clean gas onto the primary star  \citep{Waters1992A&A...262L..37W}. As a result, post-AGB binary stars exhibit extrinsic metal-poor characteristics, with [Fe/H] values ranging from approximately -5.0 to 0.5 dex \citep{Kamath2019MNRAS.486.3524K}. These objects also exhibit a range of depletion patterns. A reliable tracer for depletion efficiency is [Zn/Ti], where stars with [Zn/Ti] greater than 1.5 are classified as strongly depleted, those with [Zn/Ti] from 1.5 to 0.5 are considered mildly depleted, and [Zn/Ti] less than 0.5 are non-depleted \citep{Gezer2015MNRAS.453..133G, Oomen2018}.

Furthermore, the impact of the circumbinary disc on the post-AGB binary system leads to a wide range of orbital parameters, including eccentricities and orbital periods. Long-term radial velocity monitoring studies have shown that the typical orbital periods of post-AGB binaries range from a hundred to a few thousand days \citep{VanWinckel2009A&A...505.1221V, Oomen2018}. These values do not correspond to the theoretically predicted values \citep{Nie2012MNRAS.423.2764N}. Additionally, it has been concluded that the post-AGB binary systems did not undergo tidal circularization, and the torque exerted by the circumbinary disc on the central binary through Lindblad resonances cannot account for the high eccentricities observed in these systems \citep{Oomen2020A&A...642A.234O}.

In a recent study, \citet{Kluska2022} assembled a catalogue of all known Galactic post-AGB binaries featuring discs (85 objects). The authors found that many post-AGB binaries showed a dusty disc that starts from the dust sublimation radius and extends outward. These systems were classified as full disc candidates. Up to 12\% of post-AGB binaries show no or low near-IR excess, likely due to a larger inner cavity. These systems were classified as transition disc candidates. Additionally, \citet{Kluska2022} found that some of the transition disc candidates are 30 times more depleted than other post-AGB binaries. The authors interpreted this correlation as evidence that the exact mechanism which produces the transition disc structure also stimulates the dust and gas separation within the circumbinary disc resulting in higher depletion efficiencies. The recent mid-IR interferometric study confirmed the full and transition disc nature for 11 post-AGB binary stars \citep{Corporaal2023arXiv230412028C}. It revealed that the dust inner rims of targets lacking near-IR excess in their SEDs are 2.8-7.5 times larger than their expected dust sublimation radius. Conversely, systems without such a deficiency display inner rim sizes consistent with their predicted dust sublimation radii.

Deep interferometric studies in the near-IR wavelength regime spatially resolved the hot dust inner rim of the circumbinary disc around several Galactic post-AGB binary systems. These studies confirmed that the inner rim is predominantly located at the sublimation radius \citep[e.g.,][]{Hillen2016,Kluska2018A&A...616A.153K, Kluska2019A&A...631A.108K}. A VLTI/PIONIER ($H$-band) survey \citep{Kluska2019A&A...631A.108K} of circumbinary discs around a sample of 23 Galactic post-AGB binaries also resulted in the detection of a significant amount of over-resolved flux (more than 10\%) for 14 out of the 23 systems. This over-resolved signal was attributed to i) scattered light on the detected jets coming from the location of the secondary companion \citep[e.g.][]{Bollen2017A&A...607A..60B}, ii) scattered light on dust particles in the disc wind detected in the millimetre \citep[e.g.][]{Bujarrabal2017A&A...597L...5B} or iii) scattered light on the flared disc surface.

Radiative transfer (RT) modelling efforts for circumbinary discs around post-AGB binaries \citep[e.g.][]{Hillen2017, Kluska2018A&A...616A.153K, Corporaal2023arXiv230102622C} have shown that the high-angular resolution interferometric data is best reproduced by passively irradiated disc models developed for protoplanetary discs (PPDs). Furthermore, a high-angular resolution direct imaging study with VLT/SPHERE ($V$,\,$I$-bands) resolved the outer parts of the circumbinary disc around the post-AGB binary system AR Pup and detected the extended emission out to a separation of about 300 mas from the location of the central binary \citep[][]{Ertel2019AJ....157..110E}. The authors also detected complex structures in the extended disc (e.g., possible gaps). Such structures were also detected from high-angular resolution direct and polarimetric imaging studies of PPDs \citep[e.g.][]{Fukagawa2010PASJ...62..347F, Garufi2016A&A...588A...8G}. These results indicate similarities in the structure and properties between circumbinary discs around post-AGB binaries and PPDs around young stellar objects (YSOs). 
 
Additional investigations of these discs are crucial to gain more insights into the morphology and energetics of the circumbinary discs around post-AGB binaries and to further explore the nexus between these discs and PPDs. In the current study, we use VLT/SPHERE-IRDIS ($H$ and $Y$ bands) to conduct a systematic polarimetric differential imaging survey of the extended disc structure (with a radius of $\sim$20--200\,mas) around eight post-AGB binary systems that show the presence of over-resolved flux in near-IR interferometric observations. We aim to unravel the extended disc's morphology, detect possible disc perturbations, and investigate the origin of the observed over-resolved flux. The paper is structured as follows: we present our target sample in Section~\ref{sec:target_sel}. In Section~\ref{sec:data}, we present our observing strategy and data reduction methodology. The analysis of the VLT/SPHERE-IRDIS data and corresponding results are presented in Section~\ref{sec:analysis}. In Section~\ref{sec:discussion}, we discuss the observed diversity in disc morphologies and highlight similarities and differences between circumbinary discs around post-AGBs and PPDs. We present our conclusions in Section~\ref{sec:conclusion}.

\section{Target selection}
\label{sec:target_sel}

The target sample comprises of eight objects: U\,Mon, IRAS\,08544-4431, IW\,Car, HR\,4049, IRAS\,15469-5311, IRAS\,17038-4815, RU\,Cen, and AC\,Her. We selected these objects from a VLTI/PIONIER ($H$ band) survey \citep{Kluska2019A&A...631A.108K} of circumbinary discs around a sample of 23 Galactic post-AGB binaries. The selection criteria required that: i) the targets show the presence of a significant amount over-resolved flux in the near-IR interferometric PIONIER observations (more than 5\%) and ii) the targets were feasible to observe with VLT/SPHERE-IRDIS.

By and large, the eight target post-AGB binaries form a representative sample covering a wide range of temperatures, inclinations, orbital periods, chemical compositions and diverse SED morphologies. We also adopted the classification scheme (based on SED and IR-excess characteristics, see Section~\ref{sec:intro}) of \citet{Kluska2022}. Following this classification, six out of the eight targets fall under the full disc category, while two of them (RU\,Cen and AC\,Her) are classified as transition discs. We note that two of the eight targets (RU\,Cen and AC\,Her) have luminosities less than the tip of the RGB ($\lesssim2500$\,L$_\odot$). Therefore,  RU\,Cen and AC\,Her are likely post-RGB stars, i.e., low-luminosity analogues of post-AGB stars \citep{Kamath2016A&A...586L...5K}.

\begin{figure*} 
    \includegraphics[width=1.8\columnwidth]{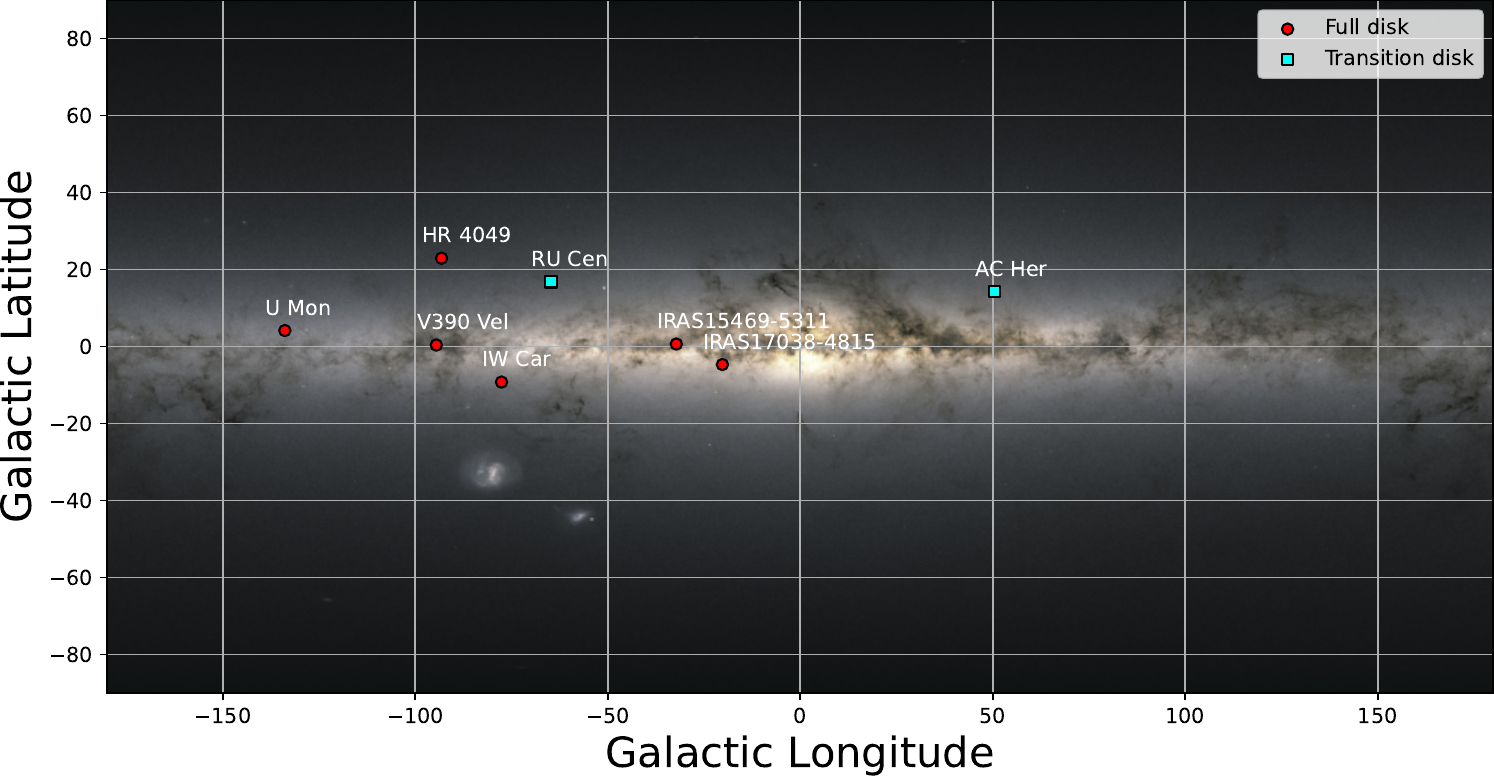}

    \caption{An edge-on Milky Way map with the positions of the target sample of post-AGB binaries marked. The colours represent disc categories based on \citet{Kluska2022}: red circles - full discs, blue squares - transition discs. See Section~\ref{sec:target_sel} for more details.}
    \label{fig:map}
\end{figure*}

In Table\,\ref{tab:sample}, we present our target sample, corresponding IRAS\,names and main properties relevant to this study. For completeness, in Appendix~\ref{sec:ap_sample}, we present an extensive literature review for each of the eight targets. Additionally, in Fig.~\ref{fig:map}, we show the distribution of the target sample in the Galaxy. The objects are field stars and distributed along the Galactic disc. 

    \begin{table*}
        \caption{Stellar and orbital properties of post-AGB binary stars in our target sample relevant to this study.}
        \begin{center}
            \begin{tabular}{ c|l|l|l|c|c|c|c|c|c||c|c| }
                \hline
            \#ID& Name &IRAS& \begin{tabular}[c]{@{}l@{}}$L$ \\ {(}$10^{3} L_\odot${)}\end{tabular} & L$_{IR}$/L$_{*}$&\begin{tabular}[c]{@{}l@{}}$P_{\rm orbital}$\\ {(}days{)}\end{tabular} & \begin{tabular}[c]{@{}l@{}}$OR$\\ {(}\%{)}\end{tabular} & [Fe/H] & [Zn/Ti] & [Zn/H] & C/O & Ref. \\
                \hline
        		\multicolumn{12}{c}{Full discs}\\
                \hline
                1& U\,Mon&07284-0940 &$5.5^{+1.8}_{-0.9}$ & 0.23 &2550$\pm$143 & 9.7$^{+0.3}_{-0.3}$ &-0.8& 0&-0.7&0.8 & 1, 2, 6, 15 \\
                2& IRAS\,08544-4431 & 08544-4431 & $13.7^{+3.8}_{-2.7}$ $^{*}$ &0.29& 501.1$\pm$1.0 & 15.3$^{+0.2}_{-0.2}$  & -0.3& 0.9&0.1&0.7 & 1, 2, 7, 14  \\
                3& IW\,Car&09256-6324 &$2.6^{+0.3}_{-0.3}$&0.87 &1449 & 10.5$^{+0.3}_{-0.3}$  & -1.1 & 2.1&-0.1&2.1 & 3, 2, 8, 15 \\
                4& HR 4049&10158-2844   & $20.3^{+14.5}_{-7.3}$ $^{*}$&0.26& 430.6 $\pm$ 0.1 & 17$^{+0.7}_{-0.7}$ &  -4.8& -&-1.3&- & 1, 9 \\
                5&IRAS 15469-5311 & 15469-5311& $17.1^{+7.2}_{-4.5}$ $^{*}$&0.95& 390.2 $\pm$ 0.7 &  12.8$^{+0.5}_{-0.6}$ & 0 & 1.8&0.3&0.9& 1, 2, 11, 13\\
                6&IRAS 17038-4815 &17038-4815 &$4.8^{+3.2}_{-1.6}$ $^{*}$&0.91& 1394 $\pm$ 12 &  4.2$^{+0.4}_{-0.4}$ & -1.5 &0.7&-1.2& - & 1, 4, 11, 13 \\
                \hline
        		\multicolumn{12}{c}{Transition discs}\\
                \hline
                7 & RU\,Cen &12067-4508& $1.1^{+0.3}_{-0.2}$&0.40& 1489 $\pm$ 10 & 4.6$^{+0.3}_{-0.3}$ & -1.9& 1&-1&0.4 & 1, 10, 15 \\
                8& AC\,Her &18281+2149 & $2.5^{+0.2}_{-0.2}$ &0.21& 1188.9 $\pm$ 1.2 & - & -1.5& 1&-0.8&0.4 & 1, 5, 12, 13, 15 \\
                \hline
            \end{tabular}
        \end{center}
        \begin{tablenotes}
        
        \small
    \item \textbf{Notes:}  The target sample is separated into full discs and transition discs based on the disc category of \citet{Kluska2022} and \citet{Corporaal2023arXiv230412028C}. See Section~\ref{sec:target_sel} for more details. $L$ represents the luminosity of the post-AGB star in units of $10^{3}\,L_\odot$. We note that $^{*}$ indicates that the luminosity value (derived using SED fitting and distances obtained from GAIA DR2) was adopted from \citep{Oomen2019A&A...629A..49O}. In other cases, the luminosity value (derived from the empirical period–luminosity–temperature–metallicity relation) was adopted from \citet{Bodi2019}. $L_{IR}/L_{*}$ represents the infrared luminosity adopted from \citep{Kluska2022}. $P_{\rm orbital}$ represents the orbital period in days. $OR$ represents the percentage of over-resolved flux from VLTI/PIONIER ($H$-band) observations \citep{Kluska2019A&A...631A.108K}. More details on the tabulated information can be found in the individual studies mentioned in column 'Ref'. The column 'Ref.' indicates the individual study: 1 - \citet{Oomen2018}, 2 - \citet{Kiss2007MNRAS.375.1338K}, 3 - \citet{Kiss2017}, 4 -\citet{Manick2017}, 5 - \citet{Samus2009yCat....102025S}, 6 - \citet{Giridhar2000ApJ...531..521G}, 7 - \citet{Maas2003A&A...405..271M}, 8 - \citet{Giridhar1994ApJ...437..476G}, 9 - \citet{VanWinckel1995PhDT........31V}, 10 - \citet{Maas2002}, 11 - \citet{Maas2005}, 12 - \citet{VanWinckel1998A&A...336L..17V}, 13 - \citet{Giridhar1998ApJ...509..366G}, 14 - \citet{Maas2003A&A...405..271M}, 15 - \citet{Bodi2019}. \\
        \end{tablenotes}
       
        \label{tab:sample}
    \end{table*}

\section{Data and Observations}
\label{sec:data}

This work focuses on high-angular-resolution polarimetric observations of eight post-AGB binaries (see Section~\ref{sec:target_sel}) at near-IR wavelengths. The data were obtained with the extreme AO instrument \citep[SPHERE,][]{Beuzit2019} using the Infra-Red Dual-beam Imaging and Spectroscopy camera \citep[IRDIS, ][]{Dohlen2008} in the $H$, $Y$-bands. In this section, we describe the observations and data reduction.

\subsection{Observations}
\label{sec:observations}

The SPHERE/IRDIS data used in this study were obtained as part of European Southern Observatory (ESO) observational programs 0101.D-0807(B) for AC\,Her and 0102.D-0696(A) for other objects in the sample (PI: Kluska). Observations were carried out in May 2018 and October 2018 - March 2019.

The data were obtained in $H$-band ($\lambda_0=1625$\,nm, $\Delta\lambda=290$\,nm; for all objects except AC\,Her) and $Y$-band ($\lambda_0=1043$\,nm, $\Delta\lambda=140$\,nm; for AC\,Her). To observe the scattered light around the post-AGB binary star, SPHERE/IRDIS was used in its dual-polarization imaging mode (IRDIS-DPI). This mode performs polarimetric differential imaging (PDI), thereby allowing to isolate the polarized scattered light from the disc while cancelling the unpolarized emission from the central star. Consequently, we can detect faint circumbinary discs at separations $\lesssim$\,1\,arcsec. We did not use a coronagraph because we require the smallest possible working angle to resolve the extended disc, which is likely the source of the over-resolved flux in the PIONIER near-IR interferometric data (see Section~\ref{sec:target_sel}). Since the targets are relatively bright in the $H$-band (see Table~\ref{tab:initialdata}), we included neutral density filters ND3.5 and ND2.0 to avoid the saturation of the detector. The detector integration time (DIT) was chosen in the range of two to eight seconds based on the target brightness and neutral density filter configuration. Observing conditions were well suited for high-angular resolution PDI, with a seeing between 0.3-2.4 arcsec for all targets. However, we note that U\,Mon was observed twice, on the 3rd and 14th of January 2019, because of improper observing conditions at the time of the first observation. Additional details on the observational conditions are presented in Appendix~\ref{sec:ap_weather}.

\subsection{Data reduction}
\label{sec:data_reduction}

Data reduction (DR) was performed using the highly-automated end-to-end pipeline IRDAP \citep[IRDIS Data reduction for Accurate Polarimetry]{Holstein2020A&A...633A..64V} complemented with our own $Python$ scripts. In the following sections, we briefly present the main steps involved in the DR procedure. We note that we use IRAS\,08544-4431 (one of the eight targets in this study) as an example to illustrate the results from the DR. The DR results for all objects in the target sample are presented in Appendix~\ref{sec:ap_pol_imag}.

\subsubsection{Polarimetric differential imaging (PDI) reduction with IRDAP}
\label{sec:pdi}

\begin{figure*}
  
    \includegraphics[width=0.5\columnwidth]{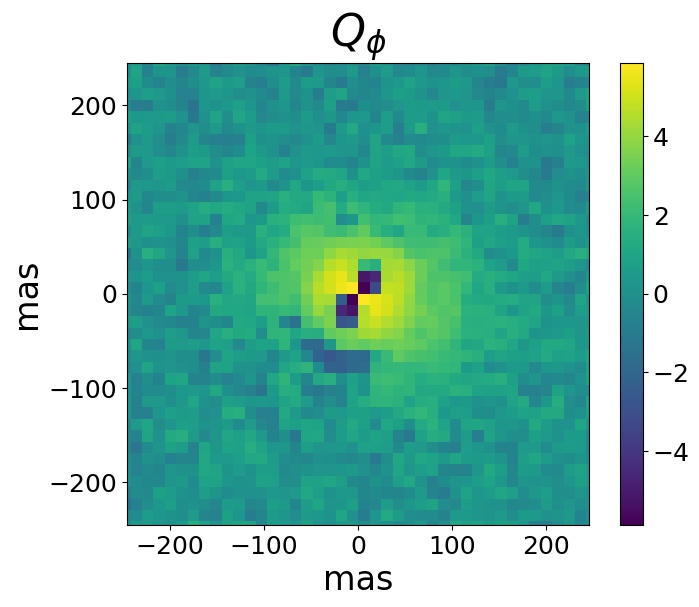}
    \includegraphics[width=0.5\columnwidth]{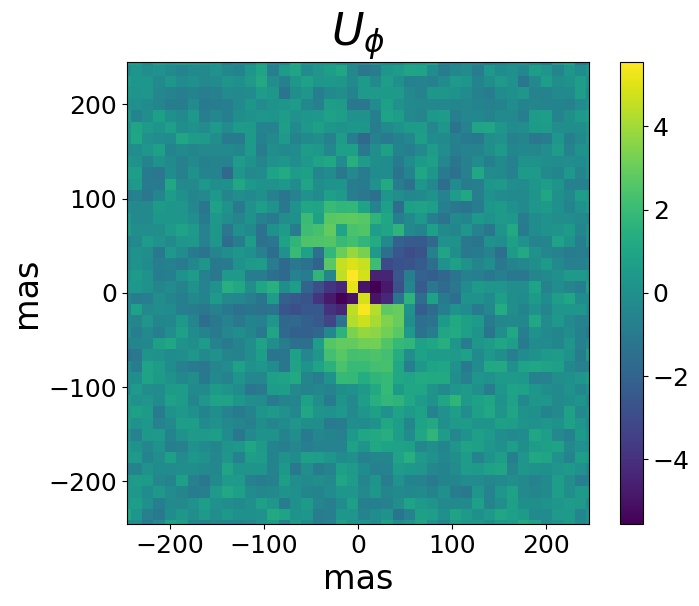}
    \includegraphics[width=0.5\columnwidth]{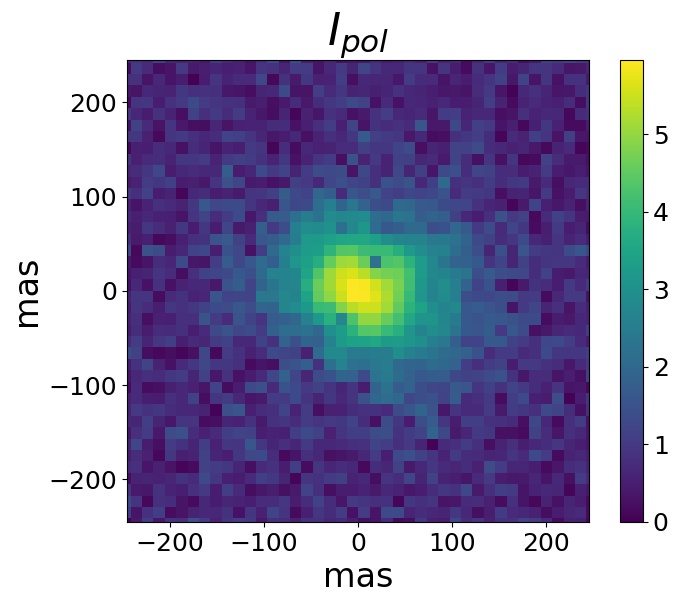}
    \includegraphics[width=0.5\columnwidth]{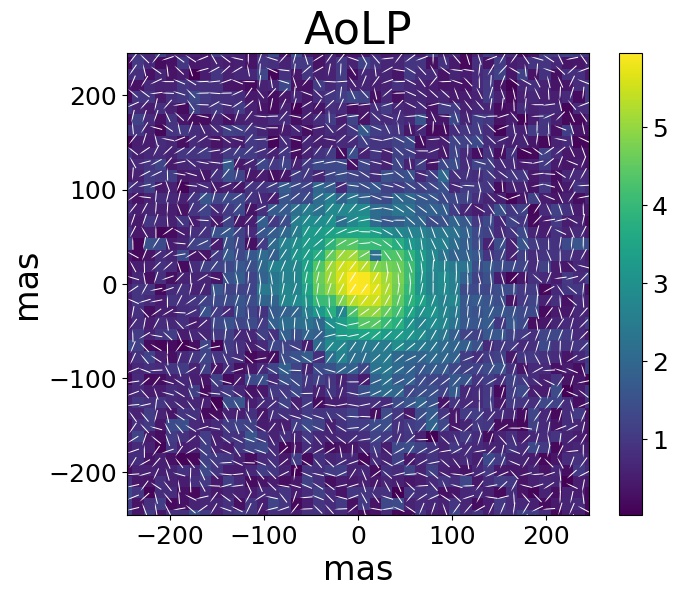}

    \includegraphics[width=0.5\columnwidth]{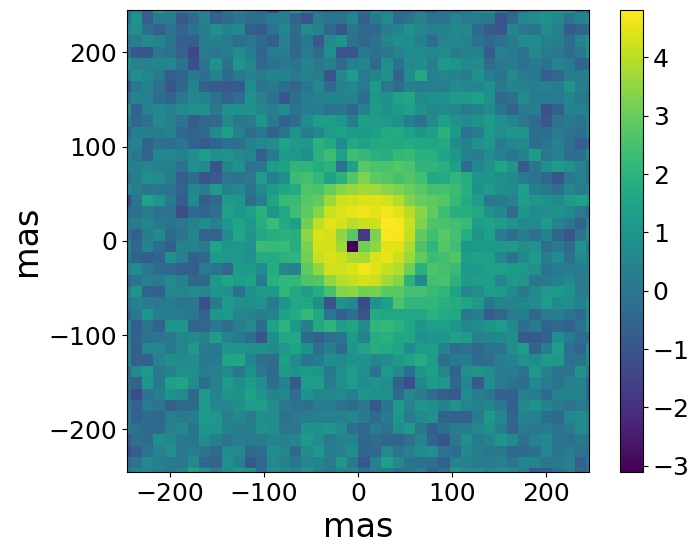}
    \includegraphics[width=0.5\columnwidth]{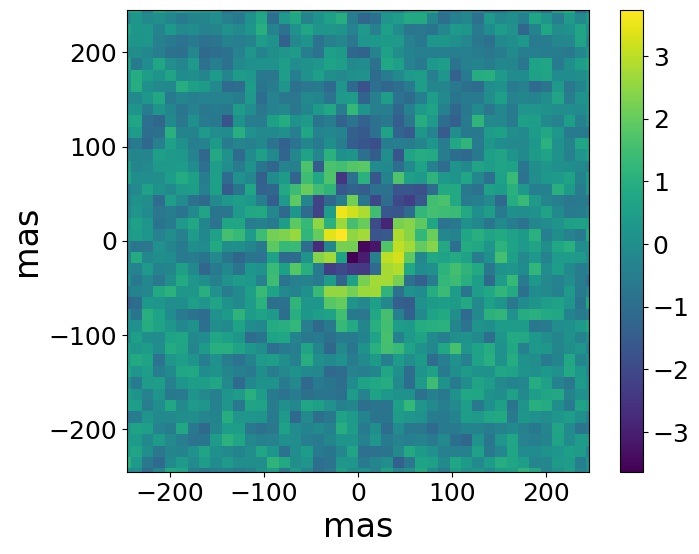}
    \includegraphics[width=0.5\columnwidth]{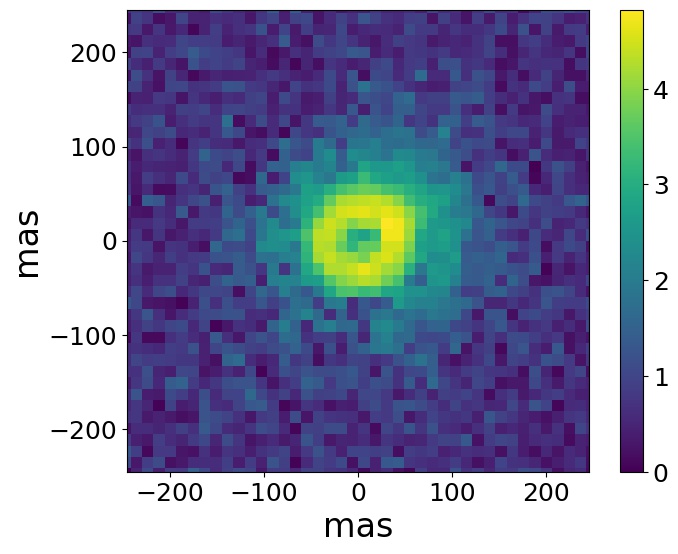}
    \includegraphics[width=0.5\columnwidth]{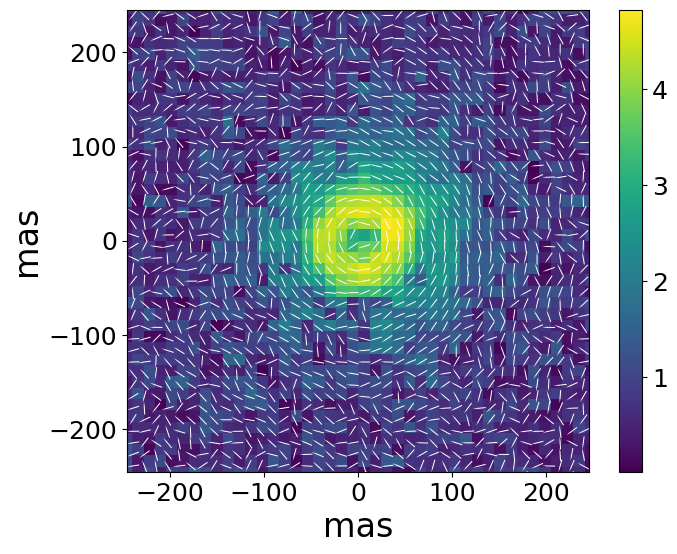}
    
    \caption{The polarized signal of IRAS\,08544-4431 before (top row) and after (bottom row) subtraction of the unresolved central polarization. The first column represents the Q$_\phi$  images, the second column represents the U$_\phi$ images, the third column represents the total polarized intensity ($I_{\rm pol}$) images and the last column represents the Angle of Linear polarization ($AoLP$) plotted over the $I_{\rm pol}$ image. All images are presented on an inverse hyperbolic scale and oriented North to up and East to the left. See Section~\ref{sec:data_reduction} for more details. Similar images for other targets are presented in Appendix~\ref{sec:ap_pol_imag}.
    \label{fig:stellar}}
\end{figure*}

To perform the PDI reduction, we predominantly used the IRDAP in its standard configuration. The core feature of IRDAP is the model-based correction method to address the crosstalk and the instrumental polarization effects of both the telescope and IRDIS. The first step of the DR is pre-processing of the data. The IRDAP pipeline does this by applying background subtraction, flat fielding and frame centring.
The pipeline then uses the double difference and double sum method to compute the classical Stokes vector components $Q$ and $U$ and the corresponding total intensities, which are later corrected for instrumental effects and crosstalk. Next, the pipeline computes the azimuthal Stokes parameters Q$_\phi$ and U$_\phi$ following the definitions of \citet{deBoer2020}, the total polarized intensity ($I_{pol}$), and the Angle of Linear polarization ($AoLP$; see Fig.\ref{fig:stellar}). Positive Q$_\phi$ indicates linear polarization in the azimuthal direction, while a negative Q$_\phi$ corresponds to linear polarization in the radial direction. U$_\phi$ shows the linear polarization shifted by $\pm45^\circ$ from the azimuthal and radial directions. Under the assumption of single-scattering on the thin disc surface and low inclinations of the disc ($i\,\lesssim 40^\circ$), we associated positive Q$_\phi$ to polarized light from the disc, and any non-azimuthal polarization (U$\phi$ and negative Q$_\phi$) to noise or calibration errors \citep{Canovas2015A&A...582L...7C}. 

Out of interest, we have also investigated a scenario where there is a polarized signal from the disc in the $U_{\phi}$ components of our dataset (see Appendix~\ref{sec:ap_sign_uphi} for details).

\subsubsection{Subtracting of the unresolved central polarization}

\label{sec:unresolved}

\begin{table}
	\centering
	\caption{Characteristics of the unresolved central polarisation for the post-AGB binaries in the sample.}
	\label{tab:unresolved}
	\begin{center}
    	    
    	\begin{tabular}{|c|l|c|c|} 
    		\hline
    		 \#ID & Name & DoLP, \% & AoLP, $^\circ$ \\
    		\hline
    		\multicolumn{3}{c}{Full discs}\\
    		\hline
            1& U\,Mon & 1.63$\pm$0.066&178$\pm$50\\
            2& IRAS\,08544-4431& 0.63$\pm$ 0.09&148$\pm$ 4.3\\
            3& IW\,Car & 0.23$\pm$ 0.17&62$\pm$ 23\\
            4 & HR 4049 & 0.17$\pm$ 0.06& 79$\pm$ 12\\
            5&IRAS 15469-5311 & 1.2$\pm$ 0.06&55$\pm$ 1.4\\
            6&IRAS 17038-4815 & 2.2$\pm$ 0.13&38$\pm$ 2.6\\
            \hline
    		\multicolumn{3}{c}{Transition discs}\\
    		\hline
            7 & RU\,Cen  &0.54$\pm$ 0.17&62$\pm$ 9.9 \\
            8& AC\,Her& 0.35$\pm$ 1.03&69$\pm$ 50\\
            \hline
    	\end{tabular}
    \end{center}
	\begin{tablenotes}
        
    \small
\item \textbf{Notes:}  'DoLP' represents the degree of linear polarisation, 'AoLP' represents the predominant angle of linear polarisation for the unresolved central polarisation (see Section~\ref{sec:unresolved}). We note that for U Mon, results were derived using the observations taken on the 14th of January 2019 as it has better observational conditions (see Table~\ref{tab:weather} in Appendix~\ref{sec:ap_weather}).\\
    \end{tablenotes}
	
\end{table}
The linearly polarized images after the DR (see Fig.\ref{fig:stellar}) show unresolved emission features. For example: i) a butterfly pattern in the central part of the Q$_\phi$ and U$_\phi$, and ii) a halo in the polarized total intensity image (see top row of Fig.\ref{fig:stellar}). These unresolved features are likely due to polarization light from the photosphere of the post-AGB star or the scattering of the stellar light on the material closer to the star than what is resolvable with SPHERE/IRDIS ($\sim$\,0.$''$04). While this unresolved emission might contain information about components of the post-AGB binary disc, it distorts the linearly polarized images. To correct the unresolved central polarization, IRDAP uses annuli from the Q, U, and double-sum total-intensity images centred on the central star position.

Typically, these annuli should only contain signal from the central star. In our case, assuming that the typical distance to the post-AGB binary system is $\gtrsim1$ kpc (see Table~\ref{tab:initialdata}), the inner rim of the disc sits at $\sim$10\,mas \citep[according to the dust sublimation radius, ][]{Kluska2019A&A...631A.108K}. Given that the scale of the IRDIS camera is 12.27\,mas/pixel, a single central pixel on the SPHERE/IRDIS image will contain emission from the post-AGB star, its main-sequence companion, the circum-companion accretion disc, and part of the circumbinary disc. Therefore, we deviated from IRDAP's standard annuli size by testing the outer radius from 1 to 15 pixels. Based on the results from these tests, we adopted a circular area with a radius of 3 pixels centred on the stellar position as the optimal annuli for the calculation of the degree and angle of the unresolved central polarization (see Table \ref{tab:unresolved}). Using these values, IRDAP performed the subtraction of the unresolved central polarization. The resulting polarized images are presented in Fig.\ref{fig:stellar} (see bottom row). 

We note that a consequence of the annuli size containing signal from the disc is an over-correction of the unresolved central polarization. This results in a lower intensity in the central 4x4 pixel region of Q$_\phi$ and I$_{\rm pol}$ images (see bottom row of Fig.~\ref{fig:stellar}). We, therefore, exclude the 4x4 pixel region centred on the binary position in our further analysis. 

\subsubsection{Deconvolution}
\label{sec:deconvolution}

\begin{figure*}
    \includegraphics[width=0.6\columnwidth]{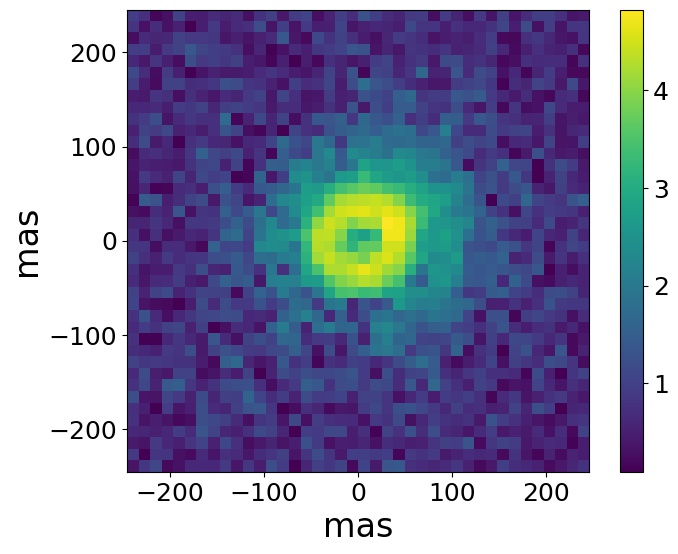}
    \includegraphics[width=0.6\columnwidth]{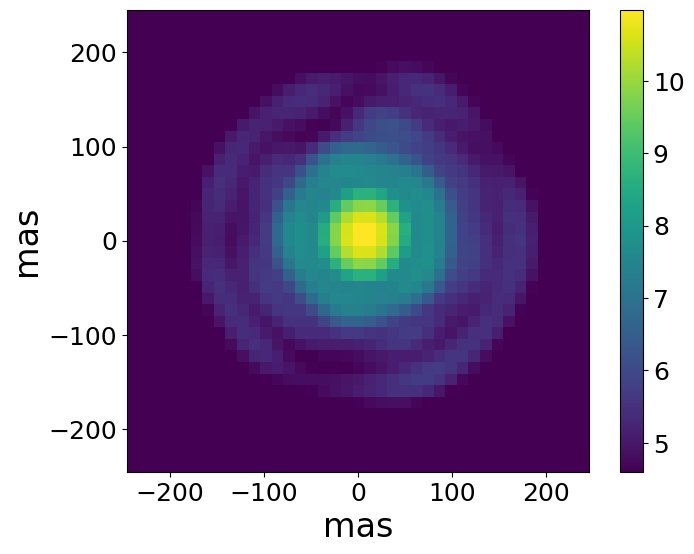}
    \includegraphics[width=0.6\columnwidth]{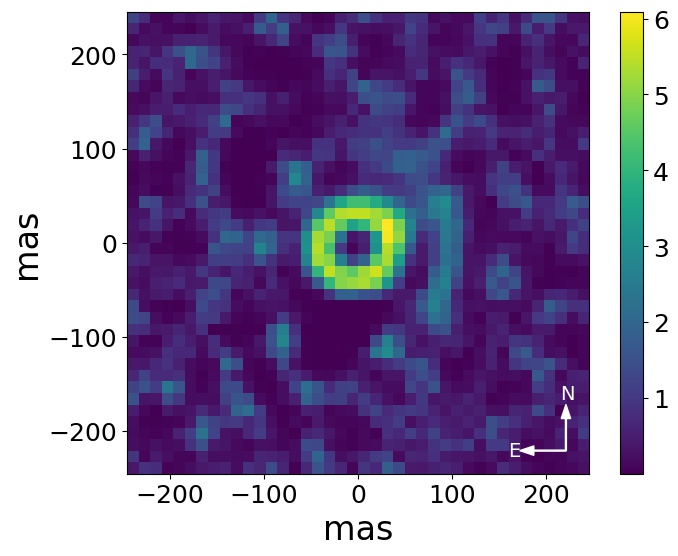}

    \caption{Illustration of the deconvolution process for IRAS\,08544-4431. The left image represents the total polarized intensity image $I_{\rm pol}$ before the deconvolution process, the middle image represents the total intensity image as a point spread function, and the right image represents the deconvolved $I_{\rm pol}$ image (see Section~\ref{fig:deconvolution}). All images are presented on an inverse hyperbolic scale. The low intensity of the central 4x4 pixel region of $I_{\rm pol}$ images is a reduction bias caused by over-correction of the unresolved central polarization (see Section~\ref{sec:unresolved}). Similar deconvolved $I_{\rm pol}$ images for other targets are presented in Appendix~\ref{sec:ap_pol_imag}.}
    \label{fig:deconvolution}
\end{figure*}

In the final step of the DR procedure, we applied the Richardson–Lucy deconvolution algorithm \citep{Richardson1972JOSA...62...55R, Lucy1974AJ.....79..745L} to deconvolve the polarized images obtained from IRDAP with the point spread function (PSF). We use the total unpolarized intensity image as a corresponding PSF for each target since the post-AGB star is the brightest source in the system, and the stellar light is mainly unpolarized. We note that the deconvolution process converged within 50 iterations for all targets in the sample (pixel-to-pixel variation between neighbouring iterations reached less than 1.5\%). As an illustration of the deconvolution process, we present the initial total polarized intensity, the PSF, and the resulting image of IRAS\,08544-4431 in Fig.\ref{fig:deconvolution}. Deconvolved azimuthally polarized images ($Q_\phi$) for the whole sample are presented in Fig.\ref{fig:snr}, while deconvolved total polarized intensity images are provided in Appendix~\ref{sec:ap_deconv_totalpol_imag}.

\begin{figure*}
    \includegraphics[width=0.5\columnwidth]{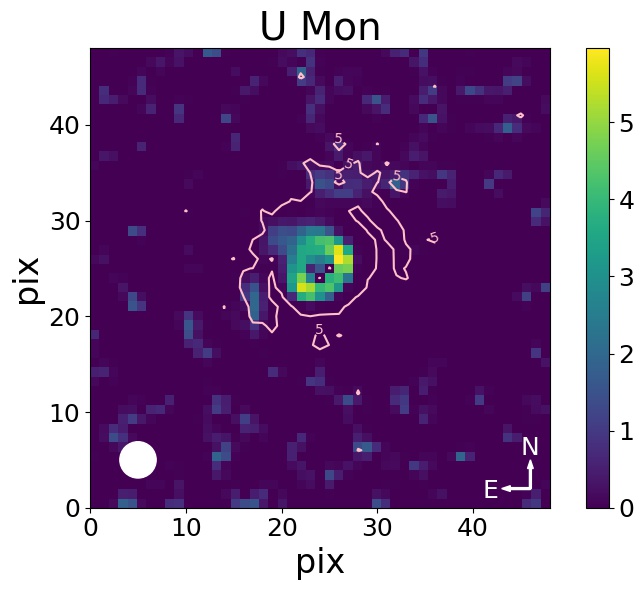}
    \includegraphics[width=0.5\columnwidth]{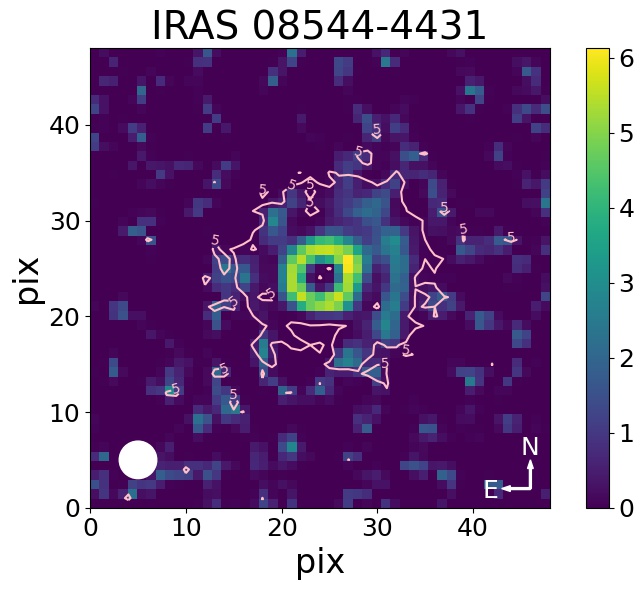}
    \includegraphics[width=0.5\columnwidth]{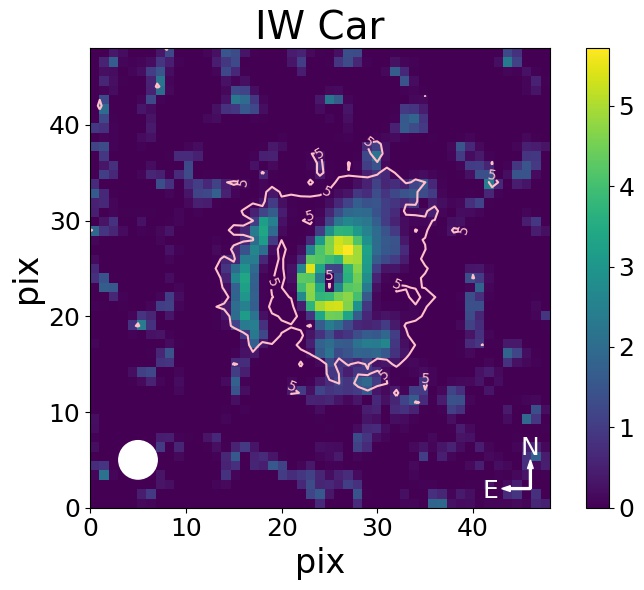}    \includegraphics[width=0.5\columnwidth]{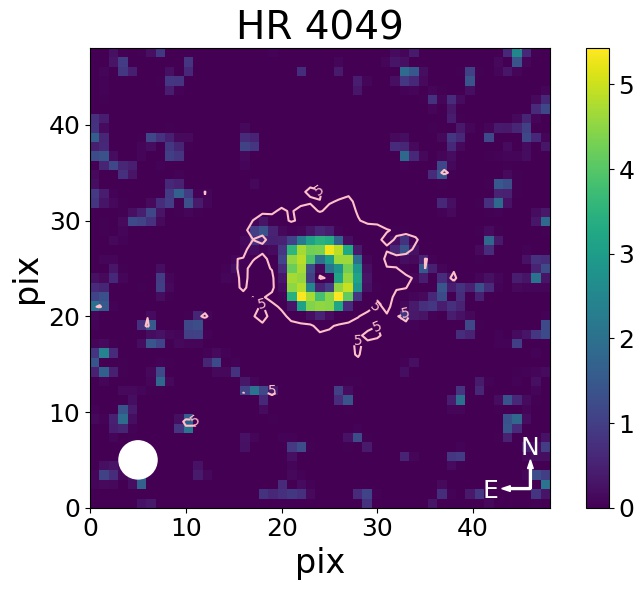}
    
    \includegraphics[width=0.5\columnwidth]{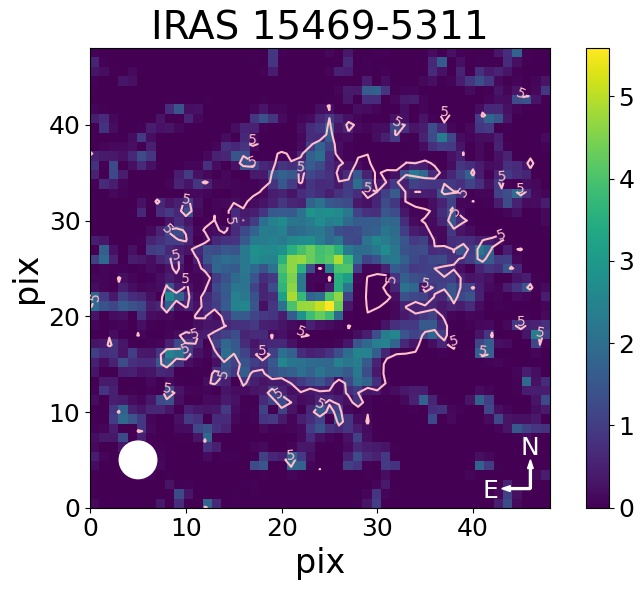}
    \includegraphics[width=0.5\columnwidth]{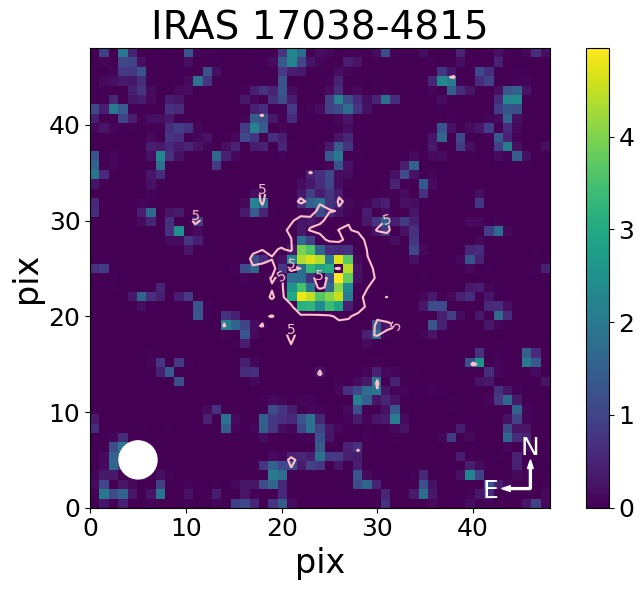}
    \includegraphics[width=0.5\columnwidth]{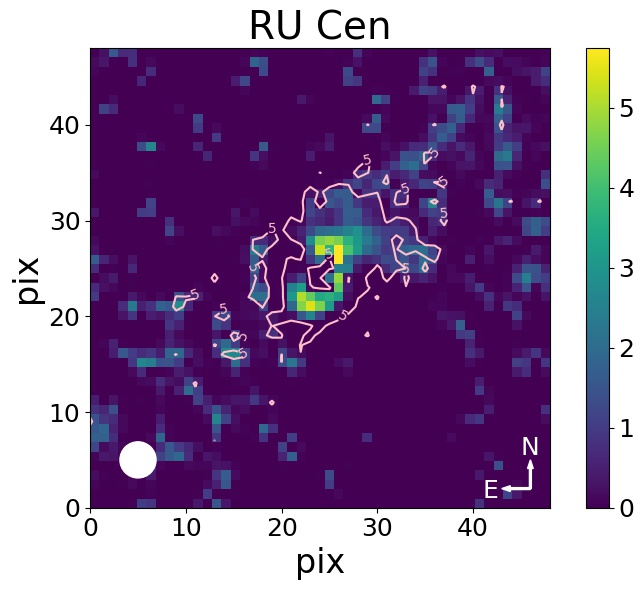}
    \includegraphics[width=0.5\columnwidth]{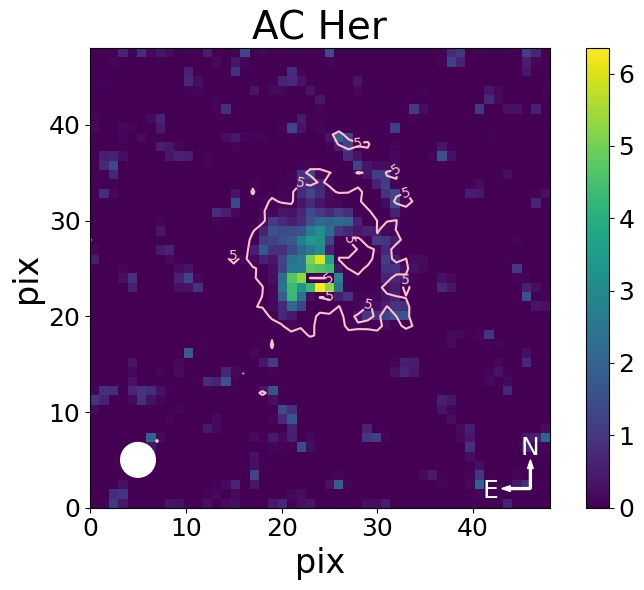}

    \caption{Deconvolved azimuthally polarized intensity (Q$_\phi$) for all targets in the sample. Solid line contours correspond to SNR$=5$, and the white circles represent the size of the resolution element (see Appendix~\ref{sec:snr}). The low intensity of the central 4x4 pixel region of each image is a reduction bias caused by over-correction of the unresolved central polarization (see Section~\ref{sec:unresolved}).}
    \label{fig:snr}
\end{figure*}

\section{Analysis of the SPHERE/IRDIS images and Results}
\label{sec:analysis}

In this section, we present the analysis of reduced linearly polarized images and briefly discuss the results for individual targets in our sample. The main analysis procedure includes determining the disc orientation, defining the relative disc brightness in polarized light, analysing azimuthal and radial brightness profiles, and characterising the most prominent substructures. We note that in our analysis,  we rely on the region of the azimuthally polarized image with signal-to-noise ratio (SNR) $\geq5$ (see Fig.\ref{fig:snr}). Details of the estimation of the SNR and the size of the resolution element are presented in Appendix~\ref{sec:snr}.

We continue to use IRAS\,08544-4431 as an example to illustrate the results from the analysis procedure. The results for all objects in the target sample are presented in Section~\ref{sec:individual_study} and Appendix~\ref{sec:ap_brprof}.

\subsection{Disc orientation}
\label{sec:orientation}

The deconvolved Q$_\phi$ images for seven out of the eight targets (i.e., all targets except AC\,Her) show a clear bright 'ring' (see Fig.~\ref{fig:snr}). These rings correspond to the closest resolved circumbinary disc part to the central binary \footnote{Although we refer to the structure observed in the SPHERE/IRDIS polarimetric images as a 'ring', it is important to note that the 'inner rim' in the polarimetric images does not correspond to a physical circumbinary disc structure. The size of the dust inner rim in the post-AGB disc is too small to be resolved by the SPHERE instrument. Instead, the 'ring' shape of the closest resolved portion of the disc is caused by over-correction of unresolved polarization during the DR (see Section \ref{sec:unresolved}).} and can be used to estimate the position angle (PA) and inclination of the disc, thereby providing insight to the disc orientation. We note that AC\,Her (discussed in Section~\ref{sec:individual_study}) does not show the presence of an elliptical ring and displays a more complex structure relative to the other targets.

\begin{figure*}
    \includegraphics[width=0.5\columnwidth]{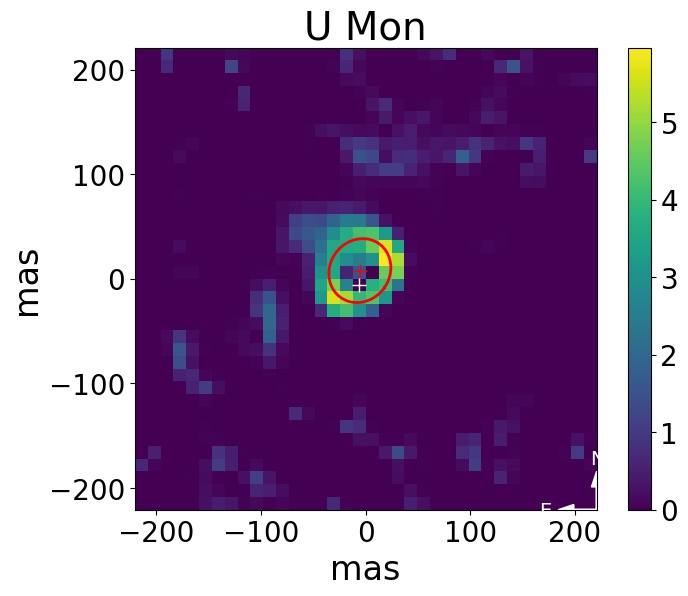}
    \includegraphics[width=0.5\columnwidth]{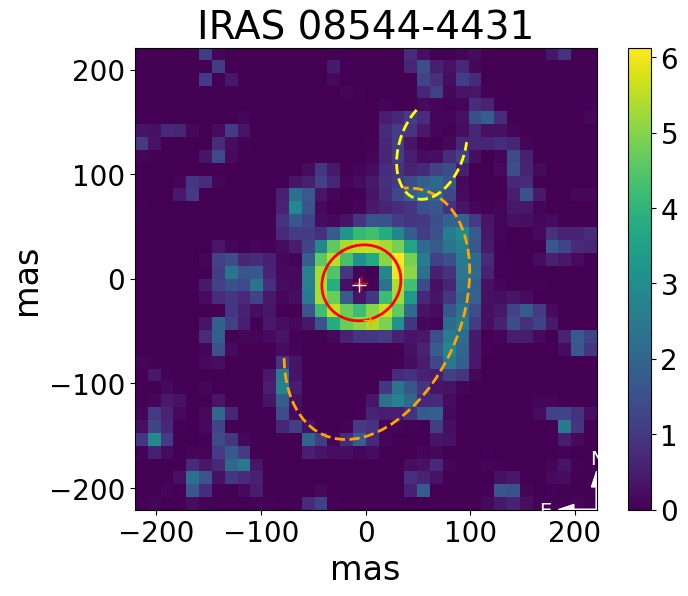}
    \includegraphics[width=0.5\columnwidth]{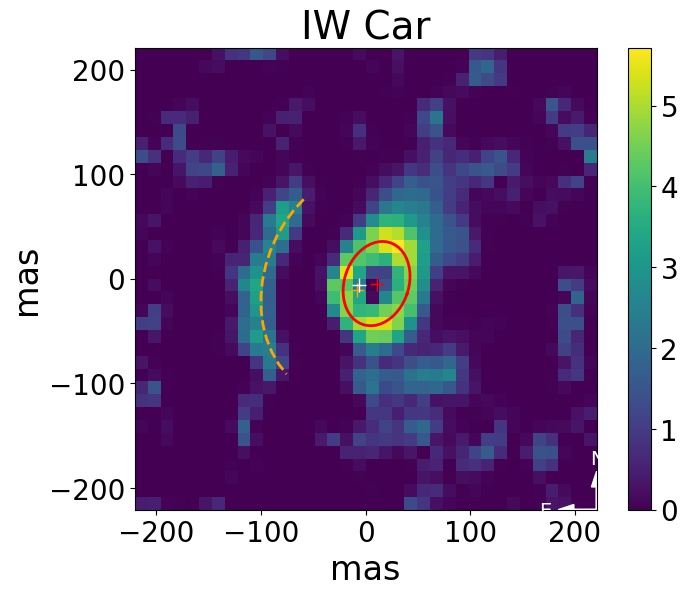}
    \includegraphics[width=0.5\columnwidth]{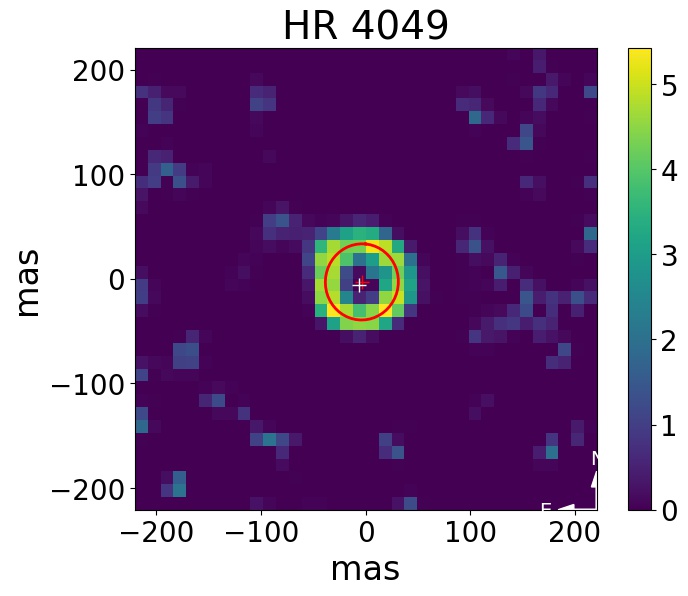}
 \includegraphics[width=0.5\columnwidth]{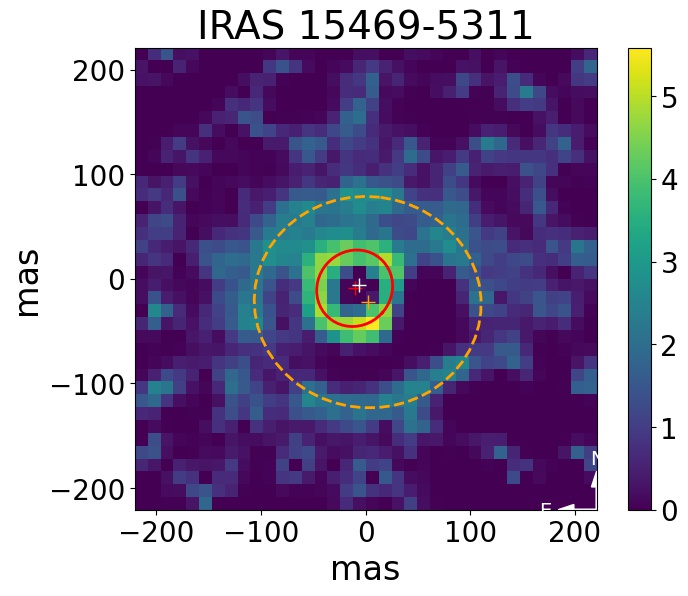}
    \includegraphics[width=0.5\columnwidth]{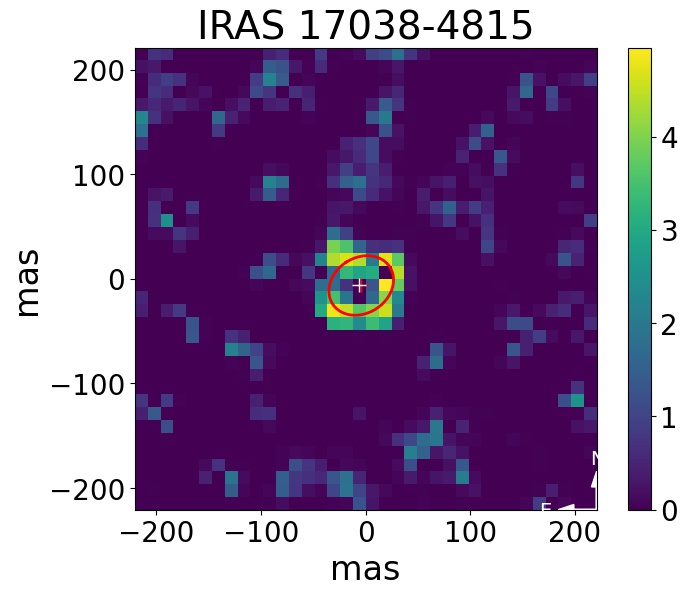}
    \includegraphics[width=0.5\columnwidth]{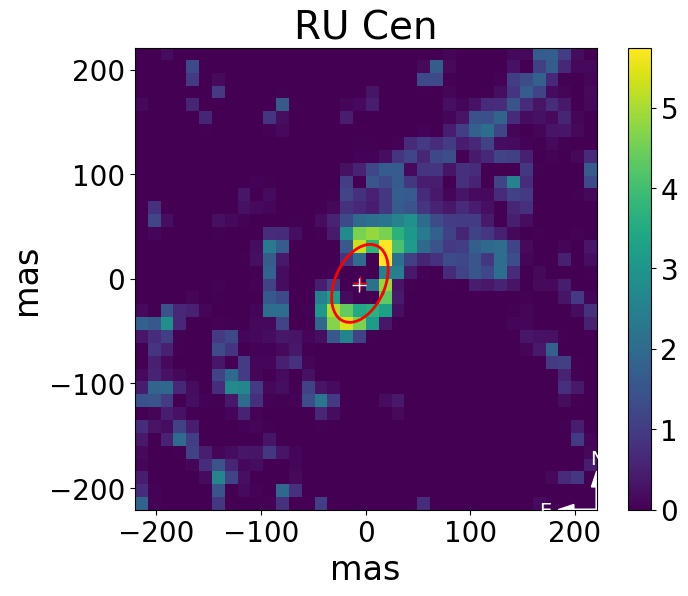}

    \caption{Disc orientation results based on the Q$_\phi$ polarized intensity images for seven out of eight target post-AGB binaries (except AC\,Her, see Section~\ref{sec:orientation}). The red ellipses illustrate the most plausible PA and inclination of discs (see Section~\ref{sec:orientation}), while dashed orange and yellow lines highlight significant substructures (see Section~\ref{sec:substructures}). The red cross in the centre of the images represents the centre of the fitted ellipse, while the white cross represents the position of the binary. The low intensity of the central 4x4 pixel region of each image is a reduction bias caused by over-correction of the unresolved central polarization (Section~\ref{sec:unresolved}). Note: all images are presented on an inverse hyperbolic scale.}
    \label{fig:ellipsefit}
\end{figure*}

To estimate the PA of the disc, we used the deconvolved Q$_\phi$ images. Firstly, we determined the positions of peak brightness along the 'ring' by fitting a Gaussian curve to the image profiles going from the central binary outwards (360 profiles). We then fitted ellipses to the resulting coordinates using the function $EllipseModel$ from $skimage.measure$ python package, based on the algorithm proposed by \citet{Halir98numericallystable}. The results are shown in Fig.~\ref{fig:ellipsefit}. As a result of this fitting, we derived the best values for the lengths of major and minor axes and the corresponding PA\footnote{The PA was defined as rising counterclockwise from the vertical axis (North) to the first principal radius (major axis). We note that our ellipse fitting procedure allows the $180^{\circ}$ ambiguity in the disc PA on the sky.} of the disc. To estimate the errors, we repeated the ellipse fitting procedure using values of 0.5 to 1.5 times the best fitting values of major and minor axes and $\pm$30 degrees around the PA. The final error bars were chosen based on the ranges of the ellipse parameters in which the sum of squared residuals was less than twice the best one.

While some of the targets in our study show eccentric orbits \citep[][see Appendix~\ref{sec:ap_sample}]{Oomen2018}, we can assume that the circumbinary disc itself should have a circular inner rim. Therefore, we can interpret the visible elliptical disc shape as an effect of the projection of the circular disc on the field of view and hence estimate the corresponding inclination using:
\begin{equation}
    \cos{i}=(\frac{b}{a}),
	\label{i}
\end{equation}
where $a, b$ are major and minor half axes of the ellipse.

Additionally, we determined the eccentricity $e$ of the 'ring' using:
\begin{equation}
    e=\sqrt{1-\frac{b^2}{a^2}}.
	\label{e}
\end{equation}

We present the most plausible positions and orientations of circumbinary discs in Fig.\ref{fig:ellipsefit} and Table~\ref{tab:fit}. We note that the correction of unresolved central polarization that we performed (Section~\ref{sec:unresolved}) affects the accuracy of our results as it might induce slight circularization of the resolved disc surface by over-subtracting fainter pixels. Thus, the inclination and eccentricity values that we tabulated in Table~\ref{tab:fit} form only lower limits of inclination for each system. However, without such a correction, we would be unable to detect any disc structures for post-AGB systems with a bright unresolved component. 

\begin{table*}
    \caption{Derived disc properties for all targets in the sample.}
    \begin{center}
            
        \begin{tabular}{ |l|c|c|c|l|c|c|c }
            \hline
        \#ID& Name & \begin{tabular}[c]{@{}l@{}}
        $a$ \\ {[}mas{]}\end{tabular} &  \begin{tabular}[c]{@{}l@{}}$b$\\ {[}mas{]}\end{tabular}&
        \begin{tabular}[c]{@{}l@{}}
        $i$ \\  {[}$^\circ${]}
        \end{tabular} & \begin{tabular}[c]{@{}l@{}}
        $PA$ \\  {[}$^\circ${]}
        \end{tabular}& $e$ & \begin{tabular}[c]{@{}l@{}} Percentage of  polarised light \\on different substructures\end{tabular} \\
            \hline
    		\multicolumn{8}{c}{Full discs}\\
            \hline
            1 & U\,Mon$^{*}$ & 31.6$^{+3}_{-2.5}$& 28.6$^{+2.4}_{-2.7}$& 25$^{+14}_{-18}$ & 144$^{+10}_{-15}$ & 0.42&80.1; 5.1\\
            2 & IRAS\,08544-4431 &38.6$^{+1.2}_{-1.2}$ & 35.5$^{+1.5}_{-1.4}$& 23$^{+7}_{-15}$ & 121$^{+27}_{-30}$ & 0.39& 78.9; 6.8; 2.2\\
            3& IW\,Car & 41.2$^{+3.3}_{-2.6}$ & 30.7$^{+2}_{-2}$& 41.7$^{+6.5}_{-7.7}$ & 161$^{+12}_{-10}$ & 0.67 & 72; 7.5; 4.5; 3.6; 1.6\\
            4 & HR 4049 & 36.5$^{+2.3}_{-1.7}$ & 34.9$^{+2}_{-2.3}$& 16.8$^{+14}_{-14}$& 174 $^{+28}_{-30}$ & 0.29&--\\
            
            4$^{\dag}$ & HR 4049 &  & & 49$^{+3}_{-3}$& 63$^{+7}_{-6}$ & &\\
            5&IRAS 15469-5311 & 37.5$^{+3.6}_{-2.4}$& 35.4$^{+2.6}_{-3.9}$& 19.6$^{+18}_{-16}$ & 139$^{+27}_{-30}$ & 0.33& 53.4; 26\\
            6&IRAS 17038-4815 &  32.6$^{+12.5}_{-6.8}$ & 26.5$^{+5.6}_{-8.5}$& 35.7$^{+29}_{-29}$ & 123$^{+28}_{-30}$ & 0.58& --\\
            \hline
    		\multicolumn{8}{c}{Transition discs}\\
            \hline
            7& RU\,Cen & 39.4& 23.9& 52.8& 156 & 0.80&--\\
            8 & AC\,Her & -- & -- & -- & -- & -- & 74.5; 13.1; 1.7\\
            \hline
        \end{tabular}
    \end{center}
    \begin{tablenotes}
     \small
    \item \textbf{Notes:} $a$ and $b$ represent the major and minor half-axes of the disc in units of mas, $i$ indicates the inclination, $e$ represents the eccentricity. The position angle ($PA$) is presented in degrees and rises counterclockwise from the vertical axis (North) to the first principal radius (major axis). Due to the limited number of pixels with a significant SNR, we could only provide a rough estimation of the parameters for RU\,Cen (see Section~\ref{sec:individual_study}). Therefore, we did not specify any uncertainties for these measurements. \\ $^{*}$ indicates that we use the mean combined result of two independent observations  of the U\,Mon (see Section~\ref{sec:observations}) to improve the signal-to-noise ratio. More details on the tabulated information can be found in Section~\ref{sec:orientation} and Section~\ref{sec:morphology}. \\
    $^{\dag}$  represents the inclination and PA values defined from the IR interferometric data \citep{Kluska2019A&A...631A.108K}. We adopt these values for the rest of the paper (see Section~\ref{sec:individual_study} for more details). 
    \end{tablenotes}
    \label{tab:fit}

\end{table*}

\subsection{Extended disc morphology}
\label{sec:morphology}

In this section, we gain insights into the extended disc morphology (with a radius of $\sim$20--200\,mas) of the individual objects by measuring their polarized disc brightness relative to the total intensity of the system and their brightness profiles. 

\subsubsection{Measuring the polarized disc brightness}
\label{sec: polarised_to_tot}

To measure the polarized disc brightness relative to the total intensity of the system, we calculated the ratio of the resolved polarized emission from the disc to the total unpolarized intensity of the target. The polarized flux and stellar intensity were measured in a circular area centred at the binary position and limited by 185\,mas. To avoid additional biasing, we used the total polarized and unpolarized images before the deconvolution with PSF (see Section~\ref{sec:deconvolution}). The resulting polarized disc brightness values form a lower limit because, during the DR process, a part of the signal from the disc was subtracted while correcting for the unresolved central polarization (Section\ref{sec:unresolved}). An upper limit can be estimated by including the unresolved polarized signal, keeping in mind that the interstellar material can partially induce this signal. We present the lower and upper limits of polarized disc brightness in Table~\ref{tab:polarisedtototal}.

The estimated polarized disc brightness is independent of the stellar luminosity and distance to the post-AGB binary, as both the stellar light and the disc intensity exhibit similar scaling with distance. Therefore, we can assume that all variations in the amount of the incident starlight reflected by the disc surface are caused only by the disc orientation, morphology and dust properties.

We identified that the brightest resolved discs in the sample (AC\,Her and IW\,Car, see Table.~\ref{tab:polarisedtototal}) are not the most extended sources in the polarized light images (see Fig.~\ref{fig:snr}). This is likely because these systems show a high inclination of the disc \citep[50$^\circ$ for AC\,Her,][and 41.7$^\circ$ for IW\,Car, see Table~\ref{tab:fit}]{Hillen2015A&A...578A..40H}, which might dim the host star and cause a decrease in stellar brightness thereby making the disc relatively brighter.

\begin{table}
	\begin{center}
    	\caption{The polarised disc brightness relative to the total intensity of the system.}
    	\label{tab:polarisedtototal}
    	\begin{tabular}{clcc} 
    		\hline
    		 \#ID& Name & Lower limit & Upper limit \\
    		\hline
    		\multicolumn{4}{c}{Full discs}\\
    		\hline
            1 & U\,Mon$^{*}$ & 0.37\%&1.58\%\\
            2 & IRAS\,08544-4431& 0.58\%&1.30\%\\
            3 & IW\,Car & 0.95\%&1.93\%\\
            4 & HR 4049 & 0.39\%&0.79\%\\
            5&IRAS 15469-5311 &0.36\%&1.60\%\\
            6&IRAS 17038-4815 & 0.31\%&2.24\%\\
            \hline
    		\multicolumn{4}{c}{Transition discs}\\
            \hline
            7 & RU\,Cen & 0.46\%&1.32\%\\
            8 & AC\,Her & 1.28\%&1.94\%\\
            \hline
    	\end{tabular}
    \end{center}
	\begin{tablenotes}
    \small
    \item \textbf{Notes:} The lower limit of the polarised disk brightness includes only resolved emission from the disc, while the upper limit also includes unresolved central polarisation (see Section~\ref{sec: polarised_to_tot}) for more details. We note that $^{*}$ indicates that we use the mean combined result of two independent observations for U Mon (see Section~\ref{sec:observations}) to improve the signal-to-noise ratio.\\
    \end{tablenotes}
\end{table}

\subsubsection{Brightness profiles}
\label{sec:profile}

To investigate the complexity of the disc, we calculated brightness profiles which reveal the spatial distribution of the polarized intensity of the disc. We calculated three types of brightness profiles: linear, radial and azimuthal. The linear brightness profiles reflect the disc symmetry along the major and minor axes of the disc. The radial brightness profile represents the relationship between the scattered polarized emission and the distance from the central binary. The azimuthal brightness profile reflects the azimuthal brightness asymmetries of the closest resolved circumbinary disc part to the central binary. 

\begin{figure}
    \includegraphics[width=\columnwidth]{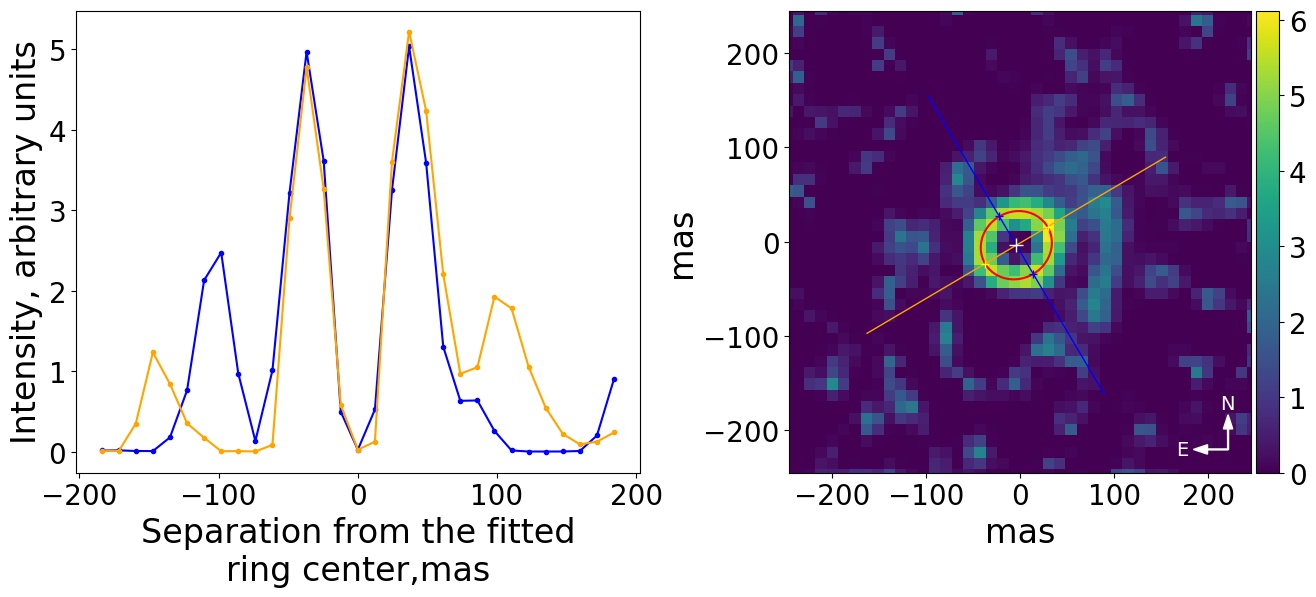}
    \includegraphics[width=1\columnwidth]{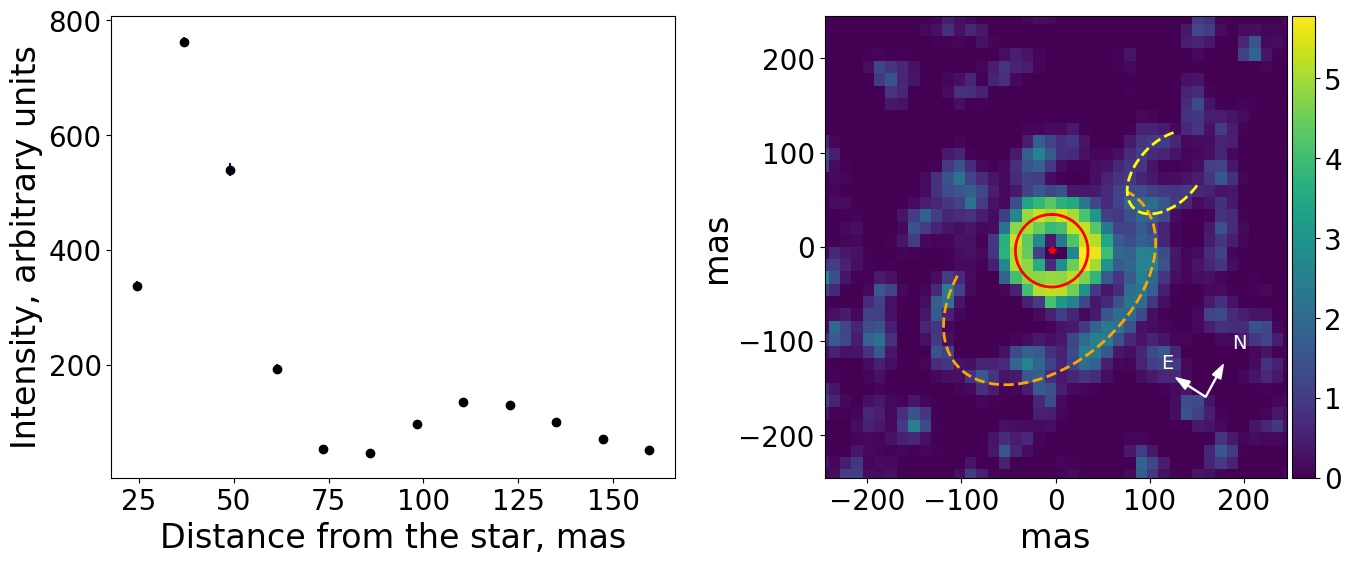}
    \includegraphics[width=1\columnwidth]{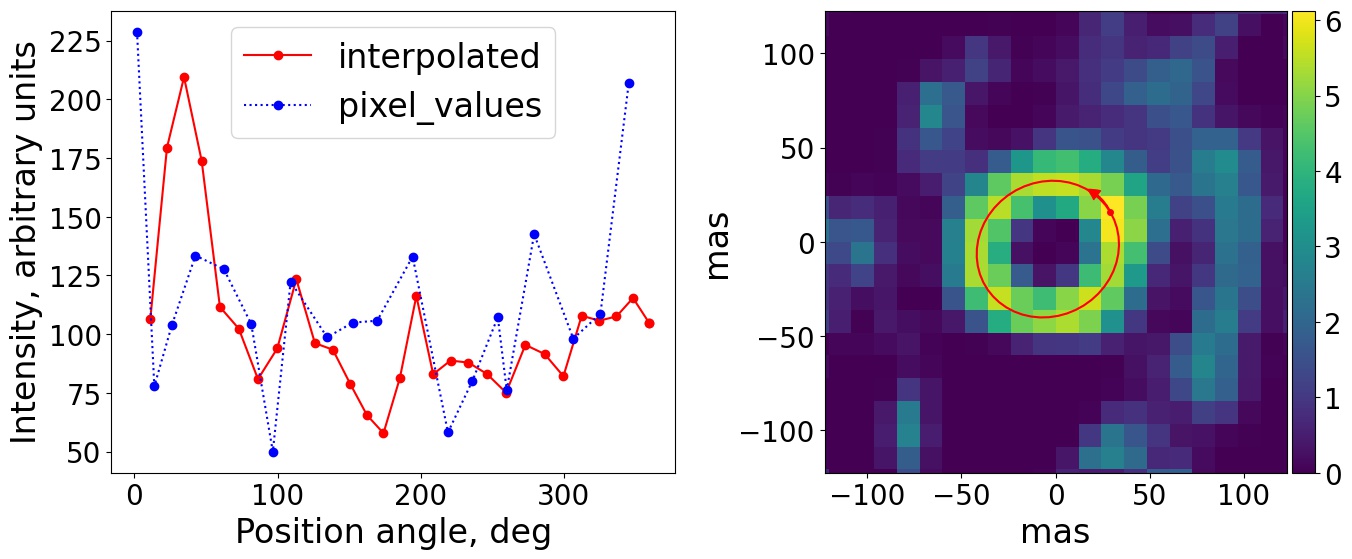}
    
    \caption{Illustration of the brightness profiles for IRAS\,08544-4431. The top row shows the linear brightness profiles of Q$_\phi$ image along the major and minor axes of the fitted ellipse. The left image of the top row represents the profile on a logarithmic scale, and the right image represents positive Q$_\phi$ intensity and positions of axes. The middle row shows the radial brightness profile of the deprojected Q$_\phi$ image. The left image of the middle row represents the profile on a linear scale, and the right image represents deprojected Q$_\phi$ image. The bottom row of the image represents the azimuthal brightness profile of Q$_\phi$ image. The left image of the bottom row shows the profile, where the blue line represents original pixel values, and the red line represents Bi-linearly interpolated intensities. The red dot and arrow in the bottom right image mark the starting point and the direction of the azimuthal brightness profile evaluation. The right column is presented on an inverse hyperbolic scale. The low intensity of the central 4x4 pixel region of each image in the right column is a reduction bias caused by over-correction of the unresolved central polarization (see Section~\ref{sec:unresolved}). Similar images for other targets are presented in Appendix~\ref{sec:ap_brprof}.}
    
    \label{fig:brprofile}
\end{figure}

To calculate the linear brightness profile, we performed Bi-linear interpolation for 36 points ($\sim$1 point per pixel) equally distributed in Q$_\phi$ image along the major or minor axis of the disc. We limited the length of the profile by the separation of 180 mas from the central binary. The positions of the axes were determined by fitting an ellipse to the bright 'ring' (see Section~\ref{sec:orientation}). The resulting linear brightness profiles are presented in the top row of Fig.\ref{fig:brprofile} for IRAS\,08544-4431 and in Fig.~\ref{fig:linearprof} for the other seven systems in the sample. For AC\,Her, we specified the positions of the axis manually, as the target does not show a clear 'ring' structure, and we could not precisely determine the positioning (see Fig.~\ref{fig:snr}).

To calculate the radial brightness profile, we estimated the mean brightness per pixel for radially tabulated annuli of the Q$_\phi$ image. We accounted for the impact of disc orientation on the radial intensity distribution by reconstructing how each disc might look from the face-on position (deprojecting). The deconvolved Q$_\phi$ image (Section~\ref{sec:deconvolution}) of each target was rotated to align the major axis with the horizontal line, and then the image was stretched along the minor axis with $cos(i)^{-1}$ to deproject the disc image to a face-on orientation. The resulting deprojected images for six out of eight targets in the sample (except for AC\,Her and RU\,Cen) are presented in Fig.\ref{fig:brprofile} and Fig.~\ref{fig:radialprof}. AC\,Her and RU\,Cen do not show a sufficiently clear circular or elliptical disc surface, and, therefore, the deprojection cannot be applied. After deprojection, we multiplied the counts in the images by the square of the projected separation from the position of the host binary to compensate for the geometric dilution of the starlight before it is scattered by dust grains. Finally, we subdivided the image into radially tabulated annuli and calculated the mean brightness per pixel ($\sim$150 mas$^{2}$) for each annulus individually. The width of the annuli was increased with a radius proportional to $\sqrt{r}$ similar to what was performed for PPD polarimetric observations by \citet{Avenhaus2018ApJ...863...44A}. The described calculations were performed only for the statistically significant region of each Q$_\phi$ image based on the SNR (see Appendix~\ref{sec:snr}). The final radial brightness profile was plotted as the mean brightness of an annulus over the separation from the post-AGB binary. To estimate the noise, we used the variance in the same annuli of U$_\phi$ image as U$_\phi$ is mainly devoid of signal (see Section~\ref{sec:data_reduction} for details). We note that this method overestimates the noise in case of any astrophysical signal in the U$_\phi$ image. However, we can still use this methodology to set the upper limit. The resulting radial brightness profile of IRAS\,08544-4431 is presented in the middle row of Fig.\ref{fig:brprofile}, similar profiles for other targets are presented in Fig.\ref{fig:radialprof}.

For six out of eight targets in the sample (except IRAS\,08544-4431 and IRAS\,15469-5311), the radial brightness profiles show only the main peak that corresponds to the bright 'ring'. For IRAS\,08544-4431, we detected a clear second bump at the separation of ~115\,mas from the central binary. This secondary bump corresponds to the position of an arc-like substructure (see Fig.\ref{fig:ellipsefit}). In the case of IRAS\,15469-5311, we found a less bright but more spatially extended secondary bump.

To calculate the azimuthal brightness profile, we initially used the original pixel values along the ellipse fit in the $Q_\phi$ image (see Section~\ref{sec:orientation}). However, the resulting shape is biased by the pixel-to-pixel fluctuations. Therefore, we estimated the polarized intensity using Bi-linear interpolation of the original values of the four closest pixels at evenly distributed points along the elliptical fit. The profile starts from the eastern end of the major axis and goes counterclockwise. Similar to radial brightness profiles, azimuthal brightness profiles are calculated for six out of eight targets (except AC\,Her and RU\,Cen). 

As an illustration, the azimuthal brightness profile of IRAS\,08544-4431 is shown in the bottom row of Fig.\ref{fig:brprofile}. Similar profiles for other targets are presented in Fig.\ref{fig:azprof}.

\subsubsection{Detection of substructures}
\label{sec:substructures}

The SPHERE study by \citet{Ertel2019AJ....157..110E} revealed complex substructures in the circumbinary disc of the post-AGB binary system AR\,Pup. The authors attributed these substructures to the disc midplane shadow, a disc wind, outflows or jets originating from the central binary due to stellar wind or accretion. To detect and characterise possible substructures in the discs of our target sample, we followed a qualitative analysis as detailed below. 

\begin{figure*}
    \includegraphics[width=0.6\columnwidth]{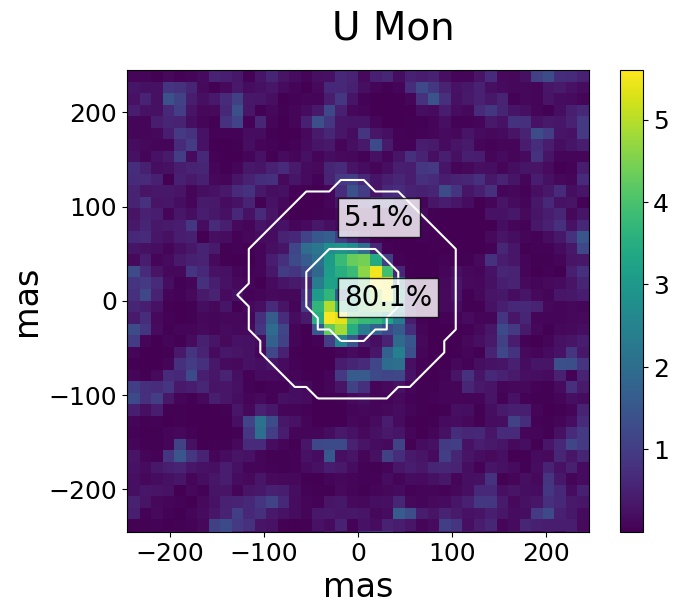}
    \includegraphics[width=0.6\columnwidth]{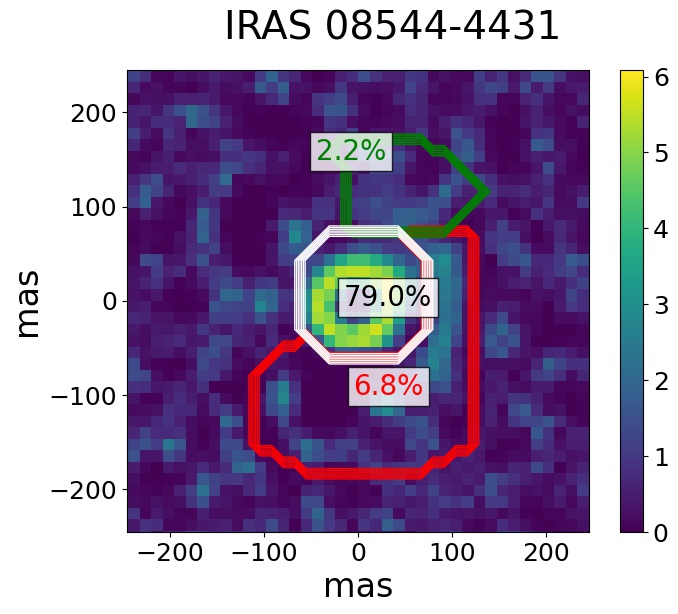}
    \includegraphics[width=0.6\columnwidth]{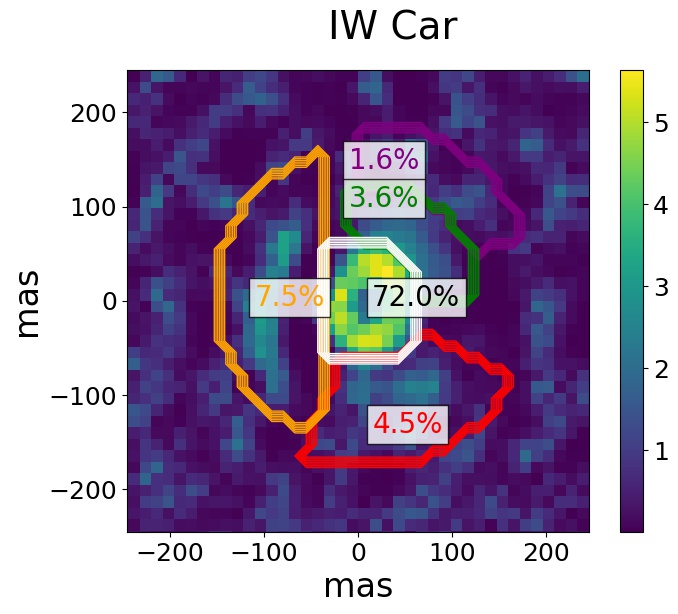}
    \includegraphics[width=0.6\columnwidth]{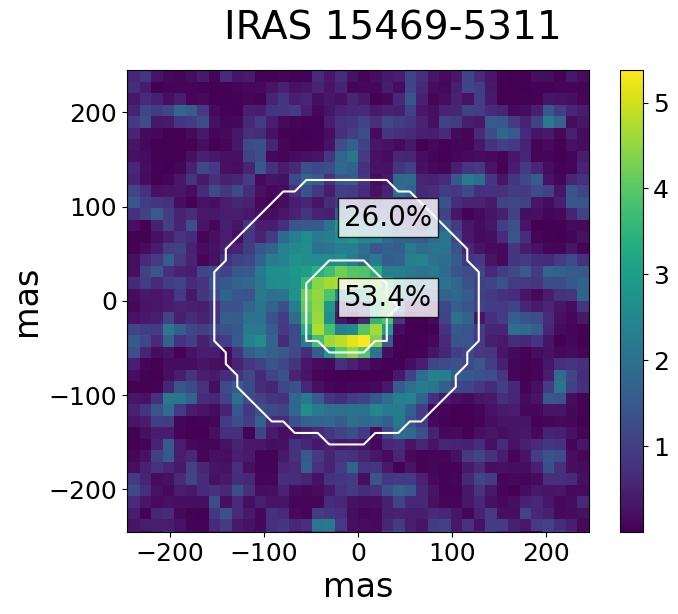}
    \includegraphics[width=0.6\columnwidth]{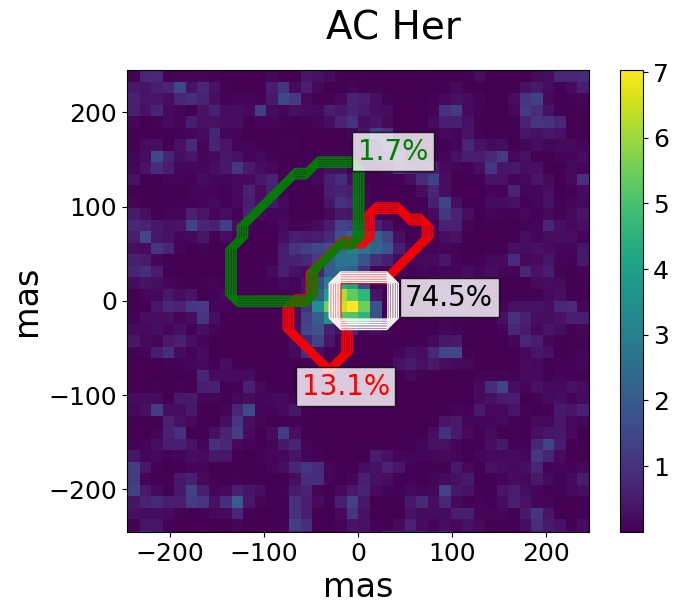}

    \caption{Percentage of total polarized intensity per resolved structure for IRAS\,08544-4431, IRAS\,15469-5311, IW\,Car, AC\,Her and U\,Mon (see Section~\ref{sec:substructures}) shown on an inverse hyperbolic scale. The low intensity of the central 4x4 pixel region of each Q$_\phi$ image is a reduction bias caused by over-subtracting of unresolved central polarization (Section~\ref{sec:unresolved}).}
    \label{fig:substructures}
\end{figure*}

Firstly, we used the calculated SNR to define the region of the Q$_\phi$ image with SNR $\geq5$ (see Appendix~\ref{sec:snr}), which reflects the statistically significant region of the polarimetric image for each object (see Fig.\ref{fig:snr}). Secondly, we calculated the percentage of total polarized intensity per resolved substructure excluding the unresolved central polarization (presented in Fig.~\ref{fig:substructures} and Table~\ref{tab:fit}), which provided the brightness of the disc substructures relative to other disc parts. Next, we examined the angle of linear polarization for the resolved substructures (AoLP, see Section~\ref{sec:data_reduction}), which reflects the orientation of the polarization vector in each region. In Fig.~\ref{fig:Aolp} we present the local angles of linear polarization over the resolved structures in the deconvolved Q$_\phi$ image for all targets in the sample. Since the polarization vectors are clearly centrosymmetric, we conclude that the residual noise in these regions is quite low, and the polarimetric signal is caused by the scattering of the stellar light on the disc surface. Finally, we combined the estimated SNR, the brightness of the disc substructures, and the orientation of the polarization vector to confirm that any features identified in our Q$_{\phi}$ and I$_{\rm pol}$ images (arcs, rings and gaps) are real.

Based on the above, we found that IRAS\,08544-4431, IW\,Car, IRAS\,15469-5311, AC\,Her show reliable substructures (see Fig.\ref{fig:substructures}). 

\subsection{Individual case study}
\label{sec:individual_study}

In this section, we present a comprehensive analysis of the SPHERE/IRDIS results from this study for each target star. Where possible, we also evaluate the results obtained within the framework of existing studies centred on geometrical modelling using near- and mid-IR interferometric observation.

\begin{figure}
    
    \includegraphics[width=0.5\columnwidth]{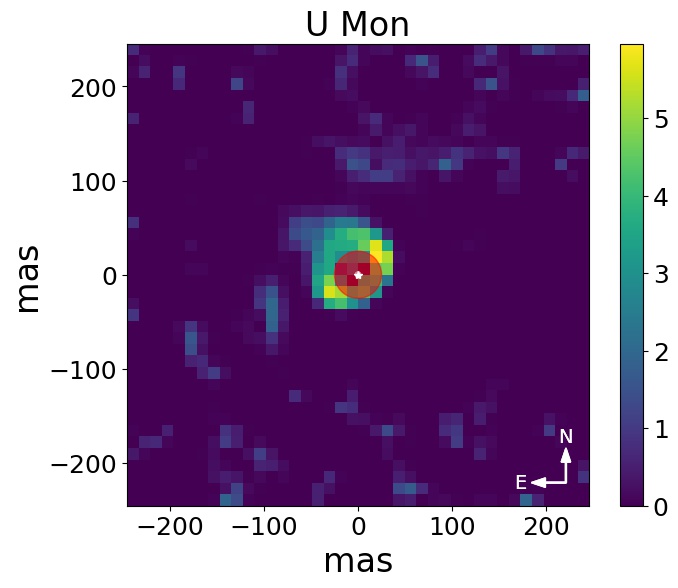}
    \includegraphics[width=0.5\columnwidth]{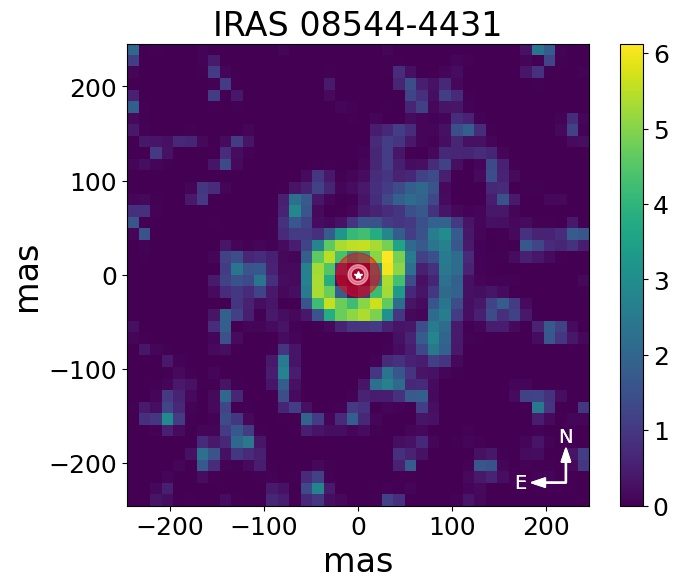}
    \includegraphics[width=0.5\columnwidth]{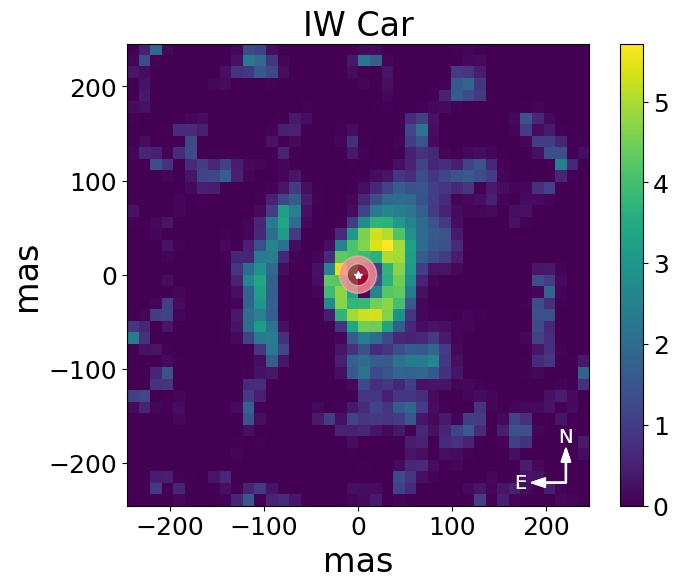}
    \includegraphics[width=0.5\columnwidth]{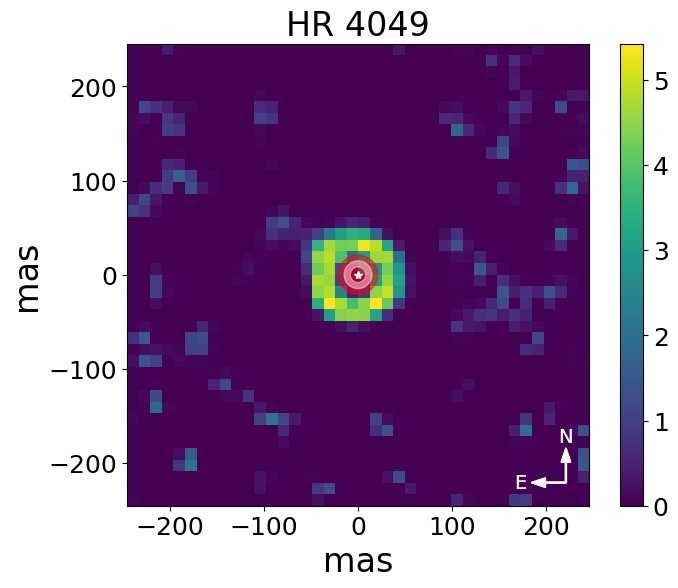}
    \includegraphics[width=0.5\columnwidth]{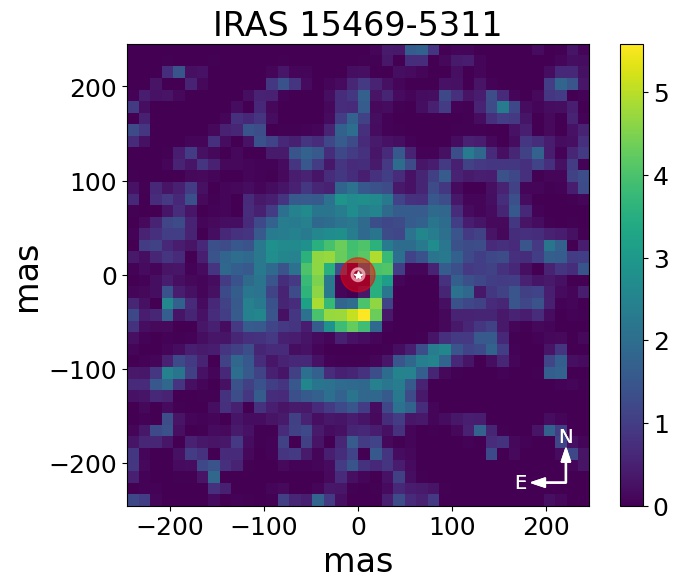}
    \includegraphics[width=0.5\columnwidth]{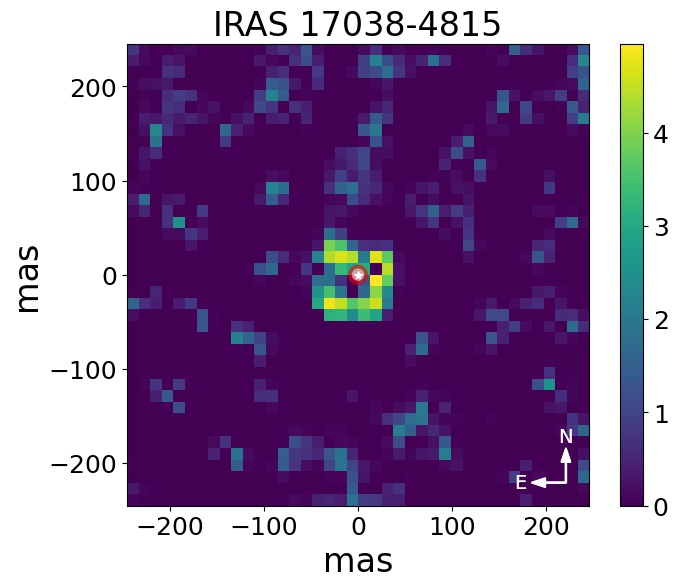}

    \caption{Comparison of the SPHERE/IRDIS polarimetric images (Q$_\phi$) with results from the geometric modelling of interferometric data for the full disc systems in our sample. The pink area represents the disc size obtained by \citet{Kluska2019A&A...631A.108K} for near-IR observations with VLTI/PIONIER instrument and the red area represents the disc size obtained by \citet{Hillen2017} for mid-IR with VLT/MIDI. The small white star indicates the position of the host binary. \label{fig:compar}}
\end{figure}

\subsubsection{Full discs}
\textit{\#ID 1: U\,Mon\footnote{Note: We observed U\,Mon twice on January 3 and 14, 2019 due to poor observing conditions during the first observation (see Section~\ref{sec:target_sel}). To increase the SNR, we separately reduced the data from each observation before combining the resulting images. For all further analysis, we use the mean combined polarimetric images.}}

We estimated the polarized disc brightness of U\,Mon (see Section~\ref{sec: polarised_to_tot}) to be relatively low among our sample, ranging from 0.37\% to 1.58\% of the total intensity of the target. U\,Mon is known to show a long-term periodic variation in mean magnitude with a period of $\sim2451$ days \citep[RVb phenomenon,][]{Kiss2017}, which is commonly explained by the presence of an inclined circumbinary disc that obscures the pulsating primary star at certain phases of its orbit \citep{Vega2017ApJ...839...48V, Manick2017, Kiss2017}. We argue that the relatively low value of the disc polarized brightness clearly indicates that the post-AGB star was not obscured at the time of observations, in agreement with the orbital phase for the observational date.

Based on SPHERE/IRDIS polarized observations of U\,Mon we found an inclined disc with a major axis of 32\,mas that complements previous studies \citep[Fig.~\ref{fig:compar},][]{Hillen2017, Kluska2019A&A...631A.108K}. However, we estimated a PA of 144$^\circ$ and an inclination of 25$^\circ$, which is twice smaller than the value obtained by geometric modelling using near-IR interferometric observations \citep{Kluska2019A&A...631A.108K}. It is likely that our obtained values are a lower limit due to the bias in the disc shape caused by the correction of unresolved central polarization during the DR (see Section~\ref{sec:data_reduction}). The idea of a higher inclination of U\,Mon circumbinary disc is also supported by the fact that the azimuthal dependence of the disc brightness shows two clear peaks at the sides of the major axis (see Fig.\ref{fig:azprof}). This phenomenon is commonly explained by the way an inclined circular disc is projected onto the sky (see Section~\ref{sec:profile}). Furthermore, both individual observations of U\,Mon (taken on the 3rd and 14th of January 2019) reveal arc-like structures (or 'outflows') on both the north-east and south-west sides that contain $\sim5\%$ of the total polarized light in the system (see Fig.\ref{fig:substructures}). It is possible that the presence of this 'outflow' causes an increase in the length of the minor axes of the projected disc resulting in an underestimation of inclination.

\textit{\#ID 2: IRAS\,08544-4431}

We estimated the polarized disc brightness of IRAS\,08544-4431 to be in the range of 0.58\% (in resolved emission) and 1.30\% (including the unresolved component) of the total intensity of the target (see Section~\ref{sec: polarised_to_tot}), which both are mid-values in the sample.

In the SPHERE/IRDIS $H$-band polarimetric data (see Fig.\ref{fig:ellipsefit}), IRAS\,08544-4431 shows a ring-like structure with an outer radius of 38.6\,mas and inclination of 23$^\circ$. Our estimated disc orientation agrees well with values obtained from IR interferometry data \citep[inclination of 19$^\circ$,][]{Hillen2016}. 

In addition to the ring-like structure, we detected two arc-like substructures that we qualitatively fit using the function $EllipseModel$ from $skimage.measure$ python package(see Fig.\ref{fig:ellipsefit}). The first extends from the northwest to the southwest, starting at a separation of about 85\,mas from the position of the binary (see Fig.\ref{fig:substructures}). This structure contains $\sim6.8\%$ of resolved polarized light from the disc and causes the second bump with a maximum at $\sim$115\,mas in the radial brightness profile (see middle panel of Fig.\ref{fig:brprofile}). The second extends to the northwest from the binary and contains $\sim2.2\%$ of the resolved polarized light.

\begin{figure}
   
    \includegraphics[width=0.49\columnwidth]{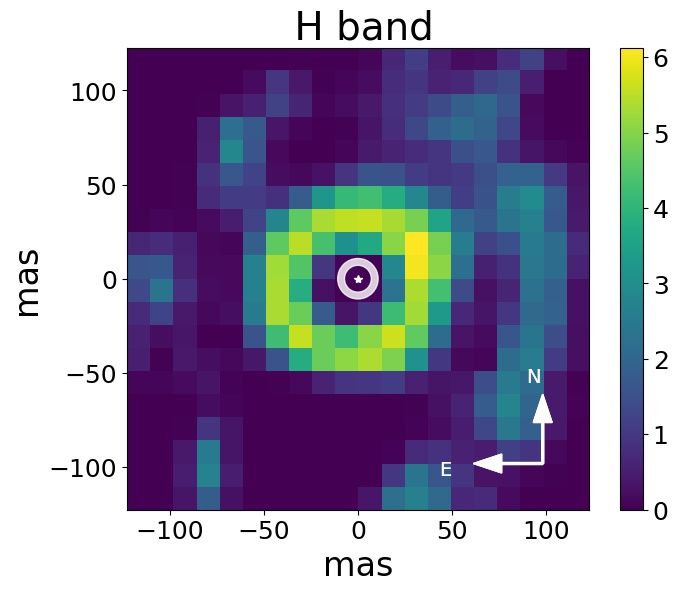}
    \includegraphics[width=0.49\columnwidth]{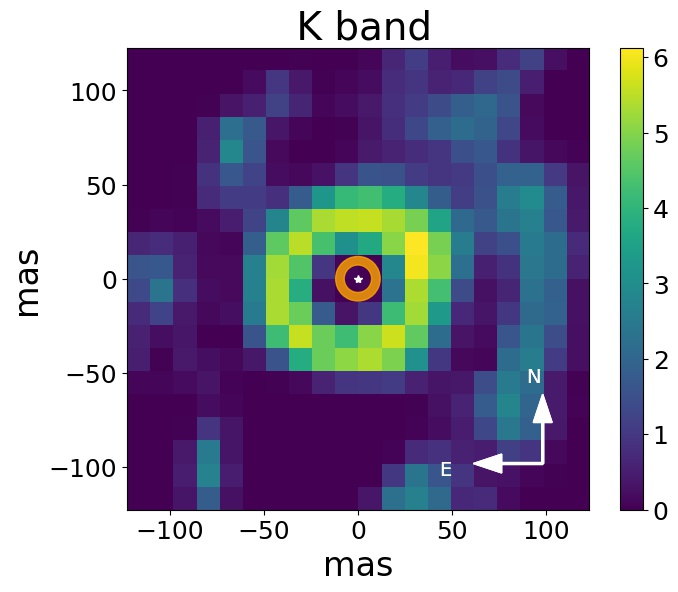}
    \includegraphics[width=0.49\columnwidth]{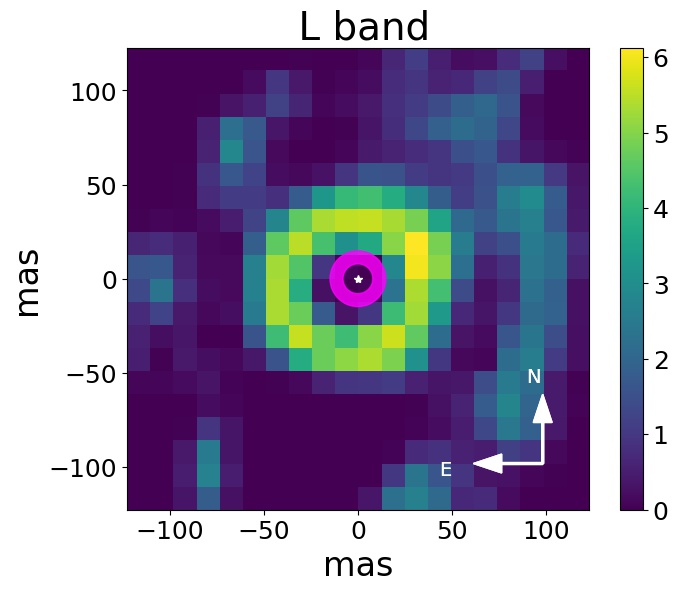}
    \includegraphics[width=0.49\columnwidth]{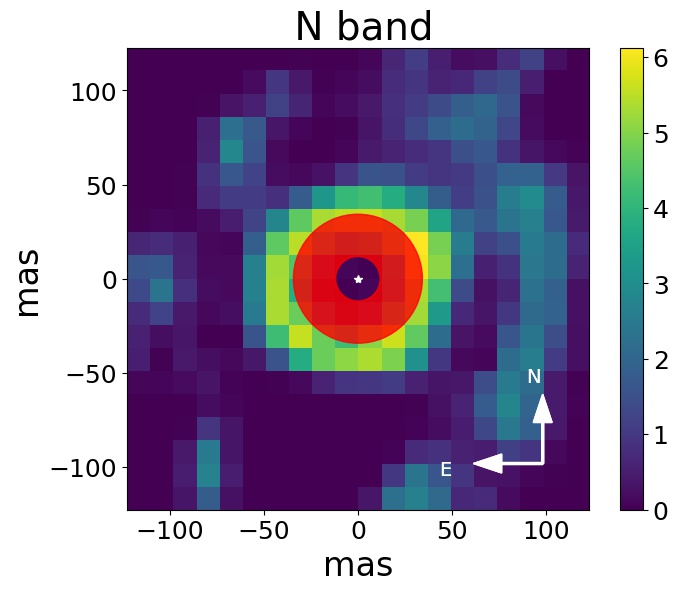}

    \caption{Comparison of our results on the disc structure of IRAS\,08544-4431 with established geometrical models, using VLTI/PIONIER (H band), VLTI/GRAVITY (K band), and VLTI/MATISSE (L, N bands) interferometric observations \citep{Corporaal2021A&A...650L..13C}. A small white star indicates the position of the central binary. See Section~\ref{sec:individual_study} for more details. \label{fig:corporaal}}
\end{figure}

We compared our results on the disc structure of IRAS\,08544-4431 with established geometrical models using VLTI/PIONIER ($H$-band), VLTI/GRAVITY ($K$-band), and VLTI/MATISSE ($L, N$-bands) interferometric observations \citep[see Fig.\ref{fig:corporaal},][]{Corporaal2021A&A...650L..13C}. The authors suggested a non-axisymmetric nature of the system. Our study complements their conclusion by revealing a complex disc structure in the polarimetric observations that cannot be replicated by a uniform disc model.

\textit{\#ID 3: IW\,Car}

IW\,Car shows high polarized disc brightness ranging from 0.95\% (in resolved polarized emission) to 1.93\% (with unresolved component) of the total intensity of the target. However, it might be misleading as IW\,Car exhibits an RVb brightness modulation. Therefore, high disc brightness might be caused by the obscuration of the primary star by the disc side at the moment of observation.

Based on the SPHERE/IRDIS polarimetric observations, we estimated the disc inclination of $42^\circ$ and PA of 161$^\circ$, which is in good agreement with IR interferometric study \citep[inclination of 45$^\circ$ and the PA of 155$^\circ$, ][]{Kluska2019A&A...631A.108K}. The azimuthal brightness profile of the disc (see Fig.\ref{fig:azprof}) shows two clear peaks confirming the high disc inclination, similar to U\,Mon.

The bright inner ring of the disc has a radius of about 41\,mas, and the arc is located on a separation of about 120\,mas to the east from the source centre. To highlight the arc, we qualitatively fit it using the function $EllipseModel$ from $skimage.measure$ python package (see Fig.\ref{fig:ellipsefit}). We note that the arc is separated from the bright ring by a dark shadow that can be interpreted as the mid-plane of the disc (see Fig.\ref{fig:ellipsefit} and Section~\ref{sec:disc_scale-height} for more details). In this case, the arc would be caused by the light scattering on the opposite disc side as was proposed for AR Pup system \citep[][]{Ertel2019AJ....157..110E}. The deprojected Q$_\phi$ image depicts the disc as viewed face-on, while the opposite disc side should not be visible. Therefore, we masked out the position of the arc in both the deprojected image and the radial brightness profile of IW\,Car (see Fig.\ref{fig:radialprof}).

In addition, we also detected an extended emission on the northwest and southwest from the bright ring that goes out to approximately 150\,mas ($\sim250$\,AU). This extended flux contains $\sim10\%$ of total polarized flux in the system while the inner disc and east-side arc have 72\% and 7.5\%, respectively.

\textit{\#ID 4: HR\,4049}

We estimated the disc brightness of HR\,4049 in polarized light to be in the range of 0.39\% (in resolved polarized emission) to 0.79\% (with unresolved component) of the total intensity of the target, which is the lowest value in the sample. 

HR\,4049 shows a clear bright 'ring' without any resolved substructures. Considering that the distance to HR\,4049 is comparable to other post-AGB binary systems (see Table~\ref{tab:initialdata}), the clear lack of the extended disc structure is likely to be intrinsic to the target. We discuss this in detail in Section~\ref{sec:discussion}. 

We obtained a disc inclination of $17^\circ$ and PA of $174^\circ$ (see Section~\ref{sec:orientation}), implying a nearly face-on disc orientation. However, this low inclination contradicts the observed RVb phenomenon of the HR\,4049 \citep{Waelkens1991A&A...242..433W} and also differs from the results of IR interferometric studies \citep[inclination of 49$^\circ$, PA of 63$^\circ$,][]{Kluska2019A&A...631A.108K}. Our obtained value of inclination is likely an underestimation due to the low number of pixels with significant SNR observed with IRDIS and over-correction of the unresolved central polarization during the DR (see Section~\ref{sec:data_reduction}). Therefore, for the calculation of the radial brightness profile (see Fig.~\ref{fig:radialprof}), we adopt the inclination and PA values determined by \citet{Kluska2019A&A...631A.108K}.

\textit{\#ID 5: IRAS\,15469-5311}

IRAS\,15469-5311 shows a polarized disc brightness ranging from 0.36\% (in resolved polarized emission) to 1.6\% (with unresolved component) of the total intensity of the target (see Table~\ref{tab:polarisedtototal}).

We estimated the inclination and PA of the disc to be $20^\circ$ and 139$^\circ$ (see Section~\ref{sec:orientation}). Our results differ from previously estimated values based on the interferometric data \citep[inclination of 53.5$^\circ$ and PA of 64$^\circ$, presented by][]{Kluska2019A&A...631A.108K}. In this case, we adopt the inclination value from our study since IRAS\,15469-5311 does not show the RVb phenomenon (implying that a low inclination value is plausible) and we resolved its extended disc structure with significant SNR (see Fig.~\ref{fig:substructures} and Table~\ref{tab:snr}).

IRAS\,15469-5311 shows two significant 'ring'-shape structures in the polarized scattered intensity. We found that the inner ring contains 53.4\% of resolved polarized emission and locates on the separation of $\sim38$\,mas from the central binary (see Fig.~\ref{fig:substructures}). The outer ring includes 26\% of resolved polarized light. We highlighted the confirmed outer ring using the ellipse fitting routine (detailed in Section~\ref{sec:orientation}). We found that the outer ring shows a radius of $\sim110$\,mas and a clear shift of $\sim18$\,mas from the stellar position. We discuss the possible origin of this shift in Section~\ref{sec:disc_scale-height}).

\textit{\#ID 6: IRAS\,17038-4815}

IRAS\,17038-4815 shows a resolved polarized disc brightness of 0.31\% of the total light from the target, which is the lowest value among the observed sample. However, it also shows the highest polarized brightness if we include the unresolved polarized emission in the measurement (2.24\%, see Table~\ref{tab:polarisedtototal}). The significant (over seven times) difference in the disc brightness with and without the unresolved component might result from a more compact and highly bright inner disc part that cannot be resolved with SPHERE/IRDIS. 

This system shows a clear bright 'ring' without any resolved substructures (see Fig.~\ref{fig:ellipsefit}). We estimated the disc inclination of $36^\circ$ and PA of $123^\circ$ (see Section~\ref{sec:orientation}), which coincides with the results from the interferometric study by \citet{Kluska2019A&A...631A.108K}. While the size of the disc is relatively small, its resolved shape is strongly affected by the DR (see Section~\ref{sec:data_reduction}), and therefore, the derived inclination should be treated with caution.

\subsubsection{Transition discs}

\textit{\#ID 7: RU\,Cen}

We defined that resolved polarized emission from the disc contains at least 0.46\% of the total light of the system. We also determined the upper limit of disc brightness to be 1.32\% (with unresolved component) of the total intensity of the target (see Section~\ref{sec: polarised_to_tot}).

We estimated a disc inclination of $53^\circ$ and PA of $156^\circ$. We note that we could only provide a rough estimation for both parameters because of the low number of pixels with significant SNR. However, the defined disc orientation coincides with results from the recent mid-IR interferometric study using VLTI/MATISSE and VLTI/MIDI instruments \citep[which resulted in an inclination of 68$^\circ$ and PA of $\sim$160$^\circ$,][]{Corporaal2023arXiv230412028C}.

We note that while our DR procedure allows us to detect faint disc substructures (see Fig.~\ref{fig:ellipsefit}), it also over-subtracts signal in a circular region of 2-3 pixel radius centred on the binary star (Section~\ref{sec:data_reduction}). Therefore, for highly inclined systems such as RU\,Cen, it causes uncertainty in disc orientation which significantly affects the disc brightness in polarized emission and prevents the accurate determination of radial and azimuthal brightness profiles. 

\textit{\#ID 8: AC\,Her}
\label{sec:ac_her}

We found the polarized disc brightness of AC\,Her to be in the range of 1.28\% (in resolved emission) and 1.94\% (including the unresolved polarization) to the total intensity of the target (Section~\ref{sec: polarised_to_tot}).

\begin{figure}
    
    \includegraphics[width=1.\columnwidth]{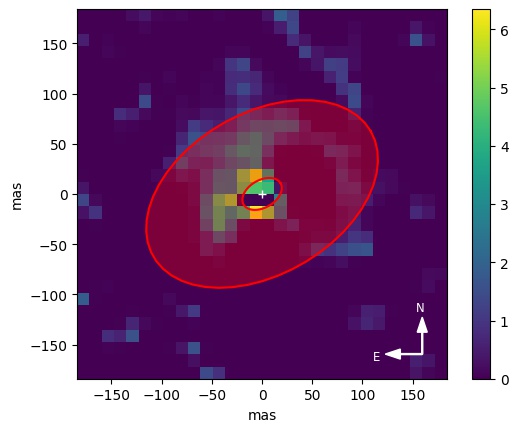}
    \caption{Comparison of SPHERE/IRDIS observations (Q$_\phi$ in $Y$-band and the disc model determined by the radiative transfer modelling for AC\,Her (in red) obtained by \citet{Hillen2015A&A...578A..40H} using VLTI/MIDI ($N$-band) instrument. See Section~\ref{sec:ac_her} for more details. \label{fig:ac_her_hillen}}
\end{figure}

AC\,Her shows the most complex and asymmetrical disc structure amongst all objects in this study (see Fig.\ref{fig:snr} and Fig.\ref{fig:I_pol}). However, we could not estimate the inclination and PA for the disc because we could not resolve a clear bright 'ring'. In Fig.~\ref{fig:substructures}, we highlighted three regions of the resolved disc based on the separation from the central binary. We found that the central region of the image with a separation up to 30\,mas contains 75.4\% of resolved polarized emission while others contain 13.1\% and 1.7\%. In Fig.\ref{fig:ac_her_hillen}, we provide a comparison of the obtained resolved disc structure for AC\,Her with the results of the radiative transfer modelling of mid-IR interferometric data \citep{Hillen2017}.

We note that the asymmetric disc morphology seen with SPHERE/IRDIS and a big inner cavity observed with mid-IR interferometry (see Section~\ref{sec:target_sel}) might be caused by still unknown disc-binary interaction mechanism (see Section~\ref{sec:origin_of_morph}).

\section{Discussion}
\label{sec:discussion}

In this section, we interpret the obtained results with the aim of better understanding the extended disc’s morphology. We do this by: i) revealing the scale-height of the post-AGB circumbinary disc, ii) investigating the possible origin of the over-resolved flux detected in near-IR interferometry \citep{Kluska2019A&A...631A.108K} using the results of SPHERE/IRDIS polarimetric imaging for all targets, iii) exploring the connection between circumbinary discs around post-AGB/RGB binaries and PPDs around YSO and iiii) investigating the diversity of disc morphologies of full disc post-AGB systems observed with SPHERE/IRDIS.

\subsection{Scale-height estimates of post-AGB circumbinary discs}
\label{sec:disc_scale-height}

SPHERE high-resolution polarimetric imaging is an essential tool to reveal the morphology of the circumbinary disc. This technique enables the investigation of the scattered light emission from the surface layers of the disc. Therefore, using SPHERE observations, we are able to directly measure the disc scale-height and investigate the flaring\footnote{A flaring disc is a disc in which the ratio of disc thickness to the distance to the star, H/r, increases with r; then, any point at the surface of such discs receives direct light from the star, and the disc intercepts a substantial part of the stellar radiation out to large distances \citep{Lagage2006Sci...314..621L}} of the disc surface.

 In the case of a highly inclined disc with inclination >40$^{\circ}$, e.g., IW\,Car (see Fig.~\ref{fig:ellipsefit}), the polarized SPHERE/IRDIS images reveal the clear substructure separated from the central bright 'ring' by a dark layer band. The dark layer band can be interpreted as the optically thick disc mid-plane while the substructure and the central bright 'ring' represent two disc surfaces scattering the starlight \citep{Ertel2019AJ....157..110E}. The scale-height of the disc can be estimated by directly measuring the width of the disc mid-plane. Using the width of the dark layer band in the polarized intensity image and the predetermined inclination of IW\,Car (see Section~\ref{sec:orientation}), we estimated the height of the disc to be $\sim90$\,AU.

Additionally, we can estimate the disc scale-height by analysing the visible offset of disc structures along the minor axis of the disc. When the observed disc surface is located above the mid-plane, the projection on the field of view results in the visible offset of any elliptical structure from the location of the binary \citep{deBoer2016A&A...595A.114D}. This method is widely used in protoplanetary disc studies \citep{Benisty2022arXiv220309991B} and is especially effective for discs with multiple rings \citep{deBoer2016A&A...595A.114D}. However, we note that this method assumes a circular disc centred on the binary's position. For any visible offset of the 'ring' centre $u$ from the host star position in our target sample, we estimated the height of the scattering surface based on the predetermined inclination of the system using:
\begin{equation}
\label{eq:height}
    H=\frac{u}{\sin{i}}.
\end{equation}
 
We did not detect any significant offset (more than 1 pixel) for the bright central 'ring' structures of all targets. However, we found a clear offset of the outer 'ring' of IRAS\,15469-5311 (see Fig.~\ref{fig:ellipsefit}), which leads to the height above the mid-plane of $\sim190$\,AU for the separation from the central binary of $\sim1100$\,AU. 

Although the dark mid-plane of the disc has previously been observed for two other post-AGB binaries \citep[AR\,Pup and Red\,Rectangle,][]{Ertel2019AJ....157..110E, Cohen2004AJ....127.2362C}, their disc scale-heights were not directly measured. In this study, we made the first estimation of the scale-height of the post-AGB circumbinary disc using SPHERE/IRDIS observations of two targets: IW\,Car and IRAS\,15469-5311. These findings show that the flaring in post-AGB circumbinary discs is significant, as predicted for passively irradiated massive discs \citep[e.g.,][]{Dullemond2001ApJ...560..957D}. Moreover, measuring the scale-height of the post-AGB disc is crucial for accurate RT modelling, as it affects the dust temperature and shape of the SED. By incorporating the measured scale-height of the disc along with other observational parameters, we can enhance the current RT modelling and develop a comprehensive model of the post-AGB binary system.

\subsection{Comparison of the SPHERE/IRDIS results with over-resolved component measured with VLTI}
\label{sec:discus_unresolved}

The essential criterion of the target selection for this study was a substantial presence of the interferometric over-resolved flux. \citet{Kluska2019A&A...631A.108K} proposed that this flux can be caused by scattering of the stellar light on the extended disc surface, on dust particles in the disc wind or on the previously detected jets \citep{Bollen2017A&A...607A..60B}. A recent study of IRAS\,08544-4431 \citep{Corporaal2023arXiv230102622C} showed that the disc geometry could explain $\sim50\%$ of the over-resolved flux component based on the radiative transfer models. However, the origin of the over-resolved emission cannot be constrained only from the interferometric observations as it allows us to access disc regions close to the dusty inner rim and dust sublimation temperature but does not resolve an extended disc surface. In contrast, SPHERE/IRDIS direct polarimetric imaging allows the investigation of the morphology of the extended flux further from the inner disc rim, unachievable with infrared interferometric techniques.

\begin{figure}
    \includegraphics[width=1.\columnwidth]{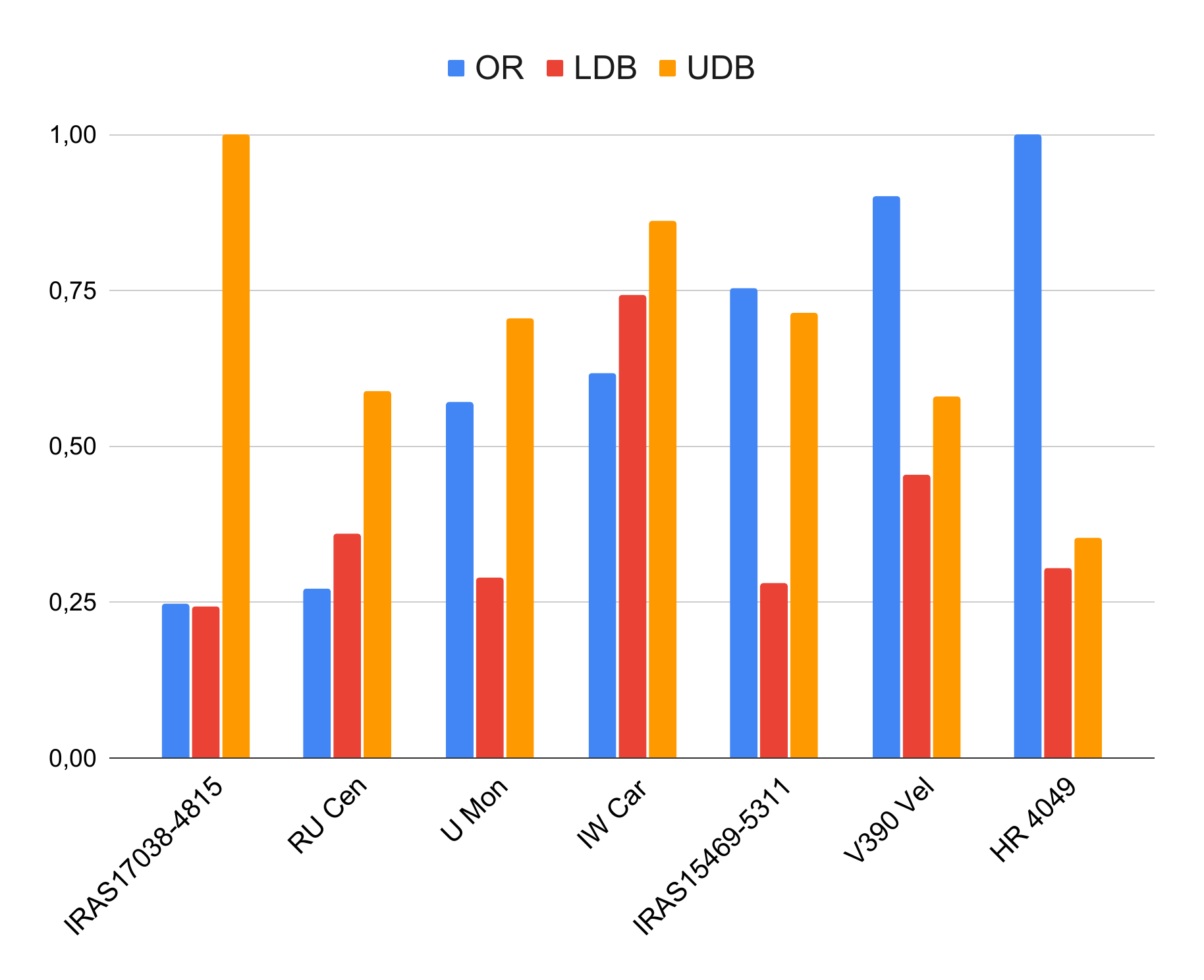}
    \caption{Comparison of the near-IR interferometric over-resolved flux with SPHERE/IRDIS $H$-band polarized disc brightness. 'OR' (in blue) represents the over-resolved flux as a percentage of the total brightness of the system, 'LDB' (in red) and 'UDB' (in orange) indicate the lower and upper limits of the polarized disc brightness relative to the total intensity of the target. To account for the different scales of these parameters, we normalized each of them to the corresponding maximum value among the target sample. See Section~\ref{sec:discus_unresolved} for more details. \label{fig:OR}}
\end{figure}

We compared the values of over-resolved flux as a percentage of the total brightness of the system (tabulate in Table~\ref{tab:sample}) obtained from the near-IR interferometric study of \citet{Kluska2019A&A...631A.108K} and results of the SPHERE/IRDIS polarimetric study, such as resolved disc size and disc polarized brightness relative to the total intensity of the target (tabulated in Table~\ref{tab:polarisedtototal}). However, counter-intuitively, we did not find any relationship between these parameters (see Fig.~\ref{fig:OR}). Although we cannot draw definitive conclusions, it is possible that the scattering of light on the disc surface is not the leading cause of the over-resolved signal seen in VLTI/PIONIER observations. In addition, we note that the total amount of scattered light significantly depends on the disc orientation and scattering efficiency \citep{Mulders2013A&A...549A.112M}. Moreover, our sample contains only eight post-AGB/RGB binaries, and more observations are needed to constrain the origin of the interferometric near-IR over-resolved flux. Conducting additional observations with SPHERE, using a larger sample size and employing the coronographic mode and longer exposure times, could provide valuable insights on the extended parts of the disc. Furthermore, polarimetric observations with ALMA are necessary to investigate scattering effects on the dust in the disc's midplane, which cannot be effectively observed with SPHERE.

\subsection{Comparison of discs around post-AGBs with protoplanetary discs around YSOs}
\label{sec:comparison}

Circumbinary discs around post-AGB binaries, also referred to as second-generation protoplanetary discs, display similar characteristics to PPDs around YSOs. Similar to circumbinary discs around post-AGBs (see Section~\ref{sec:intro}), PPDs also exhibit Keplerian rotation, a near-IR excess compatible with emission at the dust sublimation temperature \citep[e.g.,][]{DAlessio1999ApJ...527..893D}, a chemical depletion of the refractory elements in a photosphere of a host star \citep[e.g.,][and references therein]{Booth2020MNRAS.493.5079B} and an over-resolved flux in IR observations \citep[group I;][]{Meeus2001A&A...365..476M}. Moreover, post-AGB discs show signs of dust grain evolution, grain growth \citep{DeRuyter2005A&A...435..161D, Corporaal2023arXiv230102622C} and crystallisation \citep{Gielen2011A&A...533A..99G} comparable to that observed in YSO discs \citep{Follette2017AJ....153..264F}. In addition, the recent radiative transfer modelling studies \citep[e.g.,][]{Hillen2015A&A...578A..40H, Corporaal2023arXiv230102622C} showed that the dust mass of post-AGB system IRAS\,08544-4431 has to be similar to dust masses inferred from observations of PPDs around young stars. We have also found similarities with PPDs in our representative sample of 8 post-AGB binary systems. We detected nearly the same level of polarized brightness ($\sim10^{-2}$) from the resolved disc relative to the total brightness of the post-AGB system (see Section~\ref{sec: polarised_to_tot}) as it was already shown for PPDs \citep{Avenhaus2018ApJ...863...44A}.

PPDs with over-resolved flux from near-IR observations similar to that observed in post-AGB binaries (as mentioned in Section~\ref{sec:intro} and ~\ref{sec:target_sel}) display complex structures in high-angular resolution direct and polarimetric imaging \citep[e.g.][]{Fukagawa2010PASJ...62..347F, Garufi2016A&A...588A...8G}. A study by \citet{Ertel2019AJ....157..110E} first revealed that the circumbinary disc of the post-AGB binary AR\,Pup also shows complex structures (such as arcs, gaps and cavities) in SPHERE/ZIMPOL+IRDIS high-angular resolution polarimetric observations. Our current study, which is based on a larger sample of diverse post-AGB binaries, provides added confirmation of these complex substructures (see Section~\ref{sec:substructures}). This further strengthens the similarities despite an important difference that the estimated lifetime of discs around post-AGB binaries is only in the order of 10$^4$-10$^5$ years, while PPDs live up to a few Myr.

Moreover, two out of eight post-AGB systems in our sample (RU\,Cen and AC\,Her) were classified as transition discs based on their low near-IR excess and high level of chemical depletion \citep[see Section~\ref{sec:target_sel},][]{Kluska2022}. In PPDs, a lack or low near-IR excess was interpreted as a sign of inner dust-free cavities with a size of several times the dust sublimation radius \citep{Maaskant2013A&A...555A..64M, Menu2015A&A...581A.107M} and the same conclusion was proposed for post-AGB circumbinary discs by \citet{Kluska2022}. The correlation between inner disc cavities and the level of chemical depletion for PPDs was interpreted as evidence of the presence of a giant planet carving a hole in the disc \citep[e.g.,][]{Kama2015A&A...582L..10K, Booth2020MNRAS.493.5079B} and acting as a filter to trap the dust in the disc \citep{vanderMarel2016A&A...585A..58V, vanderMarel2022arXiv221005539V}. This scenario was already confirmed for a few transition discs around YSOs \citep[e.g. PDS\,70 and AB\,Aur,][]{Haffert2019NatAs...3..749H, Currie2022NatAs...6..751C}.  Based on our study, we have found that the disc structures of RU\,Cen and AC\,Her (refer to Fig.\ref{fig:snr}) exhibit complex features, including asymmetries. These complexities are consistent with the typical behaviour of transition PPDs. 

The similarities listed above provide further evidence that the morphology of transition discs around post-AGB binaries closely resembles that of transition discs around YSOs. This might be explained by various reasons, including a second or late episode of planet formation or a first-generation planet that survived the binary interaction phase, as was proposed by \citet{Kluska2022}. However, we lack more information on the post-AGB transition discs to firmly constrain these conclusions.

\subsection{Investigating the diversity of morphologies within the sample of full disc post-AGB binary systems}
\label{sec:origin_of_morph}

Within the full disc subset (U\,Mon, IRAS\,08544-4431, IW\,Car, HR\,0449, IRAS\,15469-5311 and IRAS\,17038-4815), we found that three out of six post-AGB systems (U\,Mon, HR\,4049 and IRAS\,17038-4815) show a lack of the extended disc structure in the SPHERE/IRDIS polarimetric imaging (see Fig.~\ref{fig:ellipsefit}). To investigate this lack of the extended disc structure, we examined the potential correlations between the size of the resolved disc and stellar parameters of the post-AGB star (such as effective temperature and metallicity). We also investigated the impact of orbital properties of the binary system (such as period and eccentricity), as well as the efficiency of chemical depletion. Although no clear correlation was observed between the resolved disc size and effective temperature or orbital parameters, an intriguing correlation was identified with the metallicity, as indicated by the [Zn/H] values\footnote{We note that in post-AGB binaries, refractory elements such as Fe are depleted in the post-AGB star photosphere (see Section~\ref{sec:intro}), so we cannot use [Fe/H] as a tracer of initial metallicity in these systems. However, we may be able to use the abundance of a non-depleted volatile element (such as Zn or S) as a proxy of the metallicity of the star. In this case, we need to consider the turn-off temperature, which defines the limit, and elements with higher condensation temperatures become depleted. If the turn-off temperature exceeds 800\,K, Zn is not affected by the gas re-accretion from the disc and can serve as a reliable reference for the system's initial metallicity. All of the full disc systems in our sample have turn-off temperatures greater than or equal to 800\,K, indicating that their [Zn/H] values can be used as a proxy for their initial metallicities.}.

\begin{figure}
    \includegraphics[width=0.99\columnwidth]{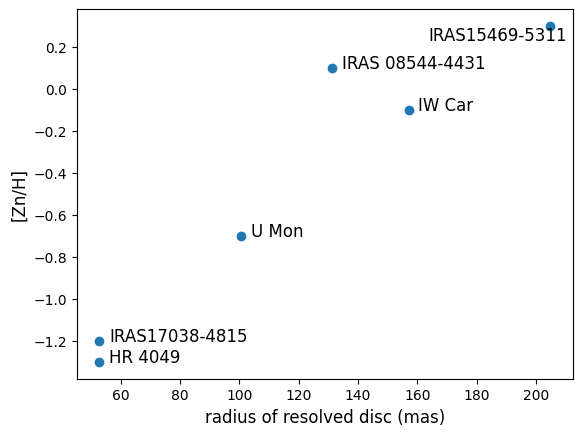}
    \caption{Comparison of the radius of resolved disc size with SPHERE/IRDIS and [Zn/H] values (see Table~\ref{tab:sample}) as an indication of the initial metallicity of post-AGB star. Typical uncertainties are $\pm10$\,mas for the disc size and $\pm0.3$ dex for abundance values.  See Section~\ref{sec:origin_of_morph} for more details.\label{fig:size_vs_met}}
\end{figure}

Exploring the correlation between the resolved disc size and initial metallicity revealed that targets, such as U\,Mon, HR\,4049, and IRAS\,17038-4815, with a radius of resolved disc\,$<$\,100\,mas (as shown in Fig.\ref{fig:size_vs_met}), have [Zn/H] values ranging from -1.3 to -0.7. Conversely, post-AGB systems with a radius of resolved disc greater than 130\,mas, such as IRAS\,08544-4431, IW\,Car, and IRAS\,15469-5311, have [Zn/H] values ranging from 0.1 to 0.3. This implies a positive correlation between resolved disc size and initial metallicity.

Recently, \citet{DellAgli2019MNRAS.486.4738D} presented low-metallicity models ($Z=10^{-4}$) of stars of low and intermediate mass, evolving through the AGB phase. They found that the amount of dust produced during this phase decreases as the metallicity decreases. Although we cannot distinguish the effect of initial metallicity from other factors that influence dust production during and after the AGB phase, we speculate that for the full-disc systems in our sample, the observed trend (i.e., systems with smaller resolved disc sizes in scattered light have lower initial metallicities) might be due to the fact that these systems have less dust production during the AGB phase.

Another possibility might be that the lack of extended disc structure could also be caused by self-shadowing of the extended structure by the puffed inner disc rim, as proposed for PPDs around YSOs \citep{Garufi2022A&A...658A.137G}. Currently, we are limited to eight post-AGB/RGB binary stars, and more observations are needed to constrain the observed correlation accurately.

\section{Conclusions}
\label{sec:conclusion}

We conducted a pilot survey using high-resolution polarimetric imaging of eight post-AGB binary systems with circumbinary discs (both full and transition discs) using the SPHERE/IRDIS instrument. The aim of the study was to investigate the morphology and orientation of the circumbinary discs in near-infrared scattered light, as well as to identify any substructures on the disc surfaces. Our study successfully resolved the extended disc structures of the circumbinary discs around the target binary systems, revealing complex morphologies and a range of disc sizes and orientations. 

The selection criteria mandated that the targets possess a substantial amount of over-resolved flux in the near-IR interferometric PIONIER observations, exceeding 5\%. This over-resolved flux could not be fully explained using current radiative transfer models. Surprisingly, we found no correlation between the over-resolved flux and the SPHERE/IRDIS polarimetric observations, suggesting that scattering of light on the disc surface may not be the primary cause of the observed over-resolved flux. 

Our analysis provided the first estimates of the disc scale-height for two of the systems, namely IW\,Car and IRAS\,15469-5311. Additionally, we found that the discs of RU\,Cen and AC\,Her, both of which have large inner cavities and are classified as transition discs, had a more intricate and asymmetric configuration. We also observed that some of the full disc systems with lower metallicities, such as U\,Mon, HR\,4049 and IRAS\,17038-4815, show lower resolved disc sizes (i.e., lacked extended scattered light), possibly due to lower levels of dust production during the AGB/RGB phase of the primary star.

Our study revealed similarities between PPDs around YSOs and circumbinary discs (also referred to as second-generation protoplanetary discs) around post-AGB binaries, such as the same level of polarized brightness from the resolved disc relative to the total brightness of the system. Our results, along with other observational parameters from spectroscopic and interferometric studies, establish post-AGB binary discs, particularly transition discs, as valuable laboratories to study the circumbinary disc evolution, including potential second-generation planet formation around evolved binaries.

Furthermore, in our following study, we will use VLT/SPHERE-ZIMPOL to investigate these discs in the visible wavelength regime ($V, I$-bands). Multi-wavelength polarimetric imaging (using SPHERE/IRDIS+ZIMPOL) will enable us to measure the wavelength-dependent dust polarization degree and define properties of the scattering particles, such as dust grain size, composition, and spatial distribution, thereby allowing us to better characterise the circumbinary second-generation protoplanetary discs around post-AGB binaries. 

\section*{Acknowledgements}

DK and KA acknowledge the support of the Australian Research Council (ARC) Discovery Early Career Research Award (DECRA) grant (DE190100813). This research was supported in part by the Australian Research Council Centre of Excellence for All Sky Astrophysics in 3 Dimensions (ASTRO 3D) through project number CE170100013. H.V.W. and A.C. acknowledge support from FWO under contract G097619N. J.K. acknowledges support from FWO under the senior postdoctoral fellowship (1281121N). We also thank Dr Rob van Holstein for providing essential information on the IRDAP data reduction software.

\section*{Data Availability}
The data underlying this article are stored online in the ESO Science Archive Facility at http://archive.eso.org, and can be accessed by program IDs 0102.D-0696(A) and 0101.D-0807(A).



\bibliographystyle{mnras}
\bibliography{andrych_main.bib} 

\begin{thebibliography}{}
\makeatletter
\relax
\def\mn@urlcharsother{\let\do\@makeother \do\$\do\&\do\#\do\^\do\_\do\%\do\~}
\def\mn@doi{\begingroup\mn@urlcharsother \@ifnextchar [ {\mn@doi@}
  {\mn@doi@[]}}
\def\mn@doi@[#1]#2{\def\@tempa{#1}\ifx\@tempa\@empty \href
  {http://dx.doi.org/#2} {doi:#2}\else \href {http://dx.doi.org/#2} {#1}\fi
  \endgroup}
\def\mn@eprint#1#2{\mn@eprint@#1:#2::\@nil}
\def\mn@eprint@arXiv#1{\href {http://arxiv.org/abs/#1} {{\tt arXiv:#1}}}
\def\mn@eprint@dblp#1{\href {http://dblp.uni-trier.de/rec/bibtex/#1.xml}
  {dblp:#1}}
\def\mn@eprint@#1:#2:#3:#4\@nil{\def\@tempa {#1}\def\@tempb {#2}\def\@tempc
  {#3}\ifx \@tempc \@empty \let \@tempc \@tempb \let \@tempb \@tempa \fi \ifx
  \@tempb \@empty \def\@tempb {arXiv}\fi \@ifundefined
  {mn@eprint@\@tempb}{\@tempb:\@tempc}{\expandafter \expandafter \csname
  mn@eprint@\@tempb\endcsname \expandafter{\@tempc}}}

\bibitem[\protect\citeauthoryear{{Acke} et~al.,}{{Acke}
  et~al.}{2013}]{Acke2013}
{Acke} B.,  et~al., 2013, \mn@doi [\aap] {10.1051/0004-6361/201219282}, \href
  {https://ui.adsabs.harvard.edu/abs/2013A&A...551A..76A} {551, A76}

\bibitem[\protect\citeauthoryear{{Avenhaus} et~al.,}{{Avenhaus}
  et~al.}{2018}]{Avenhaus2018ApJ...863...44A}
{Avenhaus} H.,  et~al., 2018, \mn@doi [\apj] {10.3847/1538-4357/aab846}, \href
  {https://ui.adsabs.harvard.edu/abs/2018ApJ...863...44A} {863, 44}

\bibitem[\protect\citeauthoryear{{Bailer-Jones}, {Rybizki}, {Fouesneau},
  {Demleitner}  \& {Andrae}}{{Bailer-Jones}
  et~al.}{2021}]{Bailer-Jones2021AJ....161..147B}
{Bailer-Jones} C.~A.~L.,  {Rybizki} J.,  {Fouesneau} M.,  {Demleitner} M.,
  {Andrae} R.,  2021, \mn@doi [\aj] {10.3847/1538-3881/abd806}, \href
  {https://ui.adsabs.harvard.edu/abs/2021AJ....161..147B} {161, 147}

\bibitem[\protect\citeauthoryear{{Benisty} et~al.,}{{Benisty}
  et~al.}{2022}]{Benisty2022arXiv220309991B}
{Benisty} M.,  et~al., 2022, arXiv e-prints, \href
  {https://ui.adsabs.harvard.edu/abs/2022arXiv220309991B} {p. arXiv:2203.09991}

\bibitem[\protect\citeauthoryear{{Beuzit} et~al.,}{{Beuzit}
  et~al.}{2019}]{Beuzit2019}
{Beuzit} J.~L.,  et~al., 2019, \mn@doi [\aap] {10.1051/0004-6361/201935251},
  \href {https://ui.adsabs.harvard.edu/abs/2019A&A...631A.155B} {631, A155}

\bibitem[\protect\citeauthoryear{{B{\'o}di} \& {Kiss}}{{B{\'o}di} \&
  {Kiss}}{2019}]{Bodi2019}
{B{\'o}di} A.,  {Kiss} L.~L.,  2019, \mn@doi [\apj] {10.3847/1538-4357/aafc24},
  \href {https://ui.adsabs.harvard.edu/abs/2019ApJ...872...60B} {872, 60}

\bibitem[\protect\citeauthoryear{{Bollen}, {Van Winckel}  \& {Kamath}}{{Bollen}
  et~al.}{2017}]{Bollen2017A&A...607A..60B}
{Bollen} D.,  {Van Winckel} H.,   {Kamath} D.,  2017, \mn@doi [\aap]
  {10.1051/0004-6361/201731493}, \href
  {https://ui.adsabs.harvard.edu/abs/2017A&A...607A..60B} {607, A60}

\bibitem[\protect\citeauthoryear{{Bollen}, {Kamath}, {Van Winckel}, {De Marco},
  {Verhamme}, {Kluska}  \& {Wardle}}{{Bollen}
  et~al.}{2022}]{Bollen2022arXiv220808752B}
{Bollen} D.,  {Kamath} D.,  {Van Winckel} H.,  {De Marco} O.,  {Verhamme} O.,
  {Kluska} J.,   {Wardle} M.,  2022, \mn@doi [\aap]
  {10.1051/0004-6361/202243429}, \href
  {https://ui.adsabs.harvard.edu/abs/2022A&A...666A..40B} {666, A40}

\bibitem[\protect\citeauthoryear{{Booth} \& {Owen}}{{Booth} \&
  {Owen}}{2020}]{Booth2020MNRAS.493.5079B}
{Booth} R.~A.,  {Owen} J.~E.,  2020, \mn@doi [\mnras] {10.1093/mnras/staa578},
  \href {https://ui.adsabs.harvard.edu/abs/2020MNRAS.493.5079B} {493, 5079}

\bibitem[\protect\citeauthoryear{{Bujarrabal}, {Castro-Carrizo}, {Alcolea}  \&
  {Van Winckel}}{{Bujarrabal} et~al.}{2015}]{Bujarrabal2015A&A...575L...7B}
{Bujarrabal} V.,  {Castro-Carrizo} A.,  {Alcolea} J.,   {Van Winckel} H.,
  2015, \mn@doi [\aap] {10.1051/0004-6361/201525742}, \href
  {https://ui.adsabs.harvard.edu/abs/2015A&A...575L...7B} {575, L7}

\bibitem[\protect\citeauthoryear{{Bujarrabal}, {Castro-Carrizo}, {Alcolea},
  {Van Winckel}, {S{\'a}nchez Contreras}  \&
  {Santander-Garc{\'\i}a}}{{Bujarrabal}
  et~al.}{2017}]{Bujarrabal2017A&A...597L...5B}
{Bujarrabal} V.,  {Castro-Carrizo} A.,  {Alcolea} J.,  {Van Winckel} H.,
  {S{\'a}nchez Contreras} C.,   {Santander-Garc{\'\i}a} M.,  2017, \mn@doi
  [\aap] {10.1051/0004-6361/201629317}, \href
  {https://ui.adsabs.harvard.edu/abs/2017A&A...597L...5B} {597, L5}

\bibitem[\protect\citeauthoryear{{Bujarrabal}, {Castro-Carrizo}, {Van Winckel},
  {Alcolea}, {S{\'a}nchez Contreras}, {Santander-Garc{\'\i}a}  \&
  {Hillen}}{{Bujarrabal} et~al.}{2018}]{Bujarrabal2018}
{Bujarrabal} V.,  {Castro-Carrizo} A.,  {Van Winckel} H.,  {Alcolea} J.,
  {S{\'a}nchez Contreras} C.,  {Santander-Garc{\'\i}a} M.,   {Hillen} M.,
  2018, \mn@doi [\aap] {10.1051/0004-6361/201732422}, \href
  {https://ui.adsabs.harvard.edu/abs/2018A&A...614A..58B} {614, A58}

\bibitem[\protect\citeauthoryear{{Canovas}, {M{\'e}nard}, {de Boer}, {Pinte},
  {Avenhaus}  \& {Schreiber}}{{Canovas}
  et~al.}{2015}]{Canovas2015A&A...582L...7C}
{Canovas} H.,  {M{\'e}nard} F.,  {de Boer} J.,  {Pinte} C.,  {Avenhaus} H.,
  {Schreiber} M.~R.,  2015, \mn@doi [\aap] {10.1051/0004-6361/201527267}, \href
  {https://ui.adsabs.harvard.edu/abs/2015A&A...582L...7C} {582, L7}

\bibitem[\protect\citeauthoryear{{Cohen}, {Van Winckel}, {Bond}  \&
  {Gull}}{{Cohen} et~al.}{2004}]{Cohen2004AJ....127.2362C}
{Cohen} M.,  {Van Winckel} H.,  {Bond} H.~E.,   {Gull} T.~R.,  2004, \mn@doi
  [\aj] {10.1086/382902}, \href
  {https://ui.adsabs.harvard.edu/abs/2004AJ....127.2362C} {127, 2362}

\bibitem[\protect\citeauthoryear{{Corporaal}, {Kluska}, {Van Winckel},
  {Bollen}, {Kamath}  \& {Min}}{{Corporaal}
  et~al.}{2021}]{Corporaal2021A&A...650L..13C}
{Corporaal} A.,  {Kluska} J.,  {Van Winckel} H.,  {Bollen} D.,  {Kamath} D.,
  {Min} M.,  2021, \mn@doi [\aap] {10.1051/0004-6361/202141154}, \href
  {https://ui.adsabs.harvard.edu/abs/2021A&A...650L..13C} {650, L13}

\bibitem[\protect\citeauthoryear{{Corporaal}, {Kluska}, {Van Winckel}, {Kamath}
   \& {Min}}{{Corporaal} et~al.}{2023a}]{Corporaal2023arXiv230102622C}
{Corporaal} A.,  {Kluska} J.,  {Van Winckel} H.,  {Kamath} D.,   {Min} M.,
  2023a, \mn@doi [arXiv e-prints] {10.48550/arXiv.2301.02622}, \href
  {https://ui.adsabs.harvard.edu/abs/2023arXiv230102622C} {p. arXiv:2301.02622}

\bibitem[\protect\citeauthoryear{{Corporaal}, {Kluska}, {Van Winckel},
  {Andrych}, {Cuello}, {Kamath}  \& {Merand}}{{Corporaal}
  et~al.}{2023b}]{Corporaal2023arXiv230412028C}
{Corporaal} A.,  {Kluska} J.,  {Van Winckel} H.,  {Andrych} K.,  {Cuello} N.,
  {Kamath} D.,   {Merand} A.,  2023b, \mn@doi [arXiv e-prints]
  {10.48550/arXiv.2304.12028}, \href
  {https://ui.adsabs.harvard.edu/abs/2023arXiv230412028C} {p. arXiv:2304.12028}

\bibitem[\protect\citeauthoryear{{Currie} et~al.,}{{Currie}
  et~al.}{2022}]{Currie2022NatAs...6..751C}
{Currie} T.,  et~al., 2022, \mn@doi [Nature Astronomy]
  {10.1038/s41550-022-01634-x10.48550/arXiv.2204.00633}, \href
  {https://ui.adsabs.harvard.edu/abs/2022NatAs...6..751C} {6, 751}

\bibitem[\protect\citeauthoryear{{D'Alessio}, {Calvet}, {Hartmann}, {Lizano}
  \& {Cant{\'o}}}{{D'Alessio} et~al.}{1999}]{DAlessio1999ApJ...527..893D}
{D'Alessio} P.,  {Calvet} N.,  {Hartmann} L.,  {Lizano} S.,   {Cant{\'o}} J.,
  1999, \mn@doi [\apj] {10.1086/308103}, \href
  {https://ui.adsabs.harvard.edu/abs/1999ApJ...527..893D} {527, 893}

\bibitem[\protect\citeauthoryear{{Dell'Agli}, {Valiante}, {Kamath}, {Ventura}
  \& {Garc{\'\i}a-Hern{\'a}ndez}}{{Dell'Agli}
  et~al.}{2019}]{DellAgli2019MNRAS.486.4738D}
{Dell'Agli} F.,  {Valiante} R.,  {Kamath} D.,  {Ventura} P.,
  {Garc{\'\i}a-Hern{\'a}ndez} D.~A.,  2019, \mn@doi [\mnras]
  {10.1093/mnras/stz1164}, \href
  {https://ui.adsabs.harvard.edu/abs/2019MNRAS.486.4738D} {486, 4738}

\bibitem[\protect\citeauthoryear{{Dohlen} et~al.,}{{Dohlen}
  et~al.}{2008}]{Dohlen2008}
{Dohlen} K.,  et~al., 2008, in {McLean} I.~S.,  {Casali} M.~M.,  eds,  Society
  of Photo-Optical Instrumentation Engineers (SPIE) Conference Series Vol.
  7014, Ground-based and Airborne Instrumentation for Astronomy II. p. 70143L,
  \mn@doi{10.1117/12.789786}

\bibitem[\protect\citeauthoryear{{Dominik}, {Dullemond}, {Cami}  \& {van
  Winckel}}{{Dominik} et~al.}{2003}]{Dominik2003A&A...397..595D}
{Dominik} C.,  {Dullemond} C.~P.,  {Cami} J.,   {van Winckel} H.,  2003,
  \mn@doi [\aap] {10.1051/0004-6361:20021478}, \href
  {https://ui.adsabs.harvard.edu/abs/2003A&A...397..595D} {397, 595}

\bibitem[\protect\citeauthoryear{{Dullemond}, {Dominik}  \&
  {Natta}}{{Dullemond} et~al.}{2001}]{Dullemond2001ApJ...560..957D}
{Dullemond} C.~P.,  {Dominik} C.,   {Natta} A.,  2001, \mn@doi [\apj]
  {10.1086/323057}, \href
  {https://ui.adsabs.harvard.edu/abs/2001ApJ...560..957D} {560, 957}

\bibitem[\protect\citeauthoryear{{Ertel} et~al.,}{{Ertel}
  et~al.}{2019}]{Ertel2019AJ....157..110E}
{Ertel} S.,  et~al., 2019, \mn@doi [\aj] {10.3847/1538-3881/aafe04}, \href
  {https://ui.adsabs.harvard.edu/abs/2019AJ....157..110E} {157, 110}

\bibitem[\protect\citeauthoryear{{Follette} et~al.,}{{Follette}
  et~al.}{2017}]{Follette2017AJ....153..264F}
{Follette} K.~B.,  et~al., 2017, \mn@doi [\aj] {10.3847/1538-3881/aa6d85},
  \href {https://ui.adsabs.harvard.edu/abs/2017AJ....153..264F} {153, 264}

\bibitem[\protect\citeauthoryear{{Fukagawa} et~al.,}{{Fukagawa}
  et~al.}{2010}]{Fukagawa2010PASJ...62..347F}
{Fukagawa} M.,  et~al., 2010, \mn@doi [\pasj] {10.1093/pasj/62.2.347}, \href
  {https://ui.adsabs.harvard.edu/abs/2010PASJ...62..347F} {62, 347}

\bibitem[\protect\citeauthoryear{{Gallardo Cava}, {G{\'o}mez-Garrido},
  {Bujarrabal}, {Castro-Carrizo}, {Alcolea}  \& {Van Winckel}}{{Gallardo Cava}
  et~al.}{2021}]{Gallardo_cava2021A&A...648A..93G}
{Gallardo Cava} I.,  {G{\'o}mez-Garrido} M.,  {Bujarrabal} V.,
  {Castro-Carrizo} A.,  {Alcolea} J.,   {Van Winckel} H.,  2021, \mn@doi [\aap]
  {10.1051/0004-6361/202039604}, \href
  {https://ui.adsabs.harvard.edu/abs/2021A&A...648A..93G} {648, A93}

\bibitem[\protect\citeauthoryear{{Garufi} et~al.,}{{Garufi}
  et~al.}{2016}]{Garufi2016A&A...588A...8G}
{Garufi} A.,  et~al., 2016, \mn@doi [\aap] {10.1051/0004-6361/201527940}, \href
  {https://ui.adsabs.harvard.edu/abs/2016A&A...588A...8G} {588, A8}

\bibitem[\protect\citeauthoryear{{Garufi} et~al.,}{{Garufi}
  et~al.}{2022}]{Garufi2022A&A...658A.137G}
{Garufi} A.,  et~al., 2022, \mn@doi [\aap] {10.1051/0004-6361/202141692}, \href
  {https://ui.adsabs.harvard.edu/abs/2022A&A...658A.137G} {658, A137}

\bibitem[\protect\citeauthoryear{{Gezer}, {Van Winckel}, {Bozkurt}, {De Smedt},
  {Kamath}, {Hillen}  \& {Manick}}{{Gezer}
  et~al.}{2015}]{Gezer2015MNRAS.453..133G}
{Gezer} I.,  {Van Winckel} H.,  {Bozkurt} Z.,  {De Smedt} K.,  {Kamath} D.,
  {Hillen} M.,   {Manick} R.,  2015, \mn@doi [\mnras] {10.1093/mnras/stv1627},
  \href {https://ui.adsabs.harvard.edu/abs/2015MNRAS.453..133G} {453, 133}

\bibitem[\protect\citeauthoryear{{Gielen} et~al.,}{{Gielen}
  et~al.}{2011}]{Gielen2011A&A...533A..99G}
{Gielen} C.,  et~al., 2011, \mn@doi [\aap] {10.1051/0004-6361/201117364}, \href
  {https://ui.adsabs.harvard.edu/abs/2011A&A...533A..99G} {533, A99}

\bibitem[\protect\citeauthoryear{{Giridhar}, {Rao}  \& {Lambert}}{{Giridhar}
  et~al.}{1994}]{Giridhar1994ApJ...437..476G}
{Giridhar} S.,  {Rao} N.~K.,   {Lambert} D.~L.,  1994, \mn@doi [\apj]
  {10.1086/175011}, \href
  {https://ui.adsabs.harvard.edu/abs/1994ApJ...437..476G} {437, 476}

\bibitem[\protect\citeauthoryear{{Giridhar}, {Lambert}  \&
  {Gonzalez}}{{Giridhar} et~al.}{1998}]{Giridhar1998ApJ...509..366G}
{Giridhar} S.,  {Lambert} D.~L.,   {Gonzalez} G.,  1998, \mn@doi [\apj]
  {10.1086/306487}, \href
  {https://ui.adsabs.harvard.edu/abs/1998ApJ...509..366G} {509, 366}

\bibitem[\protect\citeauthoryear{{Giridhar}, {Lambert}  \&
  {Gonzalez}}{{Giridhar} et~al.}{2000}]{Giridhar2000ApJ...531..521G}
{Giridhar} S.,  {Lambert} D.~L.,   {Gonzalez} G.,  2000, \mn@doi [\apj]
  {10.1086/308451}, \href
  {https://ui.adsabs.harvard.edu/abs/2000ApJ...531..521G} {531, 521}

\bibitem[\protect\citeauthoryear{{Giridhar}, {Molina}, {Arellano Ferro}  \&
  {Selvakumar}}{{Giridhar} et~al.}{2010}]{Giridhar2010MNRAS.406..290G}
{Giridhar} S.,  {Molina} R.,  {Arellano Ferro} A.,   {Selvakumar} G.,  2010,
  \mn@doi [\mnras] {10.1111/j.1365-2966.2010.16696.x}, \href
  {https://ui.adsabs.harvard.edu/abs/2010MNRAS.406..290G} {406, 290}

\bibitem[\protect\citeauthoryear{{Haffert}, {Bohn}, {de Boer}, {Snellen},
  {Brinchmann}, {Girard}, {Keller}  \& {Bacon}}{{Haffert}
  et~al.}{2019}]{Haffert2019NatAs...3..749H}
{Haffert} S.~Y.,  {Bohn} A.~J.,  {de Boer} J.,  {Snellen} I.~A.~G.,
  {Brinchmann} J.,  {Girard} J.~H.,  {Keller} C.~U.,   {Bacon} R.,  2019,
  \mn@doi [Nature Astronomy]
  {10.1038/s41550-019-0780-510.48550/arXiv.1906.01486}, \href
  {https://ui.adsabs.harvard.edu/abs/2019NatAs...3..749H} {3, 749}

\bibitem[\protect\citeauthoryear{Halir \& Flusser}{Halir \&
  Flusser}{1998}]{Halir98numericallystable}
Halir R.,  Flusser J.,  1998, Numerically Stable Direct Least Squares Fitting
  Of Ellipses

\bibitem[\protect\citeauthoryear{{Hillen}, {de Vries}, {Menu}, {Van Winckel},
  {Min}  \& {Mulders}}{{Hillen} et~al.}{2015}]{Hillen2015A&A...578A..40H}
{Hillen} M.,  {de Vries} B.~L.,  {Menu} J.,  {Van Winckel} H.,  {Min} M.,
  {Mulders} G.~D.,  2015, \mn@doi [\aap] {10.1051/0004-6361/201425372}, \href
  {https://ui.adsabs.harvard.edu/abs/2015A&A...578A..40H} {578, A40}

\bibitem[\protect\citeauthoryear{{Hillen}, {Kluska}, {Le Bouquin}, {Van
  Winckel}, {Berger}, {Kamath}  \& {Bujarrabal}}{{Hillen}
  et~al.}{2016}]{Hillen2016}
{Hillen} M.,  {Kluska} J.,  {Le Bouquin} J.~B.,  {Van Winckel} H.,  {Berger}
  J.~P.,  {Kamath} D.,   {Bujarrabal} V.,  2016, \mn@doi [\aap]
  {10.1051/0004-6361/201628125}, \href
  {https://ui.adsabs.harvard.edu/abs/2016A&A...588L...1H} {588, L1}

\bibitem[\protect\citeauthoryear{{Hillen} et~al.,}{{Hillen}
  et~al.}{2017}]{Hillen2017}
{Hillen} M.,  et~al., 2017, \mn@doi [\aap] {10.1051/0004-6361/201629161}, \href
  {https://ui.adsabs.harvard.edu/abs/2017A&A...599A..41H} {599, A41}

\bibitem[\protect\citeauthoryear{{Hinkle}, {Brittain}  \& {Lambert}}{{Hinkle}
  et~al.}{2007}]{Hinkle2007ApJ...664..501H}
{Hinkle} K.~H.,  {Brittain} S.~D.,   {Lambert} D.~L.,  2007, \mn@doi [\aj]
  {10.1086/518682}, \href
  {https://ui.adsabs.harvard.edu/abs/2007ApJ...664..501H} {664, 501}

\bibitem[\protect\citeauthoryear{{Hunziker} et~al.,}{{Hunziker}
  et~al.}{2021}]{Hunziker2021A&A...648A.110H}
{Hunziker} S.,  et~al., 2021, \mn@doi [\aap] {10.1051/0004-6361/202040166},
  \href {https://ui.adsabs.harvard.edu/abs/2021A&A...648A.110H} {648, A110}

\bibitem[\protect\citeauthoryear{{Jones}}{{Jones}}{2022}]{Jones_2022A&A...665A..21J}
{Jones} A.~P.,  2022, \mn@doi [\aap] {10.1051/0004-6361/202142718}, \href
  {https://ui.adsabs.harvard.edu/abs/2022A&A...665A..21J} {665, A21}

\bibitem[\protect\citeauthoryear{{Kama}, {Folsom}  \& {Pinilla}}{{Kama}
  et~al.}{2015}]{Kama2015A&A...582L..10K}
{Kama} M.,  {Folsom} C.~P.,   {Pinilla} P.,  2015, \mn@doi [\aap]
  {10.1051/0004-6361/201527094}, \href
  {https://ui.adsabs.harvard.edu/abs/2015A&A...582L..10K} {582, L10}

\bibitem[\protect\citeauthoryear{{Kamath} \& {Van Winckel}}{{Kamath} \& {Van
  Winckel}}{2019}]{Kamath2019MNRAS.486.3524K}
{Kamath} D.,  {Van Winckel} H.,  2019, \mn@doi [\mnras]
  {10.1093/mnras/stz1076}, \href
  {https://ui.adsabs.harvard.edu/abs/2019MNRAS.486.3524K} {486, 3524}

\bibitem[\protect\citeauthoryear{{Kamath}, {Wood}  \& {Van Winckel}}{{Kamath}
  et~al.}{2014}]{Kamath2014MNRAS.439.2211K}
{Kamath} D.,  {Wood} P.~R.,   {Van Winckel} H.,  2014, \mn@doi [\mnras]
  {10.1093/mnras/stt2033}, \href
  {https://ui.adsabs.harvard.edu/abs/2014MNRAS.439.2211K} {439, 2211}

\bibitem[\protect\citeauthoryear{{Kamath}, {Wood}  \& {Van Winckel}}{{Kamath}
  et~al.}{2015}]{Kamath2015MNRAS.454.1468K}
{Kamath} D.,  {Wood} P.~R.,   {Van Winckel} H.,  2015, \mn@doi [\mnras]
  {10.1093/mnras/stv1202}, \href
  {https://ui.adsabs.harvard.edu/abs/2015MNRAS.454.1468K} {454, 1468}

\bibitem[\protect\citeauthoryear{{Kamath}, {Wood}, {Van Winckel}  \&
  {Nie}}{{Kamath} et~al.}{2016}]{Kamath2016A&A...586L...5K}
{Kamath} D.,  {Wood} P.~R.,  {Van Winckel} H.,   {Nie} J.~D.,  2016, \mn@doi
  [\aap] {10.1051/0004-6361/201526892}, \href
  {https://ui.adsabs.harvard.edu/abs/2016A&A...586L...5K} {586, L5}

\bibitem[\protect\citeauthoryear{{Kiss} \& {B{\'o}di}}{{Kiss} \&
  {B{\'o}di}}{2017}]{Kiss2017}
{Kiss} L.~L.,  {B{\'o}di} A.,  2017, \mn@doi [\aap]
  {10.1051/0004-6361/201731876}, \href
  {https://ui.adsabs.harvard.edu/abs/2017A&A...608A..99K} {608, A99}

\bibitem[\protect\citeauthoryear{{Kiss}, {Derekas}, {Szab{\'o}}, {Bedding}  \&
  {Szabados}}{{Kiss} et~al.}{2007}]{Kiss2007MNRAS.375.1338K}
{Kiss} L.~L.,  {Derekas} A.,  {Szab{\'o}} G.~M.,  {Bedding} T.~R.,   {Szabados}
  L.,  2007, \mn@doi [\mnras] {10.1111/j.1365-2966.2006.11387.x}, \href
  {https://ui.adsabs.harvard.edu/abs/2007MNRAS.375.1338K} {375, 1338}

\bibitem[\protect\citeauthoryear{{Kluska}, {Hillen}, {Van Winckel}, {Manick},
  {Min}, {Regibo}  \& {Royer}}{{Kluska}
  et~al.}{2018}]{Kluska2018A&A...616A.153K}
{Kluska} J.,  {Hillen} M.,  {Van Winckel} H.,  {Manick} R.,  {Min} M.,
  {Regibo} S.,   {Royer} P.,  2018, \mn@doi [\aap]
  {10.1051/0004-6361/201832983}, \href
  {https://ui.adsabs.harvard.edu/abs/2018A&A...616A.153K} {616, A153}

\bibitem[\protect\citeauthoryear{{Kluska}, {Van Winckel}, {Hillen}, {Berger},
  {Kamath}, {Le Bouquin}  \& {Min}}{{Kluska}
  et~al.}{2019}]{Kluska2019A&A...631A.108K}
{Kluska} J.,  {Van Winckel} H.,  {Hillen} M.,  {Berger} J.~P.,  {Kamath} D.,
  {Le Bouquin} J.~B.,   {Min} M.,  2019, \mn@doi [\aap]
  {10.1051/0004-6361/201935785}, \href
  {https://ui.adsabs.harvard.edu/abs/2019A&A...631A.108K} {631, A108}

\bibitem[\protect\citeauthoryear{{Kluska} et~al.,}{{Kluska}
  et~al.}{2020}]{2020SPIE11446E..0DK}
{Kluska} J.,  et~al., 2020, in Society of Photo-Optical Instrumentation
  Engineers (SPIE) Conference Series. p. 114460D (\mn@eprint {arXiv}
  {2012.07448}), \mn@doi{10.1117/12.2561480}

\bibitem[\protect\citeauthoryear{{Kluska}, {Van Winckel}, {Copp{\'e}e},
  {Oomen}, {Dsilva}, {Kamath}, {Bujarrabal}  \& {Min}}{{Kluska}
  et~al.}{2022}]{Kluska2022}
{Kluska} J.,  {Van Winckel} H.,  {Copp{\'e}e} Q.,  {Oomen} G.~M.,  {Dsilva} K.,
   {Kamath} D.,  {Bujarrabal} V.,   {Min} M.,  2022, \mn@doi [\aap]
  {10.1051/0004-6361/202141690}, \href
  {https://ui.adsabs.harvard.edu/abs/2022A&A...658A..36K} {658, A36}

\bibitem[\protect\citeauthoryear{{Lagage} et~al.,}{{Lagage}
  et~al.}{2006}]{Lagage2006Sci...314..621L}
{Lagage} P.-O.,  et~al., 2006, \mn@doi [Science] {10.1126/science.1131436},
  \href {https://ui.adsabs.harvard.edu/abs/2006Sci...314..621L} {314, 621}

\bibitem[\protect\citeauthoryear{{Lucy}}{{Lucy}}{1974}]{Lucy1974AJ.....79..745L}
{Lucy} L.~B.,  1974, \mn@doi [\aj] {10.1086/111605}, \href
  {https://ui.adsabs.harvard.edu/abs/1974AJ.....79..745L} {79, 745}

\bibitem[\protect\citeauthoryear{{Maas}, {Van Winckel}  \& {Waelkens}}{{Maas}
  et~al.}{2002}]{Maas2002}
{Maas} T.,  {Van Winckel} H.,   {Waelkens} C.,  2002, \mn@doi [\aap]
  {10.1051/0004-6361:20020209}, \href
  {https://ui.adsabs.harvard.edu/abs/2002A&A...386..504M} {386, 504}

\bibitem[\protect\citeauthoryear{{Maas}, {Van Winckel}, {Lloyd Evans}, {Nyman},
  {Kilkenny}, {Martinez}, {Marang}  \& {van Wyk}}{{Maas}
  et~al.}{2003}]{Maas2003A&A...405..271M}
{Maas} T.,  {Van Winckel} H.,  {Lloyd Evans} T.,  {Nyman} L.~{\r{A}}.,
  {Kilkenny} D.,  {Martinez} P.,  {Marang} F.,   {van Wyk} F.,  2003, \mn@doi
  [\aap] {10.1051/0004-6361:20030613}, \href
  {https://ui.adsabs.harvard.edu/abs/2003A&A...405..271M} {405, 271}

\bibitem[\protect\citeauthoryear{{Maas}, {Van Winckel}  \& {Lloyd
  Evans}}{{Maas} et~al.}{2005}]{Maas2005}
{Maas} T.,  {Van Winckel} H.,   {Lloyd Evans} T.,  2005, \mn@doi [\aap]
  {10.1051/0004-6361:20041688}, \href
  {https://ui.adsabs.harvard.edu/abs/2005A&A...429..297M} {429, 297}

\bibitem[\protect\citeauthoryear{{Maaskant} et~al.,}{{Maaskant}
  et~al.}{2013}]{Maaskant2013A&A...555A..64M}
{Maaskant} K.~M.,  et~al., 2013, \mn@doi [\aap] {10.1051/0004-6361/201321300},
  \href {https://ui.adsabs.harvard.edu/abs/2013A&A...555A..64M} {555, A64}

\bibitem[\protect\citeauthoryear{{Malek} \& {Cami}}{{Malek} \&
  {Cami}}{2014a}]{Malek_1_2014ApJ...780...41M}
{Malek} S.~E.,  {Cami} J.,  2014a, \mn@doi [\apj] {10.1088/0004-637X/780/1/41},
  \href {https://ui.adsabs.harvard.edu/abs/2014ApJ...780...41M} {780, 41}

\bibitem[\protect\citeauthoryear{{Malek} \& {Cami}}{{Malek} \&
  {Cami}}{2014b}]{Malek_2_2014ApJ...794..113M}
{Malek} S.~E.,  {Cami} J.,  2014b, \mn@doi [\apj]
  {10.1088/0004-637X/794/2/113}, \href
  {https://ui.adsabs.harvard.edu/abs/2014ApJ...794..113M} {794, 113}

\bibitem[\protect\citeauthoryear{{Manick}, {Van Winckel}, {Kamath}, {Hillen}
  \& {Escorza}}{{Manick} et~al.}{2017}]{Manick2017}
{Manick} R.,  {Van Winckel} H.,  {Kamath} D.,  {Hillen} M.,   {Escorza} A.,
  2017, \mn@doi [\aap] {10.1051/0004-6361/201629125}, \href
  {https://ui.adsabs.harvard.edu/abs/2017A&A...597A.129M} {597, A129}

\bibitem[\protect\citeauthoryear{{Meeus}, {Waters}, {Bouwman}, {van den
  Ancker}, {Waelkens}  \& {Malfait}}{{Meeus}
  et~al.}{2001}]{Meeus2001A&A...365..476M}
{Meeus} G.,  {Waters} L.~B.~F.~M.,  {Bouwman} J.,  {van den Ancker} M.~E.,
  {Waelkens} C.,   {Malfait} K.,  2001, \mn@doi [\aap]
  {10.1051/0004-6361:20000144}, \href
  {https://ui.adsabs.harvard.edu/abs/2001A&A...365..476M} {365, 476}

\bibitem[\protect\citeauthoryear{{Menu}, {van Boekel}, {Henning}, {Leinert},
  {Waelkens}  \& {Waters}}{{Menu} et~al.}{2015}]{Menu2015A&A...581A.107M}
{Menu} J.,  {van Boekel} R.,  {Henning} T.,  {Leinert} C.,  {Waelkens} C.,
  {Waters} L.~B.~F.~M.,  2015, \mn@doi [\aap] {10.1051/0004-6361/201525654},
  \href {https://ui.adsabs.harvard.edu/abs/2015A&A...581A.107M} {581, A107}

\bibitem[\protect\citeauthoryear{{Mulders}, {Min}, {Dominik}, {Debes}  \&
  {Schneider}}{{Mulders} et~al.}{2013}]{Mulders2013A&A...549A.112M}
{Mulders} G.~D.,  {Min} M.,  {Dominik} C.,  {Debes} J.~H.,   {Schneider} G.,
  2013, \mn@doi [\aap] {10.1051/0004-6361/201219522}, \href
  {https://ui.adsabs.harvard.edu/abs/2013A&A...549A.112M} {549, A112}

\bibitem[\protect\citeauthoryear{{Nie}, {Wood}  \& {Nicholls}}{{Nie}
  et~al.}{2012}]{Nie2012MNRAS.423.2764N}
{Nie} J.~D.,  {Wood} P.~R.,   {Nicholls} C.~P.,  2012, \mn@doi [\mnras]
  {10.1111/j.1365-2966.2012.21087.x}, \href
  {https://ui.adsabs.harvard.edu/abs/2012MNRAS.423.2764N} {423, 2764}

\bibitem[\protect\citeauthoryear{{O'Connell}}{{O'Connell}}{1946}]{1946PRCO....2...46O}
{O'Connell} D.,  1946, Publications of the Riverview College Observatory, \href
  {https://ui.adsabs.harvard.edu/abs/1946PRCO....2...46O} {2, 46}

\bibitem[\protect\citeauthoryear{{Oomen}, {Van Winckel}, {Pols}, {Nelemans},
  {Escorza}, {Manick}, {Kamath}  \& {Waelkens}}{{Oomen}
  et~al.}{2018}]{Oomen2018}
{Oomen} G.-M.,  {Van Winckel} H.,  {Pols} O.,  {Nelemans} G.,  {Escorza} A.,
  {Manick} R.,  {Kamath} D.,   {Waelkens} C.,  2018, \mn@doi [\aap]
  {10.1051/0004-6361/201833816}, \href
  {https://ui.adsabs.harvard.edu/abs/2018A&A...620A..85O} {620, A85}

\bibitem[\protect\citeauthoryear{{Oomen}, {Van Winckel}, {Pols}  \&
  {Nelemans}}{{Oomen} et~al.}{2019}]{Oomen2019A&A...629A..49O}
{Oomen} G.-M.,  {Van Winckel} H.,  {Pols} O.,   {Nelemans} G.,  2019, \mn@doi
  [\aap] {10.1051/0004-6361/201935853}, \href
  {https://ui.adsabs.harvard.edu/abs/2019A&A...629A..49O} {629, A49}

\bibitem[\protect\citeauthoryear{{Oomen}, {Pols}, {Van Winckel}  \&
  {Nelemans}}{{Oomen} et~al.}{2020}]{Oomen2020A&A...642A.234O}
{Oomen} G.-M.,  {Pols} O.,  {Van Winckel} H.,   {Nelemans} G.,  2020, \mn@doi
  [\aap] {10.1051/0004-6361/202038341}, \href
  {https://ui.adsabs.harvard.edu/abs/2020A&A...642A.234O} {642, A234}

\bibitem[\protect\citeauthoryear{{Richardson}}{{Richardson}}{1972}]{Richardson1972JOSA...62...55R}
{Richardson} W.~H.,  1972, Journal of the Optical Society of America
  (1917-1983), \href {https://ui.adsabs.harvard.edu/abs/1972JOSA...62...55R}
  {62, 55}

\bibitem[\protect\citeauthoryear{{Sabin}, {Wade}  \& {L{\`e}bre}}{{Sabin}
  et~al.}{2015}]{Sabin2015MNRAS.446.1988S}
{Sabin} L.,  {Wade} G.~A.,   {L{\`e}bre} A.,  2015, \mn@doi [\mnras]
  {10.1093/mnras/stu2227}, \href
  {https://ui.adsabs.harvard.edu/abs/2015MNRAS.446.1988S} {446, 1988}

\bibitem[\protect\citeauthoryear{{Samus}, {Kazarovets}, {Durlevich}, {Kireeva}
  \& {Pastukhova}}{{Samus} et~al.}{2009}]{Samus2009yCat....102025S}
{Samus} N.~N.,  {Kazarovets} E.~V.,  {Durlevich} O.~V.,  {Kireeva} N.~N.,
  {Pastukhova} E.~N.,  2009, VizieR Online Data Catalog, \href
  {https://ui.adsabs.harvard.edu/abs/2009yCat....102025S} {p. B/gcvs}

\bibitem[\protect\citeauthoryear{{Scicluna}, {Kemper}, {Trejo}, {Marshall},
  {Ertel}  \& {Hillen}}{{Scicluna} et~al.}{2020}]{Scicluna2020MNRAS.494.2925S}
{Scicluna} P.,  {Kemper} F.,  {Trejo} A.,  {Marshall} J.~P.,  {Ertel} S.,
  {Hillen} M.,  2020, \mn@doi [\mnras] {10.1093/mnras/staa425}, \href
  {https://ui.adsabs.harvard.edu/abs/2020MNRAS.494.2925S} {494, 2925}

\bibitem[\protect\citeauthoryear{{Van Winckel}}{{Van
  Winckel}}{2007}]{VanWinckel2007BaltA..16..112V}
{Van Winckel} H.,  2007, Baltic Astronomy, \href
  {https://ui.adsabs.harvard.edu/abs/2007BaltA..16..112V} {16, 112}

\bibitem[\protect\citeauthoryear{{Van Winckel}}{{Van
  Winckel}}{2018}]{VanWinckel2018arXiv180900871V}
{Van Winckel} H.,  2018, arXiv e-prints, \href
  {https://ui.adsabs.harvard.edu/abs/2018arXiv180900871V} {p. arXiv:1809.00871}

\bibitem[\protect\citeauthoryear{{Van Winckel}, {Waelkens}, {Waters},
  {Molster}, {Udry}  \& {Bakker}}{{Van Winckel}
  et~al.}{1998}]{VanWinckel1998A&A...336L..17V}
{Van Winckel} H.,  {Waelkens} C.,  {Waters} L. B.~F.~M.,  {Molster} F.~J.,
  {Udry} S.,   {Bakker} E.~J.,  1998, \aap, \href
  {https://ui.adsabs.harvard.edu/abs/1998A&A...336L..17V} {336, L17}

\bibitem[\protect\citeauthoryear{{Vega}, {Stassun}, {Montez}, {Boyd}  \&
  {Somers}}{{Vega} et~al.}{2017}]{Vega2017ApJ...839...48V}
{Vega} L.~D.,  {Stassun} K.~G.,  {Montez} Rodolfo J.,  {Boyd} P.~T.,   {Somers}
  G.,  2017, \mn@doi [\apj] {10.3847/1538-4357/aa67dd}, \href
  {https://ui.adsabs.harvard.edu/abs/2017ApJ...839...48V} {839, 48}

\bibitem[\protect\citeauthoryear{{Vega} et~al.,}{{Vega}
  et~al.}{2021}]{Vega2021ApJ...909..138V}
{Vega} L.~D.,  et~al., 2021, \mn@doi [\apj] {10.3847/1538-4357/abe302}, \href
  {https://ui.adsabs.harvard.edu/abs/2021ApJ...909..138V} {909, 138}

\bibitem[\protect\citeauthoryear{{Waelkens}, {Lamers}, {Waters}, {Rufener},
  {Trams}, {Le Bertre}, {Ferlet}  \& {Vidal-Madjar}}{{Waelkens}
  et~al.}{1991}]{Waelkens1991A&A...242..433W}
{Waelkens} C.,  {Lamers} H.~J.~G.~L.~M.,  {Waters} L.~B.~F.~M.,  {Rufener} F.,
  {Trams} N.~R.,  {Le Bertre} T.,  {Ferlet} R.,   {Vidal-Madjar} A.,  1991,
  \aap, \href {https://ui.adsabs.harvard.edu/abs/1991A&A...242..433W} {242,
  433}

\bibitem[\protect\citeauthoryear{{Waters}, {Trams}  \& {Waelkens}}{{Waters}
  et~al.}{1992}]{Waters1992A&A...262L..37W}
{Waters} L.~B.~F.~M.,  {Trams} N.~R.,   {Waelkens} C.,  1992, \aap, \href
  {https://ui.adsabs.harvard.edu/abs/1992A&A...262L..37W} {262, L37}

\bibitem[\protect\citeauthoryear{{de Boer} et~al.,}{{de Boer}
  et~al.}{2016}]{deBoer2016A&A...595A.114D}
{de Boer} J.,  et~al., 2016, \mn@doi [\aap] {10.1051/0004-6361/201629267},
  \href {https://ui.adsabs.harvard.edu/abs/2016A&A...595A.114D} {595, A114}

\bibitem[\protect\citeauthoryear{{de Boer} et~al.,}{{de Boer}
  et~al.}{2020}]{deBoer2020}
{de Boer} J.,  et~al., 2020, \mn@doi [\aap] {10.1051/0004-6361/201834989},
  \href {https://ui.adsabs.harvard.edu/abs/2020A&A...633A..63D} {633, A63}

\bibitem[\protect\citeauthoryear{{de Ruyter}, {van Winckel}, {Dominik},
  {Waters}  \& {Dejonghe}}{{de Ruyter}
  et~al.}{2005}]{DeRuyter2005A&A...435..161D}
{de Ruyter} S.,  {van Winckel} H.,  {Dominik} C.,  {Waters} L.~B.~F.~M.,
  {Dejonghe} H.,  2005, \mn@doi [\aap] {10.1051/0004-6361:20041989}, \href
  {https://ui.adsabs.harvard.edu/abs/2005A&A...435..161D} {435, 161}

\bibitem[\protect\citeauthoryear{{de Ruyter}, {van Winckel}, {Maas}, {Lloyd
  Evans}, {Waters}  \& {Dejonghe}}{{de Ruyter}
  et~al.}{2006}]{DeRuyter2006A&A...448..641D}
{de Ruyter} S.,  {van Winckel} H.,  {Maas} T.,  {Lloyd Evans} T.,  {Waters}
  L.~B.~F.~M.,   {Dejonghe} H.,  2006, \mn@doi [\aap]
  {10.1051/0004-6361:20054062}, \href
  {https://ui.adsabs.harvard.edu/abs/2006A&A...448..641D} {448, 641}

\bibitem[\protect\citeauthoryear{{van Holstein} et~al.,}{{van Holstein}
  et~al.}{2020}]{Holstein2020A&A...633A..64V}
{van Holstein} R.~G.,  et~al., 2020, \mn@doi [\aap]
  {10.1051/0004-6361/201834996}, \href
  {https://ui.adsabs.harvard.edu/abs/2020A&A...633A..64V} {633, A64}

\bibitem[\protect\citeauthoryear{{van Winckel}}{{van
  Winckel}}{1995}]{VanWinckel1995PhDT........31V}
{van Winckel} H.,  1995, PhD thesis, -

\bibitem[\protect\citeauthoryear{{van Winckel}}{{van
  Winckel}}{2003}]{VanWinckel2003ARA&A..41..391V}
{van Winckel} H.,  2003, \mn@doi [\araa]
  {10.1146/annurev.astro.41.071601.170018}, \href
  {https://ui.adsabs.harvard.edu/abs/2003ARA&A..41..391V} {41, 391}

\bibitem[\protect\citeauthoryear{{van Winckel} et~al.,}{{van Winckel}
  et~al.}{2009}]{VanWinckel2009A&A...505.1221V}
{van Winckel} H.,  et~al., 2009, \mn@doi [\aap] {10.1051/0004-6361/200912332},
  \href {https://ui.adsabs.harvard.edu/abs/2009A&A...505.1221V} {505, 1221}

\bibitem[\protect\citeauthoryear{{van der Marel}}{{van der
  Marel}}{2022}]{vanderMarel2022arXiv221005539V}
{van der Marel} N.,  2022, \mn@doi [arXiv e-prints]
  {10.48550/arXiv.2210.05539}, \href
  {https://ui.adsabs.harvard.edu/abs/2022arXiv221005539V} {p. arXiv:2210.05539}

\bibitem[\protect\citeauthoryear{{van der Marel}, {van Dishoeck}, {Bruderer},
  {Andrews}, {Pontoppidan}, {Herczeg}, {van Kempen}  \& {Miotello}}{{van der
  Marel} et~al.}{2016}]{vanderMarel2016A&A...585A..58V}
{van der Marel} N.,  {van Dishoeck} E.~F.,  {Bruderer} S.,  {Andrews} S.~M.,
  {Pontoppidan} K.~M.,  {Herczeg} G.~J.,  {van Kempen} T.,   {Miotello} A.,
  2016, \mn@doi [\aap] {10.1051/0004-6361/201526988}, \href
  {https://ui.adsabs.harvard.edu/abs/2016A&A...585A..58V} {585, A58}

\makeatother
\end{thebibliography}



\appendix

\section{Relevant stellar and orbital parameters}
\label{sec:ap_sample}

Here we present an extensive literature review for each of the eight targets in the sample. For convenience, in Table~\ref{tab:initialdata}, we have also tabulated relevant observational data collected from the literature.
In Table~\ref{tab:rt-modelling}, we present the results of geometrical modelling for objects in our sample obtained by \citep{Kluska2019A&A...631A.108K} for near-IR observations with VLTI/PIONIER instrument and by \citep{Hillen2017} for mid-IR with VLT/MIDI. These results are also discussed in the text for individual targets.

\subsection{Full discs}
\textit{\#ID 1: U\,Mon}\\

U\,Mon is a widely studied post-AGB binary system with a spectral type of G0I \citep{DeRuyter2006A&A...448..641D}. \citet{Bodi2019} used period–luminosity–color (PLC) relation and $Gaia$\,DR2 data to determine the luminosity of $5480^{+1764}_{-882}$\,L$_\odot$. \citet{Oomen2019A&A...629A..49O} used SED fitting and $Gaia$\,DR2 distance to the system to estimate the value of luminosity of $3300^{+800}_{-600}$\,L$_\odot$ \footnote{However, the method based on $Gaia$\,DR2 distance to the system is less reliable than determined by period-luminosity relation as the orbital motion of the binary results in an angular displacement comparable to the parallax. Therefore, the astrometric distances for post-AGB stars still need to be determined.}. U\,Mon is the only object in our sample that shows no atmospheric depletion in refractory elements \citep[]{Giridhar2000ApJ...531..521G}. The system shows brightness variations due to a fundamental pulsation period of 45.74 days and a long-term periodic variation in mean magnitude (RVb phenomenon\footnote{It was shown for post-AGB stars that an inclined circumbinary disc could obscure the pulsating primary star at certain phases of its orbit and cause the long-term variation in mean magnitude \citep{Vega2017ApJ...839...48V, Manick2017, Kiss2017}}) with a period of $\sim2451$ days \citep{Kiss2017}. Subsequently, \citet{Oomen2018} defined the orbital properties of the U\,Mon binary system using data from the HERMES spectrograph on the 1.2 m Mercator telescope. The authors determined an orbital period of $2549\pm143$ days and an eccentricity of $0.25\pm0.06$. Assuming a typical post-AGB mass for the primary of 0.6 $M_{\odot}$ and an inclination of $75^\circ$, \citet{Oomen2018} suggested a projected semi-major axis of $3.38\pm0.31$\,AU.

\citet{Sabin2015MNRAS.446.1988S} discovered a longitudinal magnetic field at the U\,Mon surface of $10.2\pm1.7$ G. The authors suggested the magnetic field to be associated with the dynamical state of the atmosphere and perform variations with the pulsation period. Moreover, \citet{Vega2021ApJ...909..138V} recently detected X-ray emission from the system. The nature and origin of the X-ray emission remain unclear and U\,Mon is currently a unique case among all known post-AGB objects. Furthermore, \citet{Bollen2022arXiv220808752B} established the presence of jets in this system by analysing and modelling the orbital phase-dependent changes in the Balmer lines. The authors defined the outflow inclination of $61.4\pm0.2^{\circ}$, outer and inner opening angles of $60.4\pm0.2^{\circ}$ and $58.1\pm0.6^{\circ}$, respectively. Although U\,Mon shows a lack of near-IR emission and might be classified as a system with a transition disc, it is more likely to host a full disc \citep[][see Section~\ref{sec:target_sel} for details]{Kluska2022}.

\citet{Hillen2017} performed geometric modelling for U\,Mon using  VLTI/MIDI mid-infrared observations (mid-IR). The authors proposed a uniform disc with $\sim50$\,mas diameter. Subsequently, \citet{Kluska2019A&A...631A.108K} reproduced VLTI/PIONIER near-IR observations for U\,Mon using a binary and inner ring model. The authors estimated an inner-disc diameter for the ring of $5.49\pm0.03$\,mas.

\textit{\#ID 2: IRAS\,08544-4431}\\
IRAS\,08544-4431 is a post-AGB binary with an F3 spectral type. \citet{Oomen2019A&A...629A..49O} used SED fitting and $Gaia$\,DR2 distance to the system to estimate the luminosity value of $13700^{+3800}_{-2700}$\,L$_\odot ^5$. This system has an orbital period of $501.1\pm1.0$ days, a projected semi-major axis of $0.398\pm0.008$\,AU, a mass function of $0.033\pm0.002 M_\odot$, and a minimum mass of $0.31 M_\odot$ for the companion, by assuming a typical post-AGB star mass for the primary of $0.6 M_\odot$ and an inclination of 75° \citep{Oomen2018}.

\citet{Bujarrabal2018} presented ALMA maps of $^{12}CO$ and $^{13}CO$ J = 3-2 lines and showed that the system contains a relatively extended disc in rotation and slowly expanding gas that is escaping from the disc. It was found that about 90\% of the nebular material would be placed in the disc in a purely Keplerian rotation for the central part and rotates sub-Keplerian outwards. 

IRAS\,08544-4431 is also extensively studied in the near-IR and mid-IR with PIONIER (H-band), GRAVITY (K-band) and MATISSE (L and N-band) interferometers \citep[ see Table~\ref{tab:rt-modelling}]{Hillen2016, Kluska2018A&A...616A.153K, Kluska2019A&A...631A.108K, 2020SPIE11446E..0DK, Corporaal2021A&A...650L..13C, Corporaal2023arXiv230102622C}. These studies show that the inner rim of the disc is well resolved. The components that contribute to the flux are the central star, the accretion disc around the companion, the circumbinary disc, and an over-resolved component. It was found that $\sim$15\% of the H-band flux comes from an over-resolved emission, while only half of this flux was accounted for by the radiative transfer model, including a disc in hydrostatic equilibrium and scattered light \citep{Kluska2018A&A...616A.153K}. \citet{Corporaal2021A&A...650L..13C} combined observations in all available bands and concluded that the inner rim of the circumbinary disc is puffed-up, and the mid-IR data is better fitted with the temperature gradient model, probing disc structures beyond the inner edge. These authors found that the disc inner rim is located at a radius of 7.1\,mas, which coincides with the theoretical dust sublimation radius. Recently, \citep{Corporaal2023arXiv230102622C} showed that the disc geometry can explain only $\sim50\%$ of the over-resolved flux while the missing component has a temperature of the order of 1400-3600\,K.

\textit{\#ID 3: IW\,Car}\\
IW\,Car is a spectral type F7I post-AGB binary star. \citet{Bodi2019} used PLC relation and $Gaia$\,DR2 data to determine the luminosity of $2622^{+338}_{-296}$\,L$_\odot$. \citet{Oomen2019A&A...629A..49O} used SED fitting and $Gaia$\,DR2 distance to the system to estimate the luminosity value of $9200^{+1100}_{-900}$\,L$_\odot ^5$. Moreover, IW\,Car is strongly depleted in refractory elements (see Table~\ref{tab:sample}) with [Zn/Ti]= 2.1 \citep{Giridhar1994ApJ...437..476G}. \citet{1946PRCO....2...46O} first noted the presence of an RVb phenomenon for this object as the pulsations had larger amplitudes when the star was brighter. \citet{Kiss2017} defined the orbital period for the system to be 1449\,days and pulsating period of 71.98\,days based on visual observations of the American Association of Variable Star Observers (The AAVSO International Database). 

\citet{Bujarrabal2017A&A...597L...5B} obtained ALMA maps of $^{12}$CO and $^{13}$CO $J = 3-2$ lines that clearly show both rotating and expanding gas. The authors defined an hourglass-like nebula in expansion, with an axis oriented along the PA of $\sim75^\circ$, and a disc in Keplerian rotation, probably in the equator of the nebula. They also suggested that the mass of the outflow is approximately eight times smaller than the disc's mass.

Geometric modelling for IW\,Car was performed by \citet{Hillen2017} to reproduce mid-IR interferometric observations obtained using the VLTI/MIDI instrument. The authors determined the outer disc diameter in mid-IR to be $38\pm3$\,mas. Subsequently, \citet{Kluska2019A&A...631A.108K} provided geometric modelling results for near-IR observations with VLTI/PIONIER. The best model in H-band contains a ring with an inner diameter of $23\pm0.7$\,mas, a width of $16.1\pm0.46$\,mas and an inclination of $45^\circ$. \citet{Kluska2022} classified the system as a full-disc model (category 1, see Section~\ref{sec:target_sel}) pointing the inner disc rim at the dust sublimation radius.

\textit{\#ID 4: HR\,4049}\\
HR\,4049 is a spectral type A4Ib/II post-AGB binary star. \citet{Oomen2019A&A...629A..49O} used SED fitting and $Gaia$\,DR2 distance to the system to estimate the luminosity value of $20300^{+14500}_{-7300}$\,L$_\odot ^5$. The target shows an orbital period of $430.6\pm0.1$\,days and an eccentricity of $0.30\pm0.01$ \citep{Oomen2018}. 

\citet{Malek_1_2014ApJ...780...41M, Malek_2_2014ApJ...794..113M} revealed the optically thick molecular gas in the radially extended circumbinary disc of HR\,4049 at infrared wavelengths using the Spitzer-IRS observations. The presence of strong near-IR excess (1/3 of the stellar flux) was established by \citet{Dominik2003A&A...397..595D}. In addition, HR\,4049 shows strong emission features due to polycyclic aromatic hydrocarbons, however, signs of silicates are absent in the infrared spectrum \citep{Acke2013}. \citet{Hinkle2007ApJ...664..501H} suggested that HR\,4049 has an oxygen-rich circumbinary disc in a carbon-rich circumstellar shell. In addition, \citet{Jones_2022A&A...665A..21J} elaborated the presence and properties of nano-diamonds in the optically thin parts of the circumbinary disc (outer layers) through the identification of their characteristic CH and CH$_{2}$ stretching modes at 3.43 and 3.53 $\mu$m, respectively.

\citet{Hillen2017} detected a significant over-resolved flux component with VLTI/MIDI interferometric observations in mid-IR. The authors reproduced HR\,4049 data with a disc of $42\pm2$\,mas diameter. The object was also observed in near-IR using VLTI/PIONIER by \citet{Kluska2019A&A...631A.108K}. The best-fitted model included a disc with an inner rim diameter of $16.4\pm0.5$\,mas, a width of $13\pm1.6$\,mas and an inclination of $49^\circ$. Recently, \citet{Kluska2022} classified HR\,4049 as a full-disc system (see Section~\ref{sec:target_sel}).

\textit{\#ID 5: IRAS\,15469-5311}\\
IRAS\,15469-5311 is a spectral type F3 object.  \citet{Oomen2019A&A...629A..49O} used SED fitting and $Gaia$\,DR2 distance to the system to estimate the luminosity value of $17100^{+7200}_{-4500}$\,L$_\odot ^5$. \citet{Maas2005} found an effective temperature of the post-AGB star to be 7500 K. In addition, IRAS\,15469-5311 shows the strong depletion pattern of refractory elements in the atmosphere [Zn/Ti]=1.8 \citep{Maas2005}. More recently, the orbit of the binary system was established by \citet{Oomen2018}: orbital period is $390.2\pm0.7$ days, eccentricity is $0.08\pm0.02$, and the projected semi-major axis is $0.438\pm0.015$.

Geometric modelling for IRAS\,15469-5311 was performed by \citet{Hillen2017} using mid-IR interferometric data. The authors found a disc of $36\pm4$\,mas diameter. Subsequently, \citet{Kluska2019A&A...631A.108K} performed the same type of modelling for near-IR H-band interferometric observations of the system. The best-suited geometric model from the latter study contains two stars and a ring with an inner diameter of $10.4\pm0.3$\,mas, a width of $4\pm0.3$\,mas and an inclination of $53.5^\circ$. The system was recently classified as a full disc system by \citet{Kluska2022} based on the spectral energy distribution and the characteristics of the infrared excess (see Section~\ref{sec:target_sel}).

\textit{\#ID 6: IRAS\,17038-4815}\\
IRAS\,17038-4815 shows a spectral type G2p. \citet{Oomen2019A&A...629A..49O} used SED fitting and $Gaia$\,DR2 distance to the system to estimate the luminosity value of $4800^{+3200}_{-1600}$\,L$_\odot ^5$. \citet{Maas2005} derived an effective temperature for the post-AGB star of 4750 K, a metallicity of [Fe/H] = -1.5 and [Zn/Ti] = 0.7, which is an indicator of a depleted atmosphere. IRAS\,17038-4815 was initially classified as an RVa photometric type based on the only found period of $75.9 \pm1.9$ days using the ASAS photometric time series \citep{Kiss2007MNRAS.375.1338K}. The RVb nature of IRAS\,17038-4815 was found by the long-term photometric variation in the light curves on the time scale of the orbital period of 1356.2 days and the fundamental pulsation period of 37.9 days \citep{Manick2017}. The preliminary orbit of IRAS\,17038-4815 was already published by \citet{VanWinckel2007BaltA..16..112V} and refined by \citet{Oomen2018} using data from the CORALIE and HERMES spectrographs. The orbital period is found to be $1394\pm12$ days, eccentricity is $0.63\pm0.06$, and the projected semi-major axis is $1.52\pm0.08$. 

\citet{Hillen2017} modelled this post-AGB binary system by using geometric modelling for mid-IR interferometric observations. The authors determined the outer disc diameter to be $17\pm1$\,mas. Subsequently, \citet{Kluska2018A&A...616A.153K} performed the same technique for the near-IR observations with VLTI/PIONIER data. The resulting model for H-band contains a ring with an inner diameter of $5.3\pm0.3$\,mas, a width of $6.4\pm0.5$\,mas and an inclination of $36^\circ$. However, such low inclination contradicts the observed RVb phenomenon.

\subsection{Transition discs}

\textit{\#ID 7: RU\,Cen}\\
 RU\,Cen is a binary system with spectral type F6I. \citet{Oomen2019A&A...629A..49O} used SED fitting and $Gaia$\,DR2 distance to the system to estimate the luminosity value of $1100\pm200$\,L$_\odot ^5$. However, \citet{Bodi2019} defined the luminosity for the system of $1054^{+304}_{-159}$\,L$_\odot$ using empirical PLC relation and $Gaia$\,DR2 data.  Based on the luminosity, which is less than the tip-of-the-RGB ($\sim2500$\,L$_\odot$), RU\,Cen is very likely to be a dusty post-RGB binary star \citep{Kamath2016A&A...586L...5K}. \citet{Maas2002} defined two dominant pulsation periods for the object: 32.30 days and 64.64 days. The authors also performed a detailed chemical analysis and determined a strong depletion pattern in the photosphere (see Table~\ref{tab:initialdata}) and an effective temperature of $6000\pm250$ K. RU\,Cen is a wide binary (orbital period of $1489\pm10$ days) with one of the highest known eccentricities \citep[$e=0.62\pm0.07$, see][]{Oomen2018}. In a recent study by \citet{Oomen2020A&A...642A.234O}, the authors attempted to reproduce the orbit and chemical depletion pattern of RU\,Cen using the effect of accretion during the post-AGB phase with the MESA code. However, they found that it can only be reproduced if the initial orbit is allowed to be already eccentric. This formation scenario requires the presence of an unknown binary interaction mechanism that causes orbital eccentricity.
 
In addition, \citet{Kluska2019A&A...631A.108K} performed the geometric modelling for VLTI/PIONIER near-IR observations, and the best-suited model took into account only binary and a background flux. \citet{Kluska2022} classified RU\,Cen as a system with a large cavity in the disc based on colour diagrams and synthetic radiative transfer modelling (category 2, see Section~\ref{sec:target_sel}). In a recent mid-IR interferometric study with VLTI/MATISSE \citet{Corporaal2023arXiv230412028C} found the inner disc rim for the RU\,Cen to be ~3.5 times bigger than the theoretical dust sublimation radius.

\textit{\#ID 8: AC\,Her}\\
AC\,Her is a spectral type F2I post-AGB binary with an effective temperature of $5800\pm250$ K \citep{Oomen2018}. \citet{Bodi2019} defined the luminosity for the system of $2475^{+183}_{-209}$\,L$_\odot$ using empirical PLC relation and $Gaia$\,DR2 data. The luminosity of the system corresponds to the tip-of-the-RGB ($\sim2500$\,L$_\odot$) and thereby AC\,Her might be a dusty post-RGB binary star \citep{Kamath2016A&A...586L...5K}.

Gas in the rotation has been first directly observed for this object by \citet{Bujarrabal2015A&A...575L...7B}. \citet{Gallardo_cava2021A&A...648A..93G} also presented $^{12}$CO J = 2-1 mm-wave interferometric results and noticed the presence of a low-mass outflow surrounding the Keplerian disc. The authors estimated the outflow mass to be $<~ 3\%$ of the total nebulae mass. Furthermore, \citet{Bollen2022arXiv220808752B} defined the presence of the jet in the system with an  inclination of  $49.8\pm0.3^{\circ}$ and the jet opening angle of $29.8\pm0.1^{\circ}$ by modelling the orbital phase-dependent changes in the $H_\alpha$ lines.

The inner radius of the disc does not coincide with the dust sublimation radius despite the typical behaviour for post-AGB discs \citep{DeRuyter2006A&A...448..641D, Hillen2016}. \citet{Hillen2015A&A...578A..40H} showed that the disc in the AC\,Her system has a small gas/dust ratio and the inner disc rim is located much further than the dust sublimation radius (an inner rim of the disc was defined to be 35\,AU, the outer radius was fixed to 200\,AU, $i = 50\pm8^\circ$ and $PA = 305 \pm 10^\circ$). More recently, \citet{Kluska2019A&A...631A.108K} also performed geometric modelling of the near-IR interferometric observations. However, the inner rim of the disc appeared to be over-resolved, which is compatible with the most likely model: a star with an over-resolved flux. In a recent mid-IR interferometric study with VLTI/MATISSE \citet{Corporaal2023arXiv230412028C} found the inner disc rim for the AC\,Her to be ~7.5 times bigger than the theoretical dust sublimation radius.

\begin{landscape}
    
    \begin{table}
        \caption{Additional stellar and orbital parameters for each target in the sample.}
        \begin{tabular}{ c|l|l|c|c|c|c|c|c|c|c|c|c|c|l }
            \hline
        \#ID& Name & \begin{tabular}[c]{@{}l@{}}$\alpha$ 2000\\ {[}h m s{]}\end{tabular} & \begin{tabular}[c]{@{}l@{}}$\delta$ 2000\\ {[} $^{\circ}$ ’ ”{]}\end{tabular} & \begin{tabular}[c]{@{}l@{}}$H$ \\ {[}mag{]}\end{tabular}& \begin{tabular}[c]{@{}l@{}}distance \\ {[}pc{]}\end{tabular} & \begin{tabular}[c]{@{}l@{}}Spectral \\ type\end{tabular}& \begin{tabular}[c]{@{}l@{}}$T_{\rm eff}$ \\ {[}K{]}\end{tabular}&  \begin{tabular}[c]{@{}l@{}}$P_{\rm puls}$\\ {[}days{]}\end{tabular} & RVb & [Zn/Fe] &[C/H]& \begin{tabular}[c]{@{}l@{}}$T_{\rm turn-off}$ \\ {(}K{)}\end{tabular}&  \begin{tabular}[c]{@{}l@{}}Depletion \\ profile\end{tabular}& Ref. \\
            \hline
    		\multicolumn{15}{c}{Full discs}\\
            \hline
            1& U\,Mon & 07 30 47 & -09 46 37 &3.9& 800$^{+117}_{-87}$ & G0I & 5000 & 92$\pm$3  & y & 0.1&-0.18&-&N& 1, 2, 6, 15 \\
            2& IRAS 08544-4431 & 08 56 14.1 & -44 43 10.7 &4.7& 1575$^{+124}_{-126}$ & F3 & 7250  & 72 & n &0.4&-0.02&1200&S & 1, 2, 7,14  \\
            3& IW\,Car   & 09 26 53 & -63 37 49 &5.1& 1314$^{+55}_{-71}$&F7I&6700& 72$\pm$1 & y & 1&0.32&1100&S& 3, 2, 8, 15 \\
            4& HR 4049  & 10 18 07.5 & -28 59 31.2 &4.2& 1449$^{+310}_{-186}$&A4Ib/II&7600&-&  y & 3.5&-&800&S & 1, 9 \\
            5&IRAS 15469-5311 & 15 50 44 & -53 20 43 &6.2& 3471$^{+375}_{-320}$     &F3&7500& 50 & n &0.3&0.3&1300&S & 1, 2, 11,13\\
            6&IRAS 17038-4815& 17 07 37 & -48 19 08 &7.1& 4718$^{+1014}_{-504}$     &G2p&4750& 37.9$\pm$1.5 &   y  &0.3&0.3&1400&S & 1, 4, 11, 13 \\
            \hline
    		\multicolumn{15}{c}{Transition discs}\\
            \hline
            7& RU\,Cen   & 12 09 23 & -45 25 35 &7.2& 4618$^{+1311}_{-1084}$ &F6I&6000&  &  n & 0.9&-0.42&800&P& 1, 10, 15 \\
            8& AC\,Her & 18 30 16.24   & 21 52 00.6  &5.3& 1627$^{+99}_{-90}$ &F2Iep&5500& 75$\pm$2 &  y &  0.7&-0.35&1200&U& 1, 5, 12, 13, 15 \\
            \hline
        \end{tabular}
        \begin{tablenotes}
        
        \small
        \item \textbf{Notes:}  The target sample is separated into full discs and transition discs based on the disc category of \citet{Kluska2022}. See Section~\ref{sec:target_sel} for more details. RA and Dec. coordinates are given for the J2000 epoch. $H$ represents the observed photometric $H$-band magnitude. The distances to binary post-AGB stars were adopted from $Gaia$\,EDR3 \citep{Bailer-Jones2021AJ....161..147B}. However, we note that these distances are uncertain because: i) they are too far away and therefore not flagged as astrometric binaries ii) the orbital motion of the binary results in an angular displacement comparable to the parallax. $T_{\rm eff}$ represents the spectroscopically determined effective temperature. $P_{\rm puls}$ represents the pulsating period of the post-AGB star in days. RVb represents the presence of RVb phenomenon with 'y' indicating 'yes' and 'n' indicating 'no'. $T_{\rm turn-off}$ represents the turn-off temperature (which defines the limit for which elements with higher condensation temperatures become depleted). Depletion profile shapes were adopted from \citet{Oomen2019A&A...629A..49O}. Saturated profiles are assigned by 'S', plateau profiles are assigned by 'P', in cases where the distinction is not clear - 'U', and the stars that are not depleted - 'N'. More details on the tabulated information can be found in the individual studies mentioned in column 'Ref': 1 - \citet{Oomen2018}, 2 - \citet{Kiss2007MNRAS.375.1338K}, 3 - \citet{Kiss2017}, 4 -\citet{Manick2017}, 5 - \citet{Samus2009yCat....102025S}, 6 - \citet{Giridhar2000ApJ...531..521G}, 7 - \citet{Maas2003A&A...405..271M}, 8 - \citet{Giridhar1994ApJ...437..476G}, 9 - \citet{VanWinckel1995PhDT........31V}, 10 - \citet{Maas2002}, 11 - \citet{Maas2005}, 12 - \citet{VanWinckel1998A&A...336L..17V}, 13 - \citet{Giridhar1998ApJ...509..366G}, 14 - \citet{Maas2003A&A...405..271M}, 15 - \citet{Bodi2019}. \\
       
        \end{tablenotes}
       
        \label{tab:initialdata}
    \end{table}   
    
    \begin{table}
        \centering
         \caption{Geometrical modelling results for VLTI/PIONIER and VLTI/MIDI  surveys}
     	\label{tab:rt-modelling}
        \begin{tabular}{c|l|c|l|l|l|l|l|c|} 
            \hline
            & & \multicolumn{6}{c}{Geometrical modelling for VLTI/PIONIER survey} & \multicolumn{1}{c}{A mid-IR interferometric survey with VLTI/MIDI}\\ 
            & & \multicolumn{6}{c}{Kluska et al. 2019} & \multicolumn{1}{c}{Hillen et. al. 2017}\\ 
    
            \hline
            \#ID& Name & Model & Tring & $\theta$, mas & $\delta\theta$ & $i$ & PA  & $\theta$, mas \\ 
            \hline
            1& U\,Mon & br5 & 2619$^{+138}_{-132}$&5.49$^{+0.03}_{-0.03}$& 0.02$^{+0.02}_{-0.01}$ & 57.9$^{+1.6}_{-1.5}$ & 45$^{+1}_{-1}$ & 50$^{+0.5}_{-0.5}$ $^{*}$ \\ 
           
            2&IRAS 08544-4431 & br5 & 875$^{+10}_{-9}$&14.3$^{+0.1}_{-0.1}$& 0.48$^{+0.01}_{-0.01}$& 21.3$^{+0.8}_{-0.8}$ & 12$^{+3}_{-3}$& 46$^{+2}_{-2}$\\ 
            
            3& HR 4049 & br5 & 711$^{+24}_{-23}$& 16.4$^{+0.5}_{-0.5}$& 0.8$^{+0.1}_{-0.1}$& 49.3$^{+3.2}_{-3.3}$ & 63$^{+7}_{-6}$& 42$^{+2}_{-2}$\\ 
    
            4& IW\,Car & br5 & 1047$^{+15}_{-15}$& 23.0$^{+0.7}_{-0.6}$& 0.71$^{+0.02}_{-0.02}$& 44.6$^{+1.7}_{-1.8}$ & 155$^{+2}_{-2}$&38$^{+3}_{-3}$ \\

            5&IRAS 15469-5311 &br4 & 818$^{+17}_{-17}$&10.4$^{+0.3}_{-0.3}$&0.39$^{+0.03}_{-0.03}$& 53.5$^{+1.8}_{-2.2}$ & 64$^{+2}_{-2}$ &36$^{+3}_{-4}$  \\ 
    
            6&IRAS 17038-4815& sr6& 2132$^{+118}_{-104}$ &5.3$^{+0.3}_{-0.3}$ & 1.2$^{+0.1}_{-0.1}$ & 36.7$^{+5.1}_{-7.8}$ & 158$^{+7}_{-8}$&  
        19.8$^{+0.8}_{-0.9}$ $^{*}$  \\ 
            7& RU\,Cen & b2 & - & -& -& -& -&-\\ 
            8& AC\,Her & s0& 5812$^{+2713}_{-2239}$ & - &-& -& -&65.7$^{+0.7}_{-0.7}$\\
            \hline
         
        \end{tabular}
        \begin{tablenotes}
            \small
            \item \textbf{Notes:} \citet{Kluska2019A&A...631A.108K}: $\theta$ represents an inner rim diameter for the fitted ring, $\delta\theta$ represents the width of the ring in the units of the ring radius. Column 'Model' represents components of the model that were used for the analysis: 'br5' - the primary star, the secondary star, a ring, and a background, 'br4' -  the primary star, the secondary star, a ring, and a background, 'b2' - two stars and a background flux, 'sr6' - the primary star, the ring, and the background, 's0' - a single star and background flux.\\
            \citet{Hillen2017}: $\theta$ represents an outer diameter of the disc. This parametric modelling assumed that the dust disc starts at the sublimation radius, which is not applicable for AC\,Her as it has confirmed inner disc radius is much larger than the dust-sublimation radius \citep[][]{Hillen2015A&A...578A..40H}.\\
            $^{*}$ - observations were made during the maximum brightness of the system.
        \end{tablenotes}
    \end{table}

\end{landscape}

\section{Observation details}
\label{sec:ap_weather}

In Table~\ref{tab:weather}, we list weather and seeing conditions for the observations. We note that U\,Mon was observed twice, on the 3rd and 14th of January 2019, because of improper observing conditions at the time of the first observation.

\begin{table*}
	
	\caption{Weather and seeing conditions for the SPHERE/IRDIS observations (data from ESO NightLogs).}
	\label{tab:weather}
	\begin{tabular}{|c|c|c|c|c|c|p{2.5cm}|c|}
		\hline
		\#ID& Name & Seeing (arcsec) & Humidity (\%) & Wind (m/s) & Temperature (C) & Comments & Date of observation\\
		\hline
    	\multicolumn{8}{c}{Full discs}\\
		\hline
		1& U\,Mon$^{a}$ & 0.32 - 1.78 & 12.0 - 19.0 & 0.0 - 8.1 & 15.5 - 18.4 &Thick clouds at the east thin clouds passing during the night on 3.01.2019 & 03.01.2019\\
            1& U\,Mon$^{b}$ & 0.3 - 1.46 & 6.0 - 30.0 & 3.98 - 11.0 & 10.8 - 13.9 && 14.01.2019\\
	    2&IRAS 08544-4431& 0.47 - 1.38 & 8.0 - 12.0 & 0.4 - 5.78 & 12.8 - 13.9 & &  25.10.2018\\
	    3&IW\,Car & 0.56 - 2.39 & 27.0 - 31.0 & 4.3 - 11.5 & 6.1 - 7.9 & &  13.10.2018\\
	    4&HR 4049 & 0.46 - 1.31 & 13.0 - 26.0 & 0.0 - 8.65 & 10.4 - 12.0 & &  23/29.11.2018 \\
		5&IRAS 15469-5311 & 0.48 - 1.16 & 6.0 - 14.0 & 0.0 - 7.25 & 14.7 - 16.3 & &  21/24.03.2019\\
		6&IRAS 17038-4815 & 0.39 - 1.41 & 3.0 - 42..0 & 6.63 - 12.7 & 12.7 - 15.3 & &  14/16.03.2019\\
		\hline
    	\multicolumn{8}{c}{Transition discs}\\
		\hline
		7&RU\,Cen & 0.48 - 1.31 & 28.0 - 41.0 & 6.08 - 13.5 & 8.16 - 10.7 & &  12.01.2019\\
	    8& AC\,Her & 0.36 -1.05 &4 -14 & 1.3 - 6.43 & 12.8 - 15.5 & &  16.05.2018\\
		\hline
		
	\end{tabular}
	\begin{tablenotes}
        
    \small
    \item \textbf{Notes:} U\,Mon was observed twice, on the 3rd ($^{a}$) and 14th ($^{b}$) of January 2019, because of improper observing conditions at the time of the first observation.\\
    \end{tablenotes}
\end{table*}

\section{The polarized signal of individual targets}
\label{sec:ap_pol_imag}

\begin{figure*}
  
    \includegraphics[width=0.5\columnwidth]{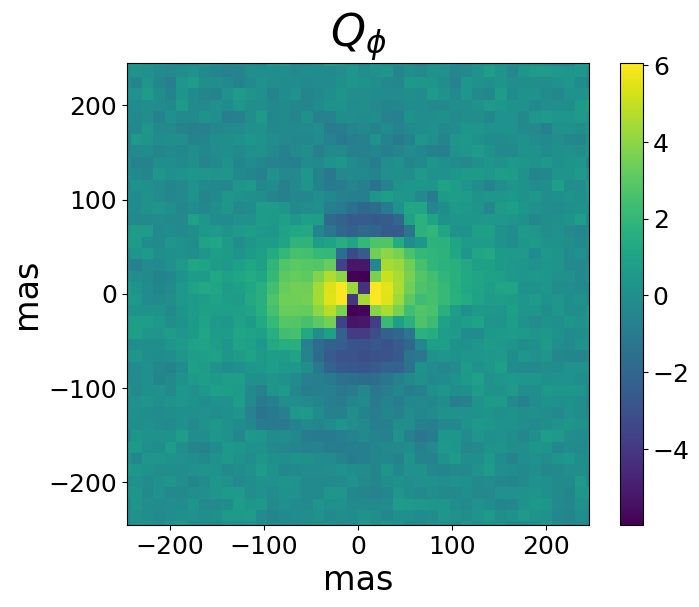}
    \includegraphics[width=0.5\columnwidth]{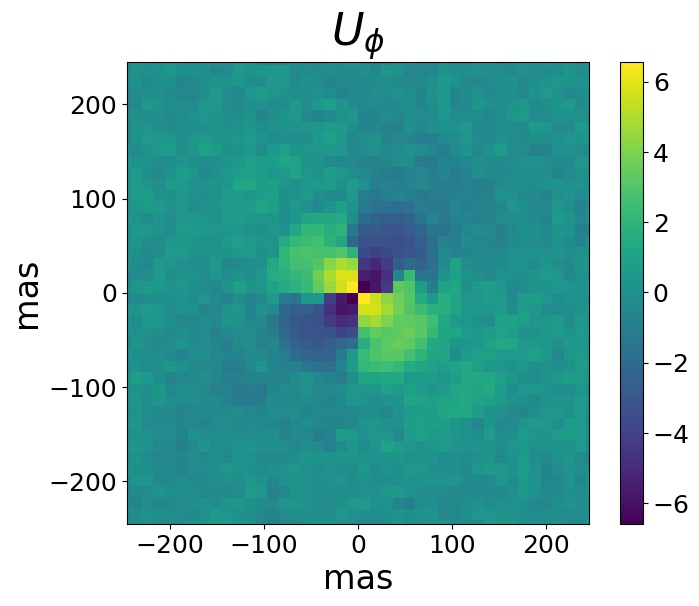}
    \includegraphics[width=0.5\columnwidth]{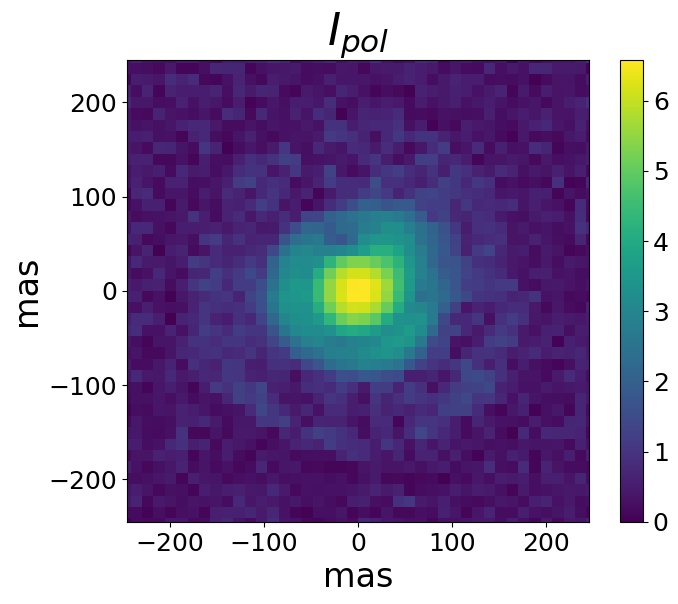}
    \includegraphics[width=0.5\columnwidth]{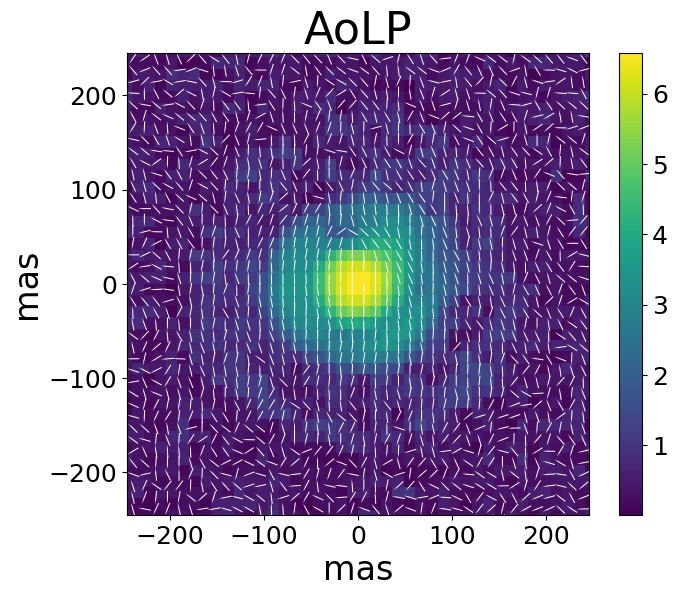}

    \includegraphics[width=0.5\columnwidth]{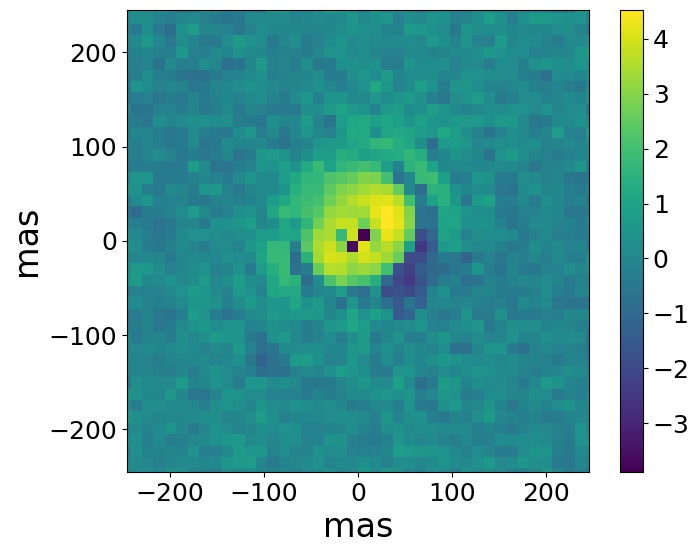}
    \includegraphics[width=0.5\columnwidth]{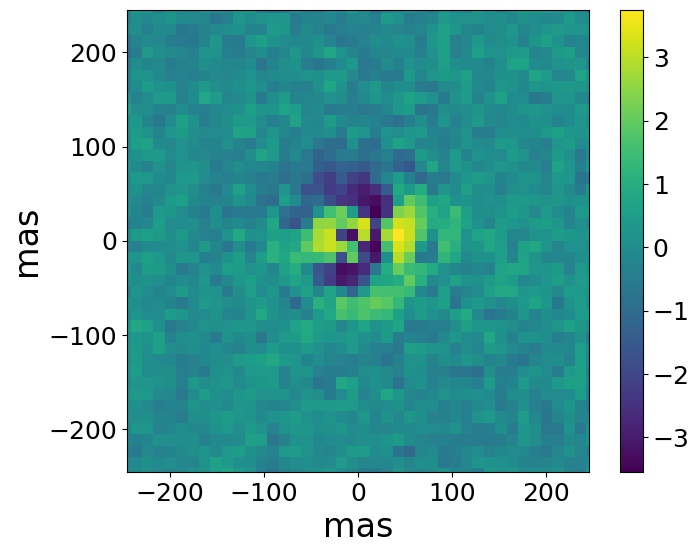}
    \includegraphics[width=0.5\columnwidth]{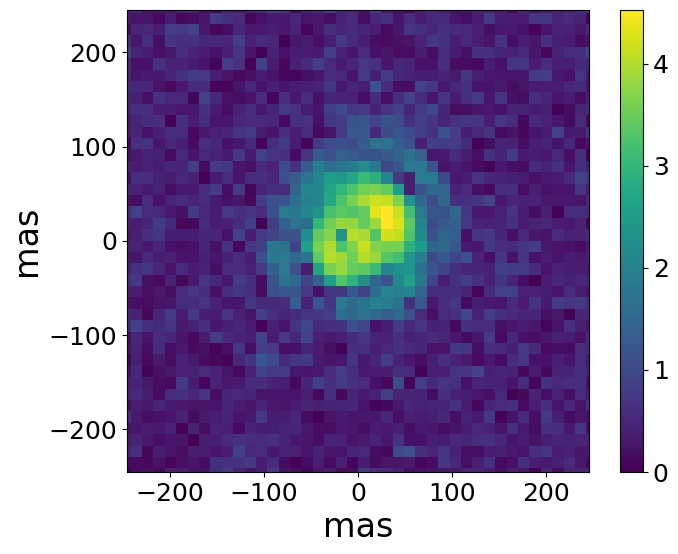}
    \includegraphics[width=0.5\columnwidth]{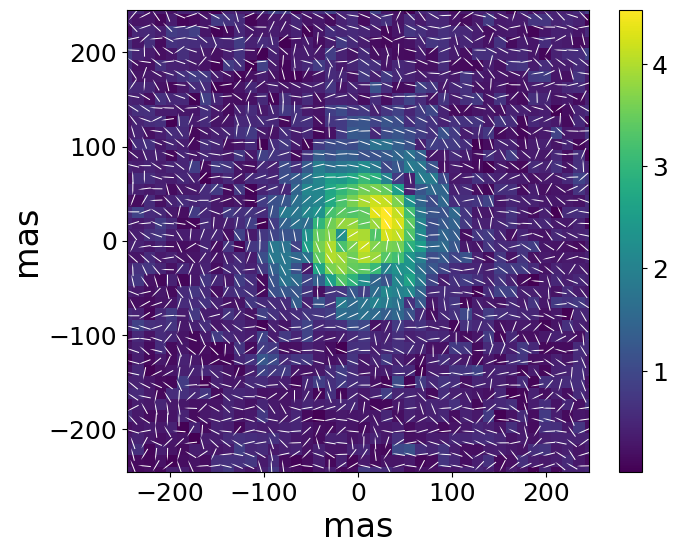}
    
    \caption{The polarized signal of U\,Mon before (top row) and after (bottom row) subtraction of the unresolved central polarization. To improve the signal-to-noise ratio, we provide results for the mean combined reduced data of two independent observations (see Section~\ref{sec:target_sel}). The first column represents the Q$_\phi$  images, the second column represents the U$_\phi$ images, the third column represents the total polarized intensity ($I_{\rm pol}$) images, and the last column represents the Angle of Linear polarization ($AoLP$) plotted over the $I_{\rm pol}$ image. All images are presented on an inverse hyperbolic scale and oriented North to up and East to the left. See Section~\ref{sec:data_reduction} for more details. 
    \label{fig:u_mon}}
\end{figure*}

\begin{figure*}
  
    \includegraphics[width=0.5\columnwidth]{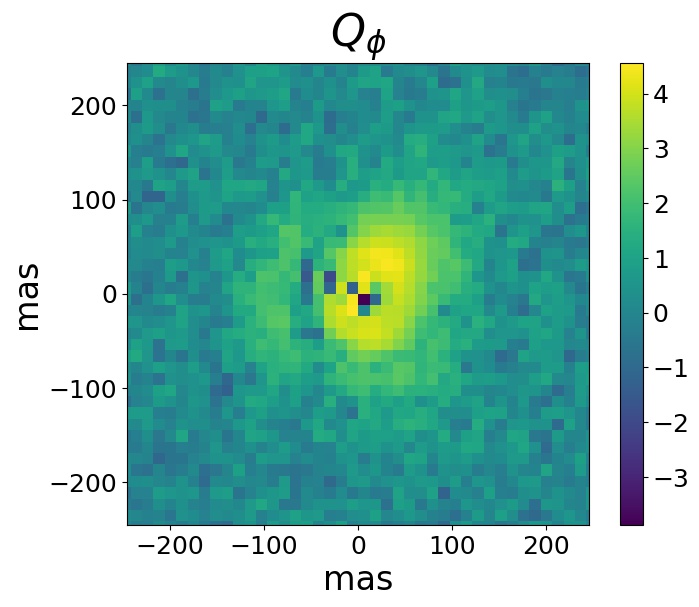}
    \includegraphics[width=0.5\columnwidth]{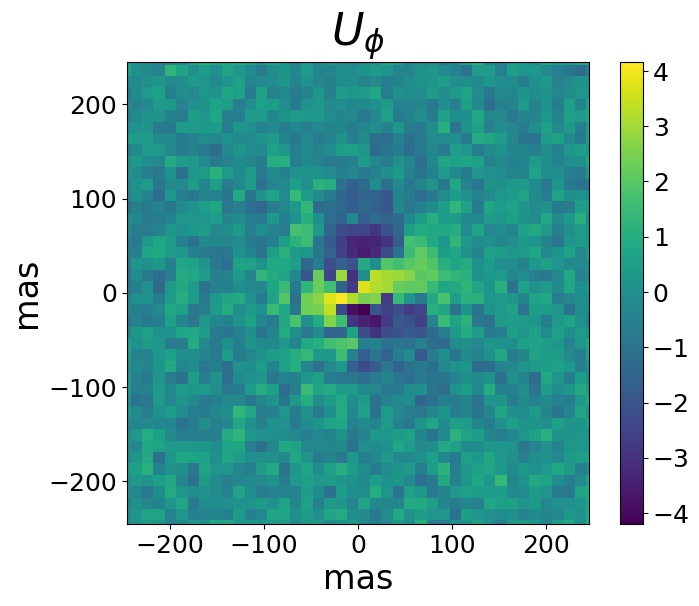}
    \includegraphics[width=0.5\columnwidth]{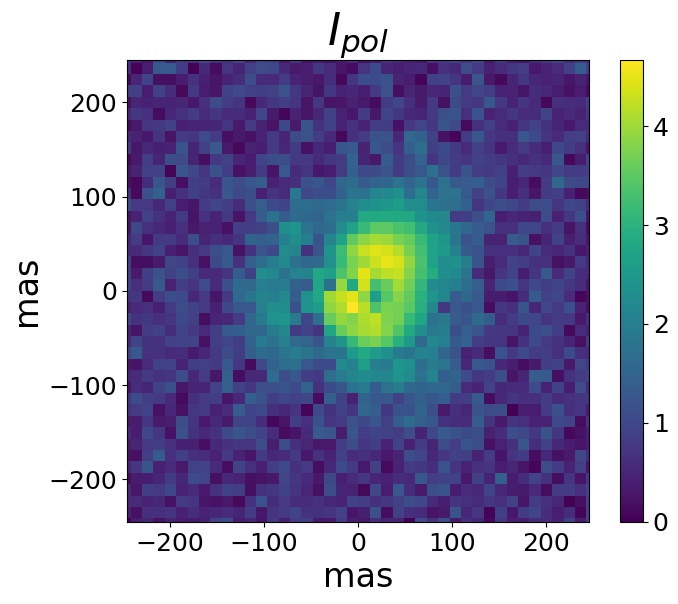}
    \includegraphics[width=0.5\columnwidth]{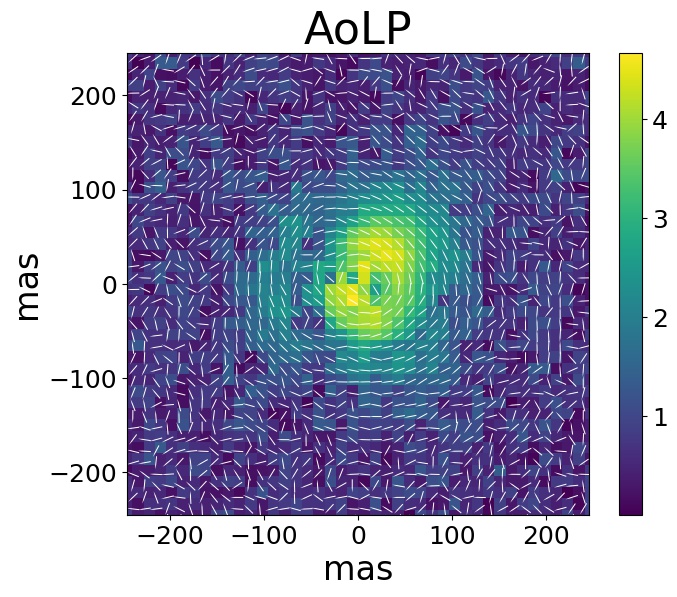}

    \includegraphics[width=0.5\columnwidth]{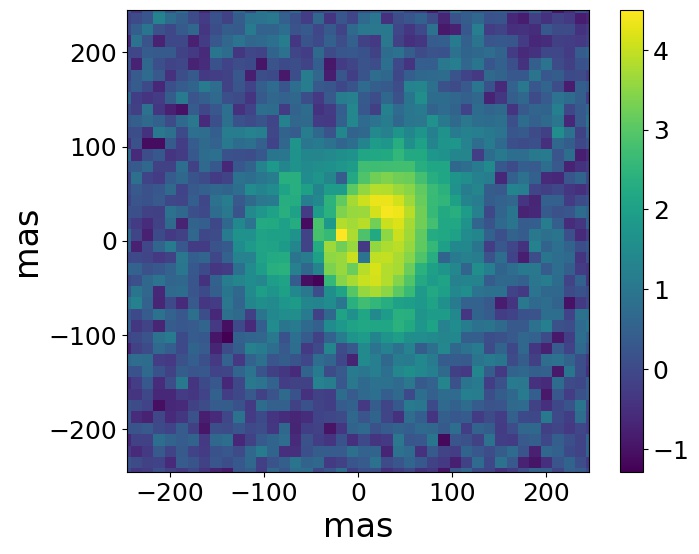}
    \includegraphics[width=0.5\columnwidth]{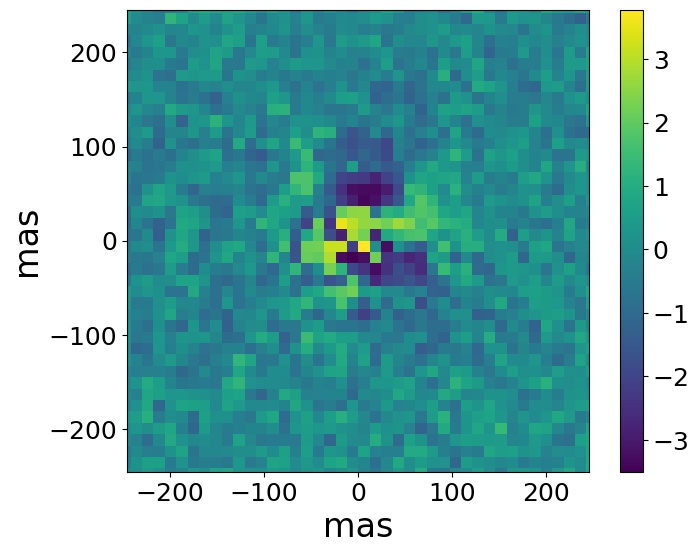}
    \includegraphics[width=0.5\columnwidth]{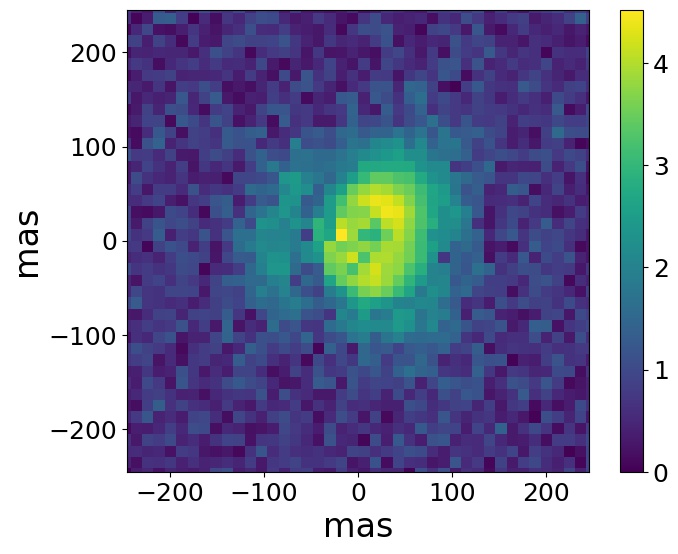}
    \includegraphics[width=0.5\columnwidth]{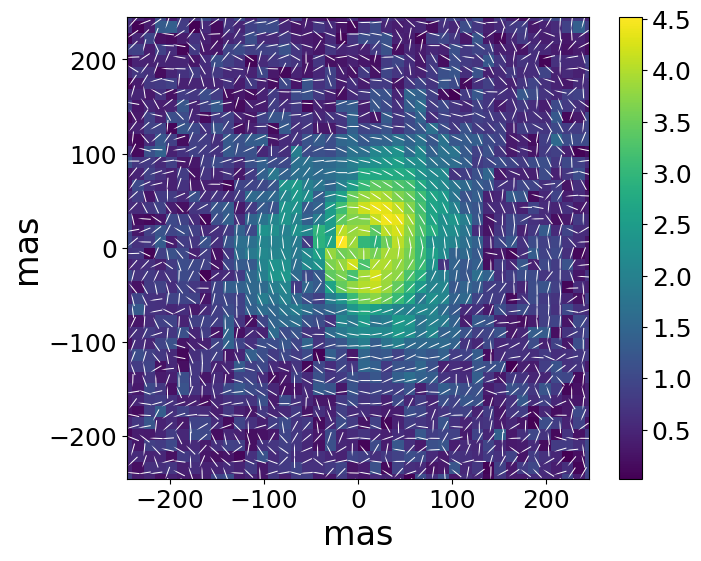}
    
    \caption{Same as Fig.~\ref{fig:u_mon} for IW\,Car.
    \label{fig:iw_car}}
\end{figure*}

\begin{figure*}
  
    \includegraphics[width=0.5\columnwidth]{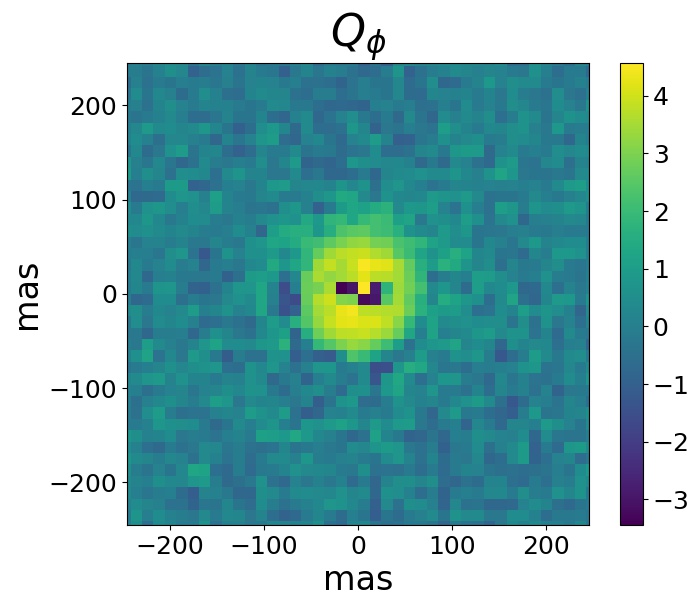}
    \includegraphics[width=0.5\columnwidth]{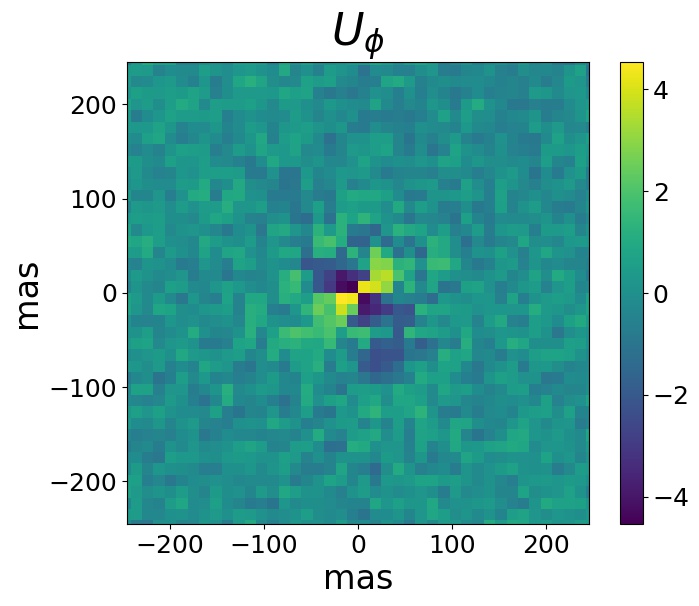}
    \includegraphics[width=0.5\columnwidth]{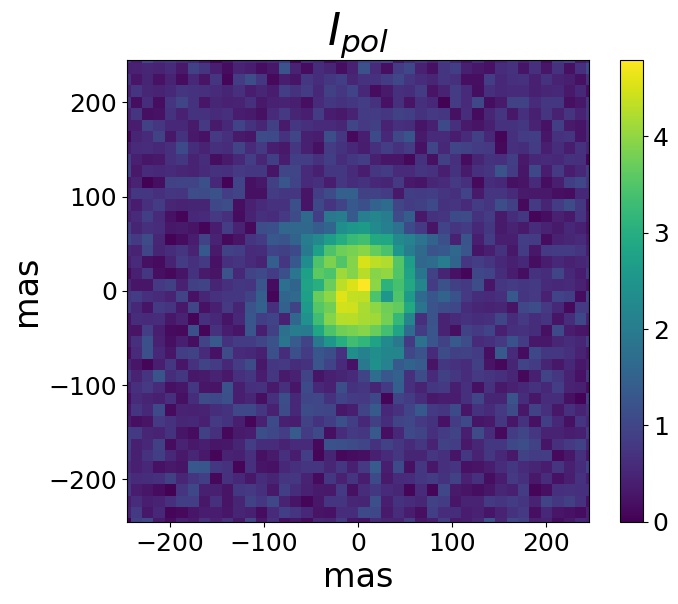}
    \includegraphics[width=0.5\columnwidth]{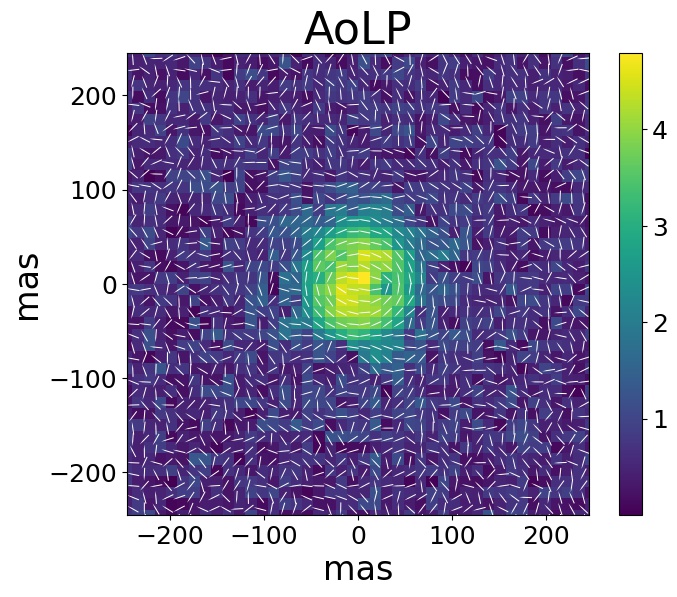}

    \includegraphics[width=0.5\columnwidth]{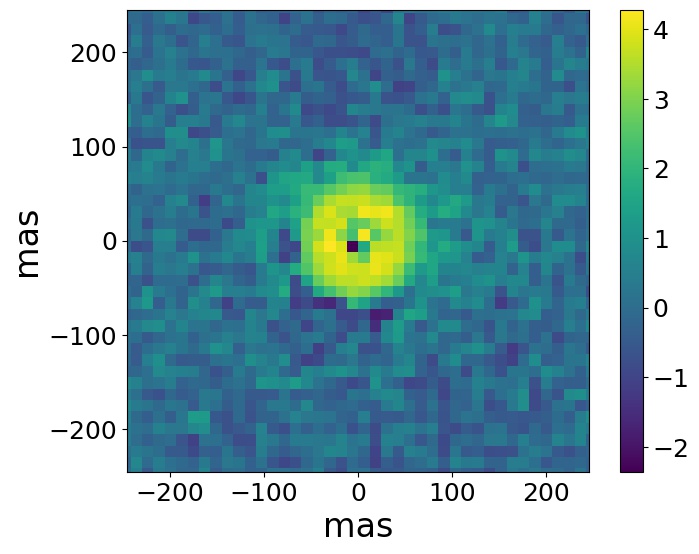}
    \includegraphics[width=0.5\columnwidth]{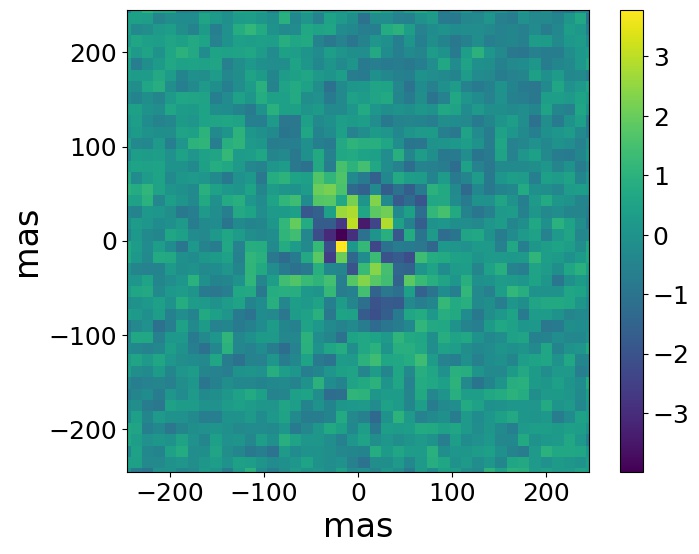}
    \includegraphics[width=0.5\columnwidth]{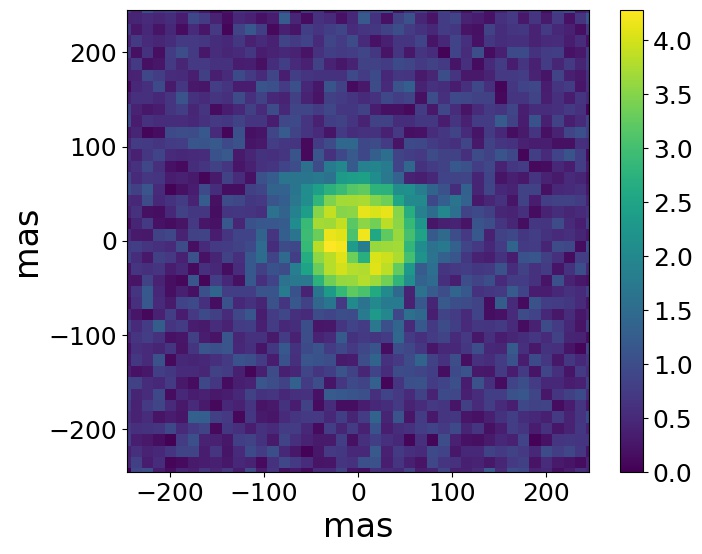}
    \includegraphics[width=0.5\columnwidth]{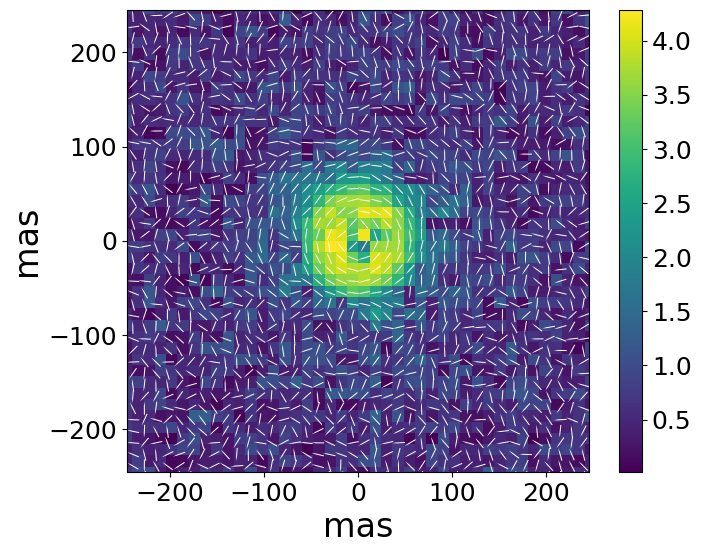}

    \caption{Same as Fig.~\ref{fig:u_mon} for HR\,4049.
    \label{fig:hr_4049}}
\end{figure*}

\begin{figure*}
  
    \includegraphics[width=0.5\columnwidth]{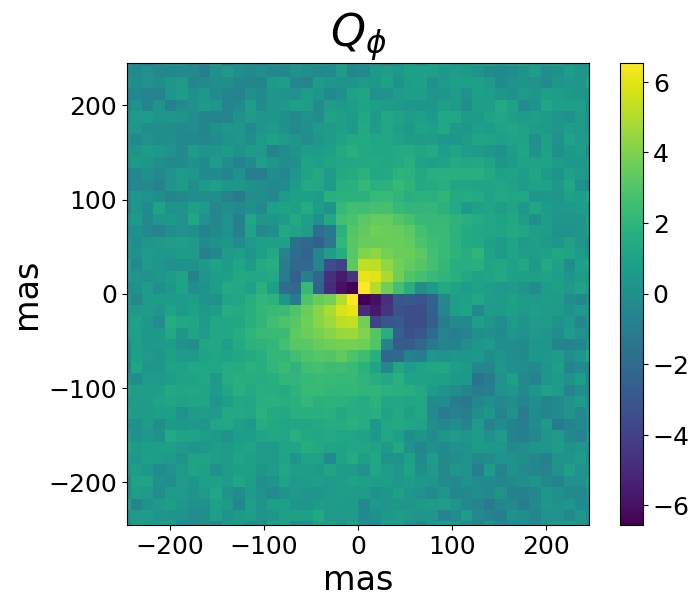}
    \includegraphics[width=0.5\columnwidth]{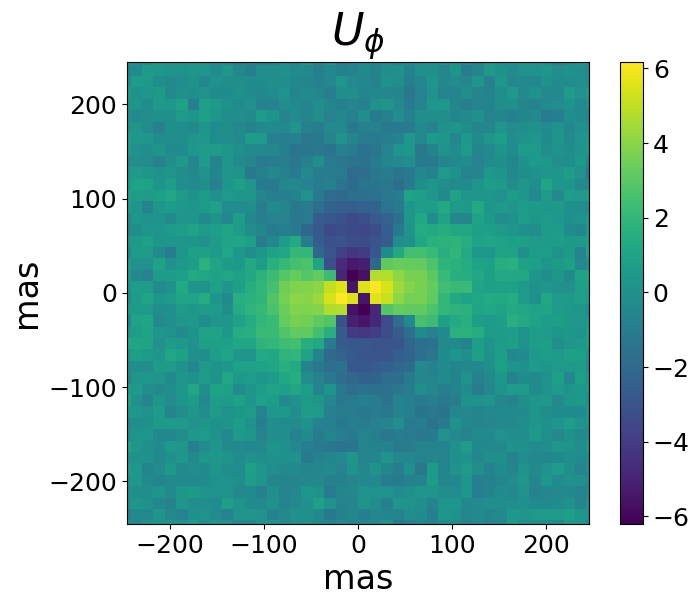}
    \includegraphics[width=0.5\columnwidth]{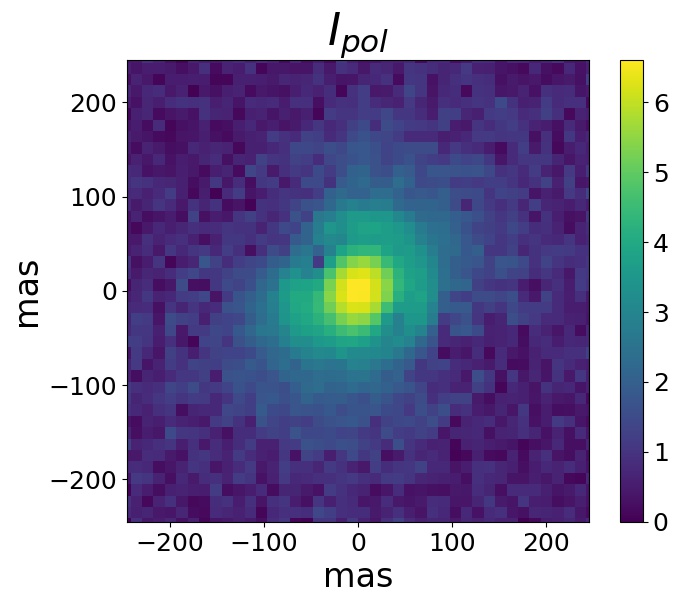}
    \includegraphics[width=0.5\columnwidth]{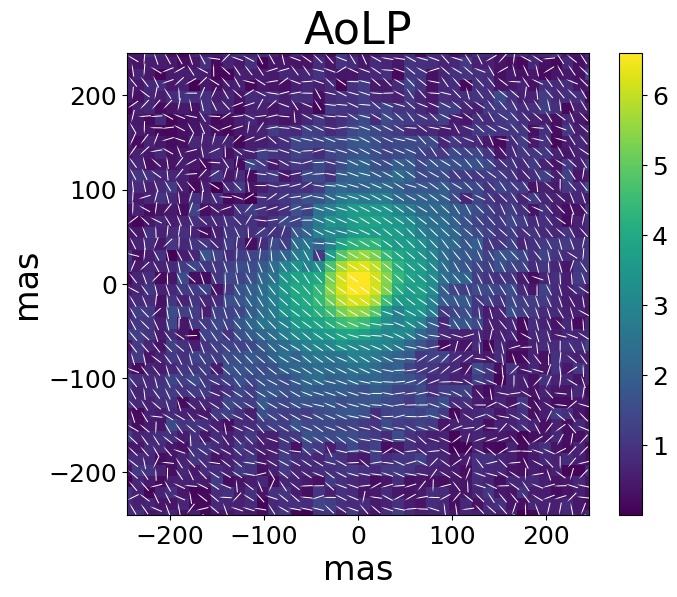}

    \includegraphics[width=0.5\columnwidth]{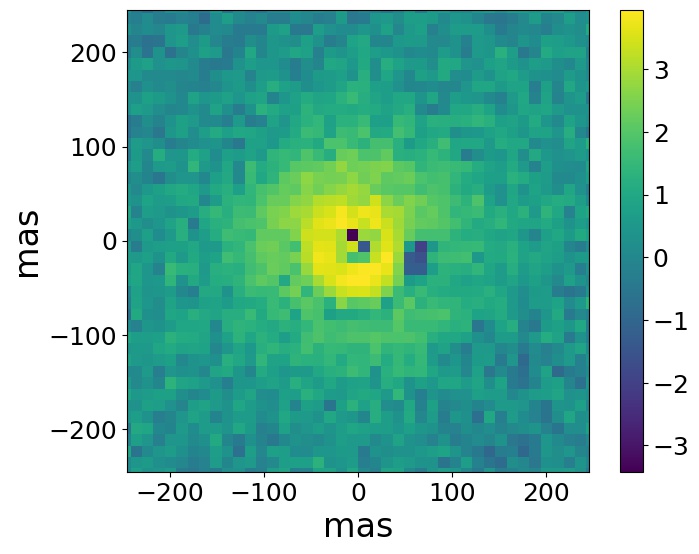}
    \includegraphics[width=0.5\columnwidth]{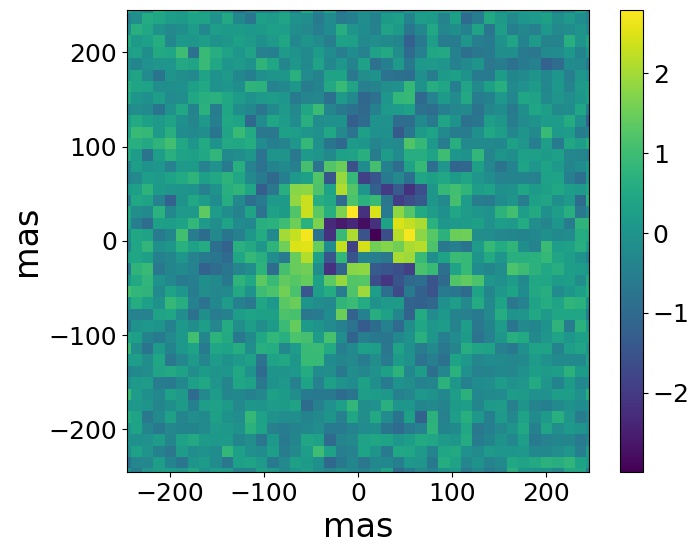}
    \includegraphics[width=0.5\columnwidth]{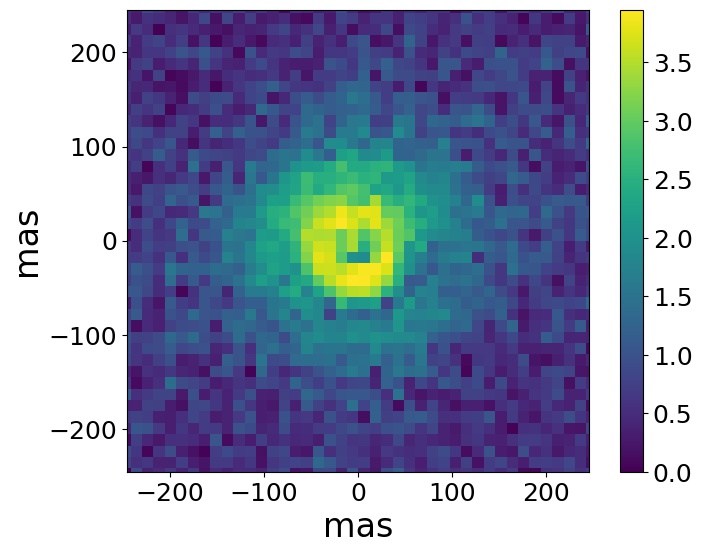}
    \includegraphics[width=0.5\columnwidth]{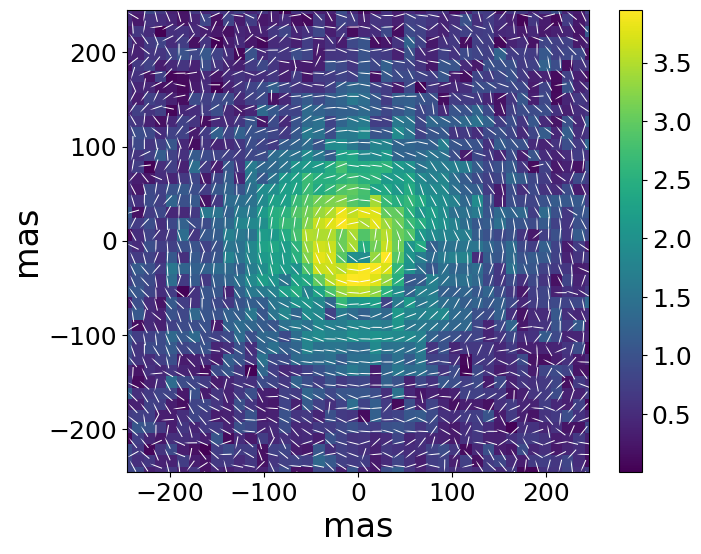}

    \caption{Same as Fig.~\ref{fig:u_mon} for IRAS\,15469-5311.
    \label{fig:iras15469-5311}}
\end{figure*}

\begin{figure*}
  
    \includegraphics[width=0.5\columnwidth]{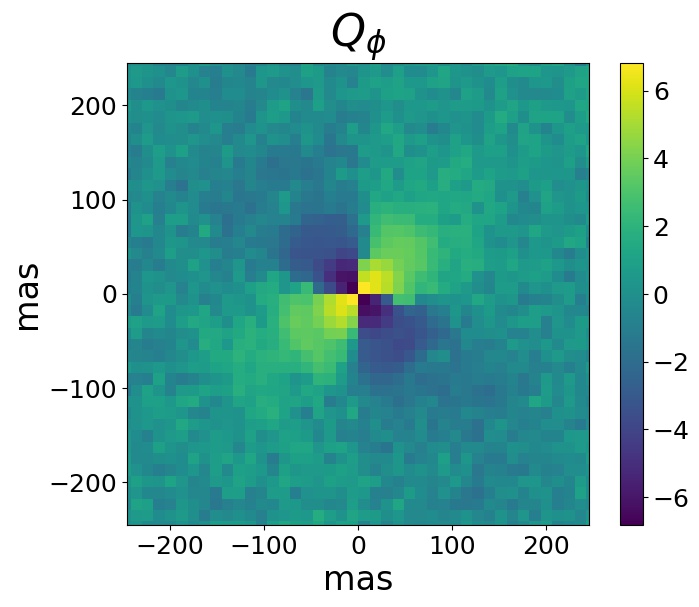}
    \includegraphics[width=0.5\columnwidth]{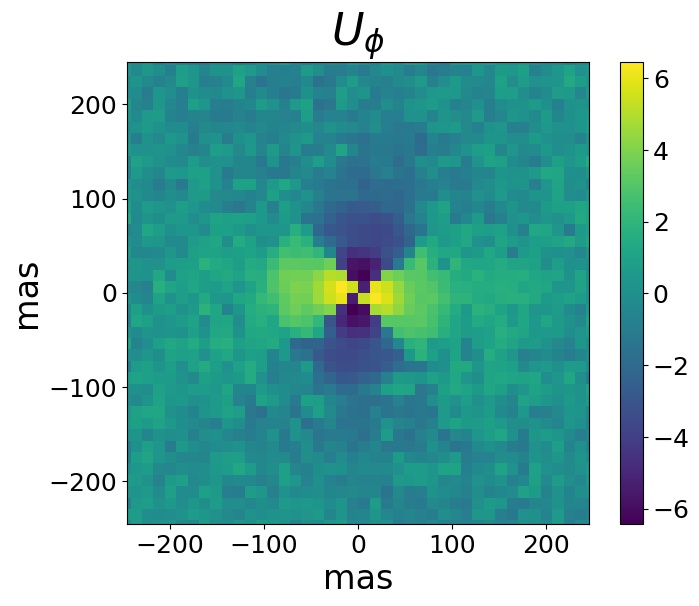}
    \includegraphics[width=0.5\columnwidth]{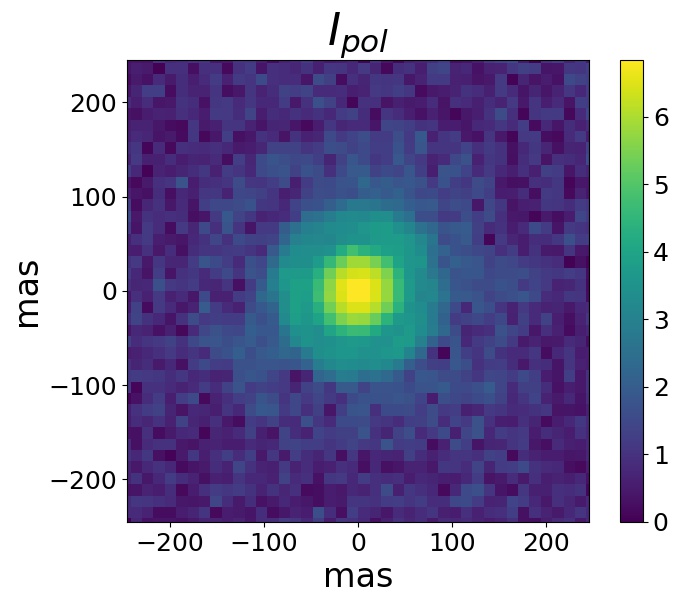}
    \includegraphics[width=0.5\columnwidth]{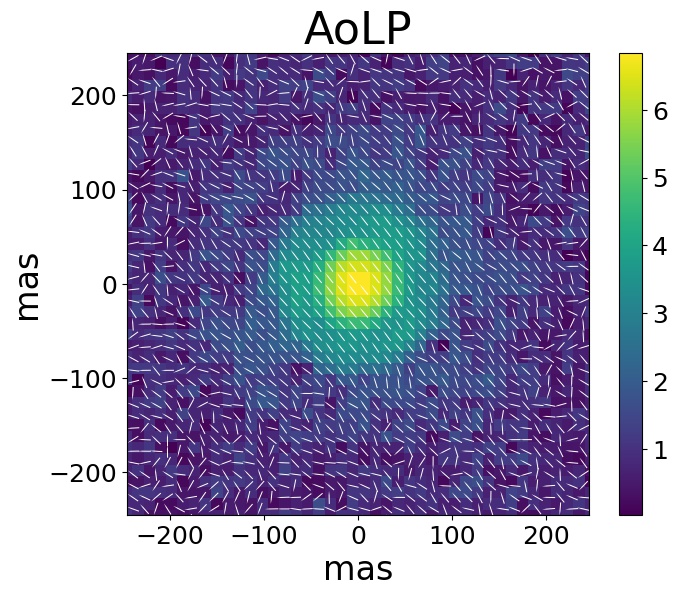}

    \includegraphics[width=0.5\columnwidth]{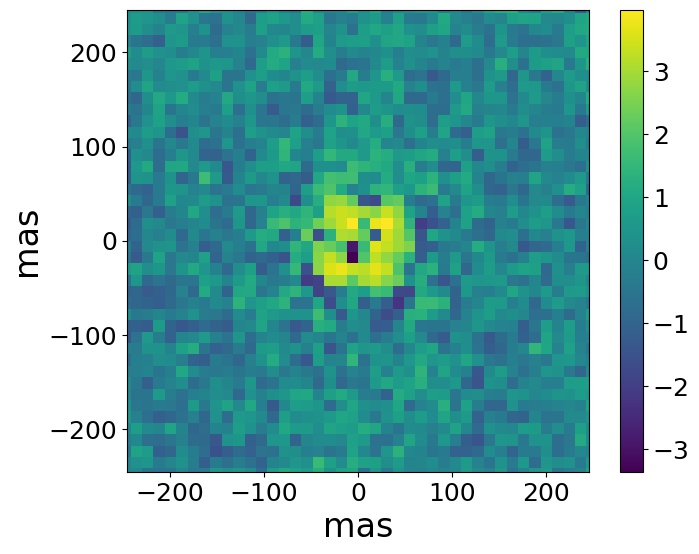}
    \includegraphics[width=0.5\columnwidth]{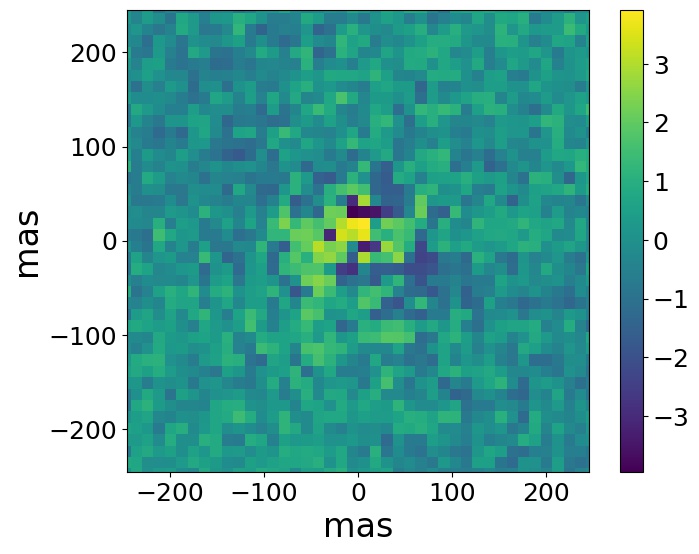}
    \includegraphics[width=0.5\columnwidth]{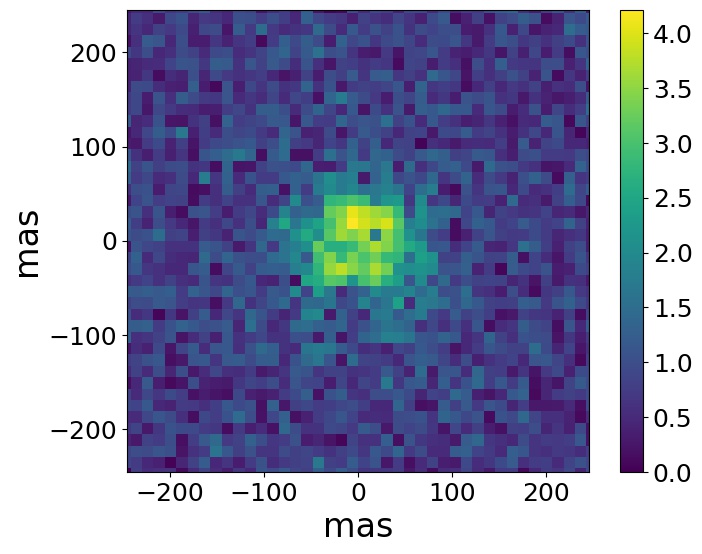}
    \includegraphics[width=0.5\columnwidth]{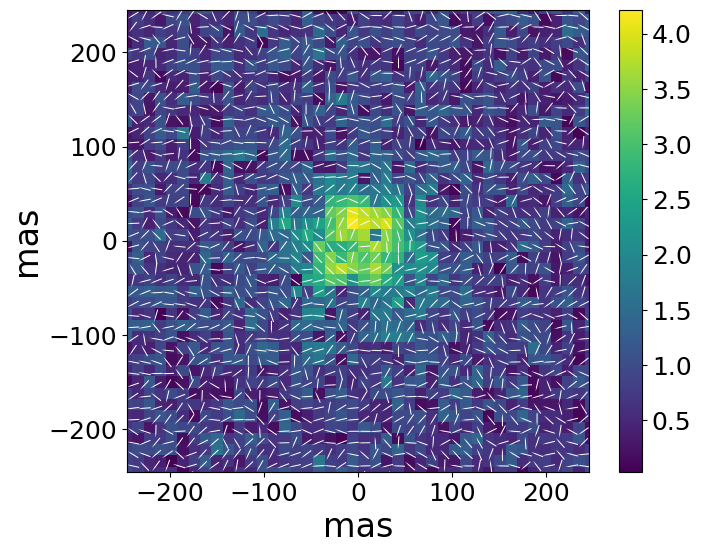}

    \caption{Same as Fig.~\ref{fig:u_mon} for IRAS\,17038-4815.
    \label{fig:iras17038-4815}}
\end{figure*}

\begin{figure*}
  
    \includegraphics[width=0.5\columnwidth]{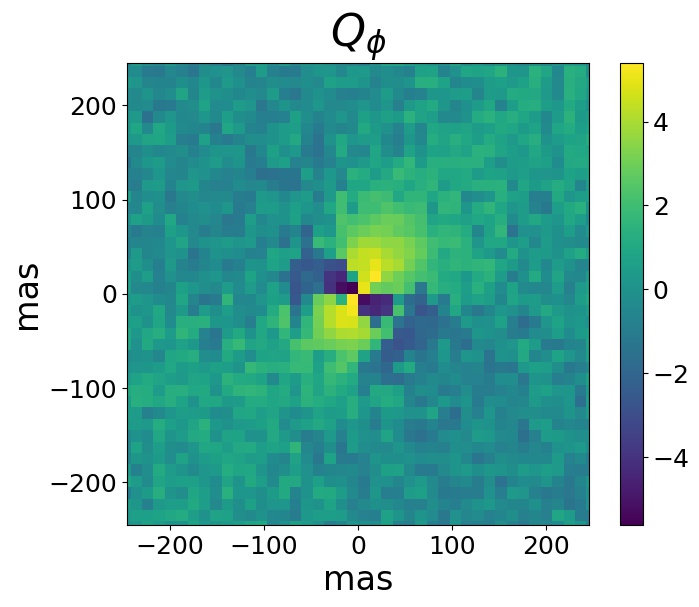}
    \includegraphics[width=0.5\columnwidth]{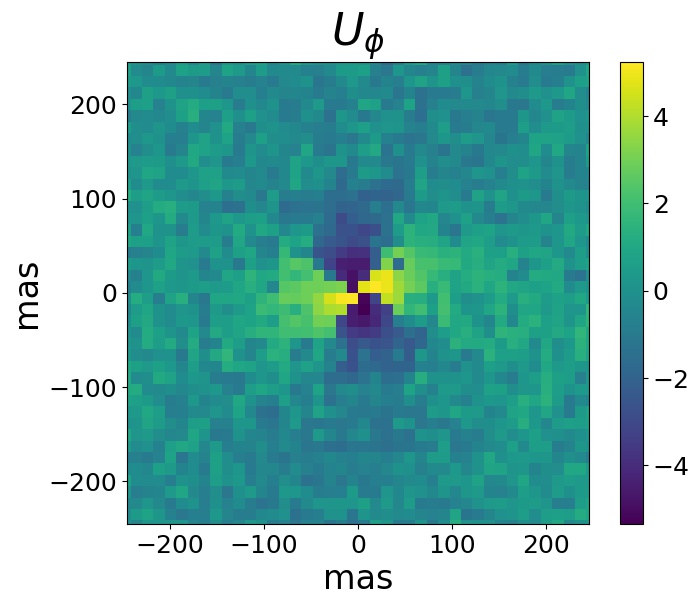}
    \includegraphics[width=0.5\columnwidth]{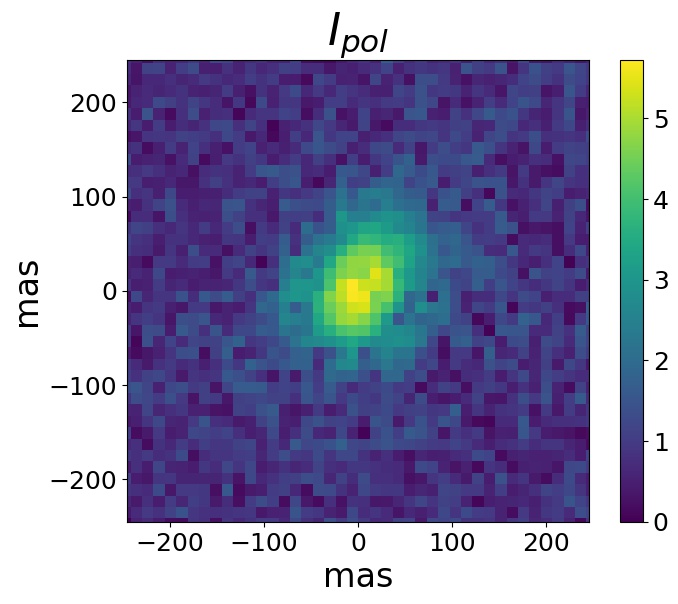}
    \includegraphics[width=0.5\columnwidth]{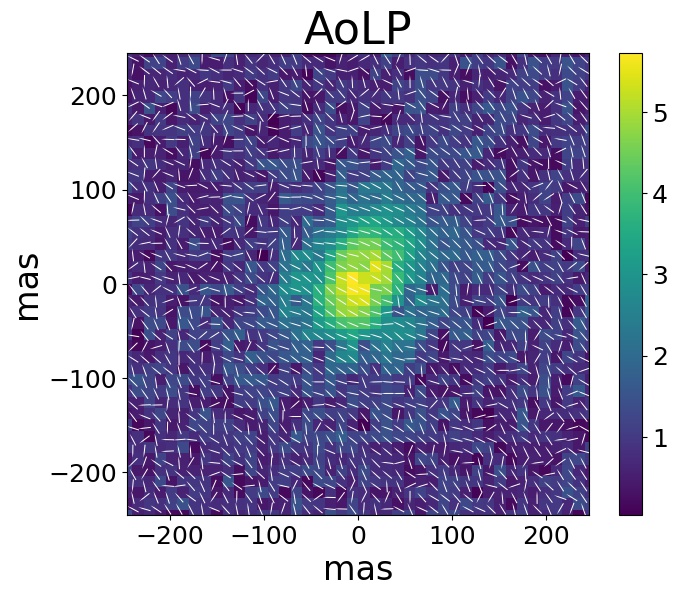}

    \includegraphics[width=0.5\columnwidth]{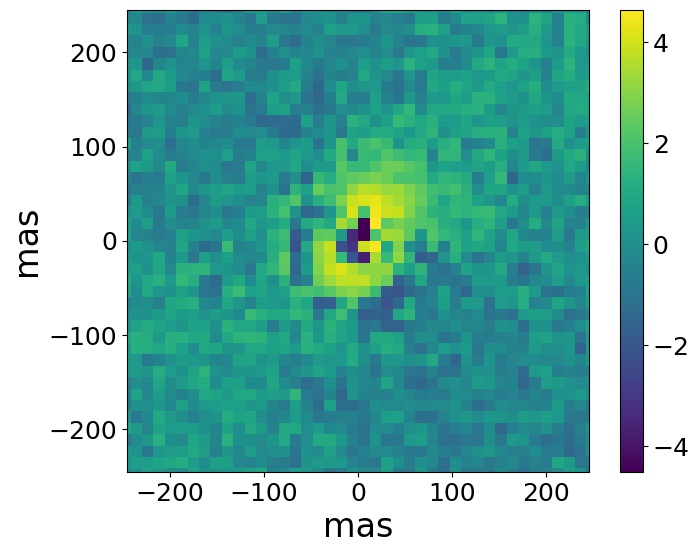}
    \includegraphics[width=0.5\columnwidth]{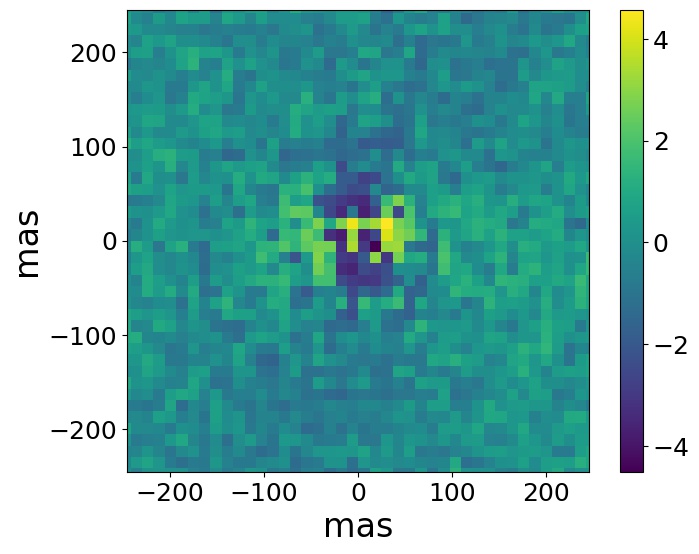}
    \includegraphics[width=0.5\columnwidth]{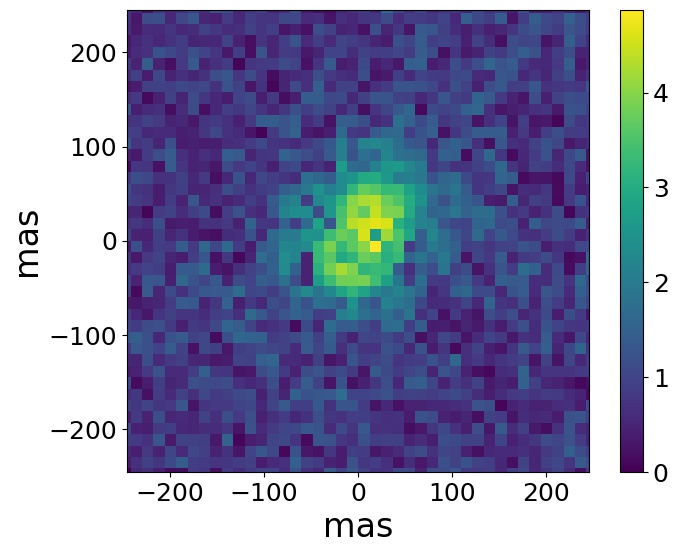}
    \includegraphics[width=0.5\columnwidth]{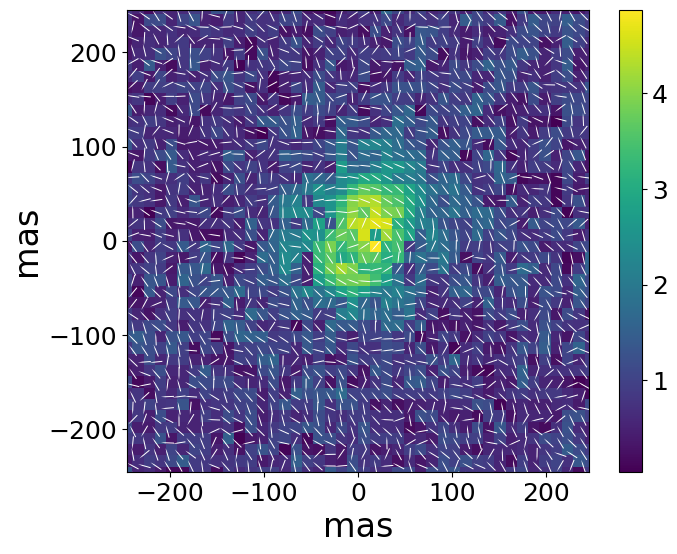}

    \caption{Same as Fig.~\ref{fig:u_mon} for RU\,Cen.
    \label{fig:ru_cen}}
\end{figure*}

\begin{figure*}
  
    \includegraphics[width=0.5\columnwidth]{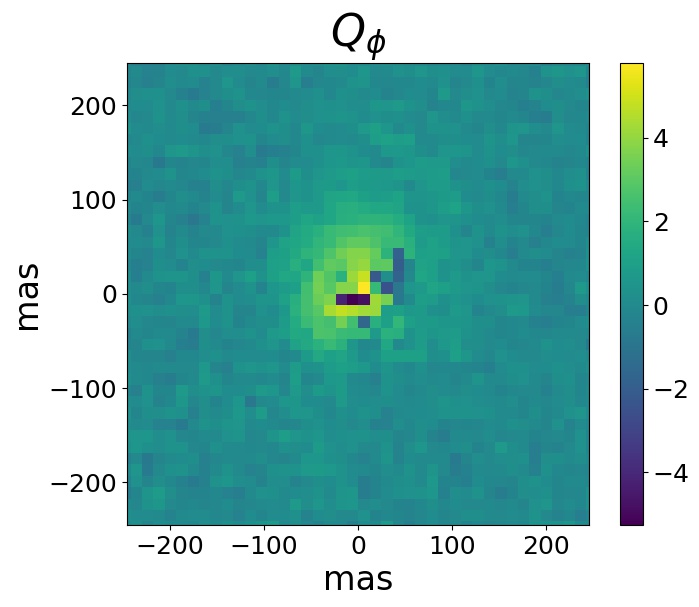}
    \includegraphics[width=0.5\columnwidth]{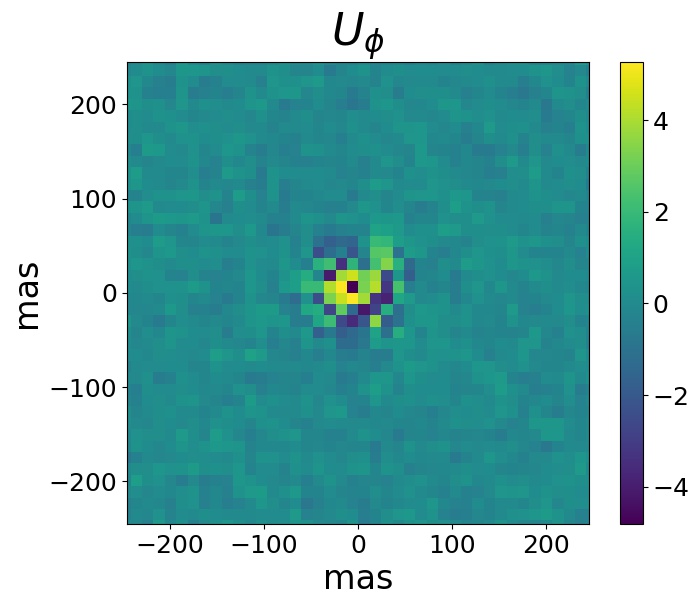}
    \includegraphics[width=0.5\columnwidth]{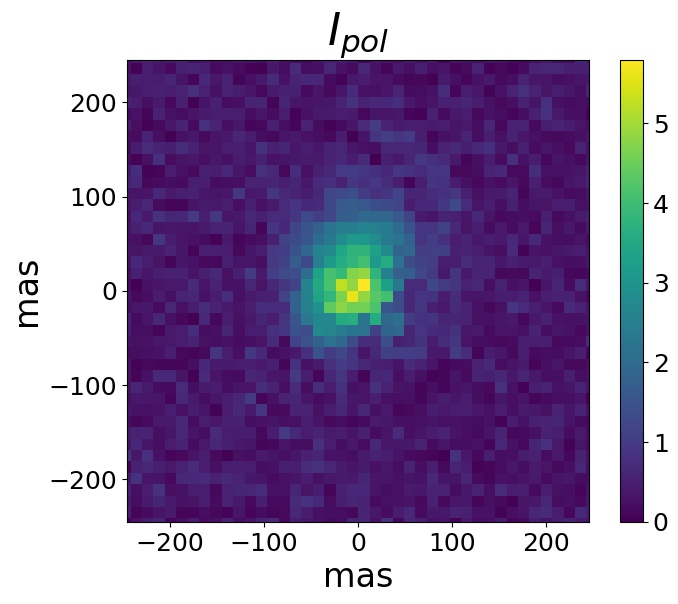}
    \includegraphics[width=0.5\columnwidth]{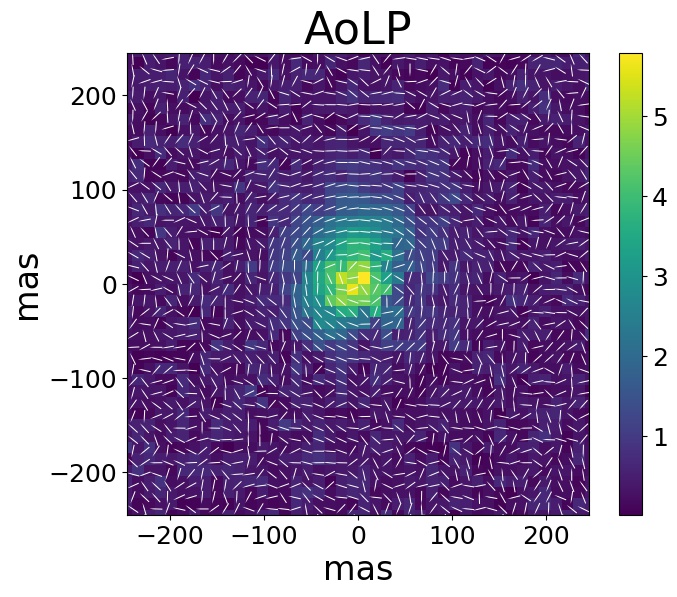}

    \includegraphics[width=0.5\columnwidth]{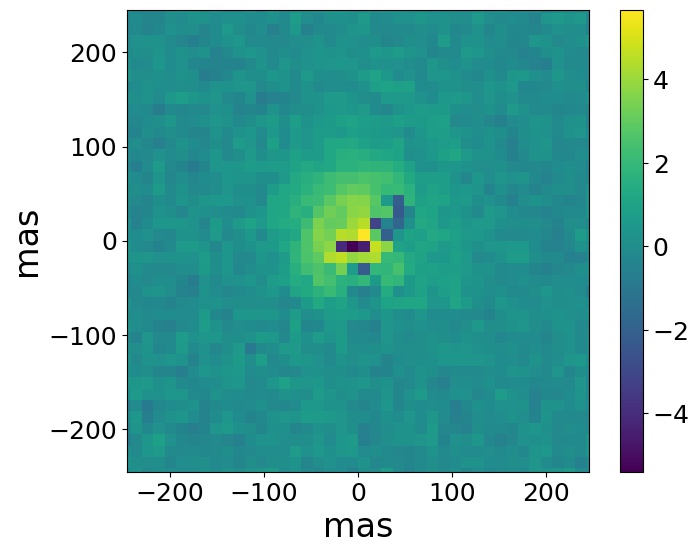}
    \includegraphics[width=0.5\columnwidth]{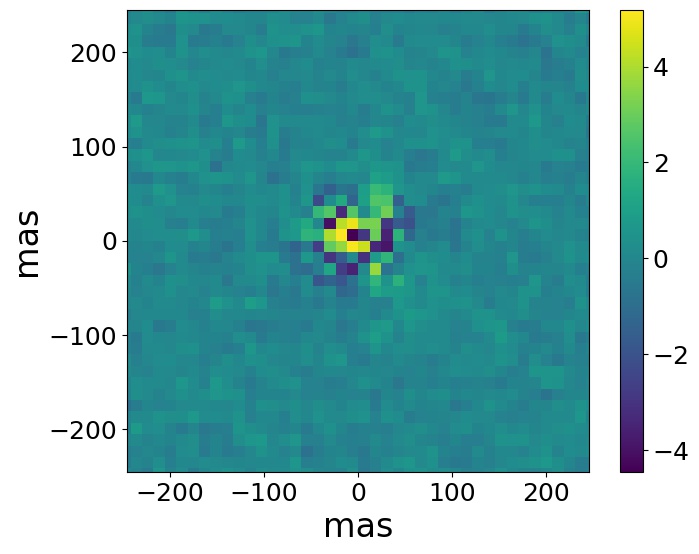}
    \includegraphics[width=0.5\columnwidth]{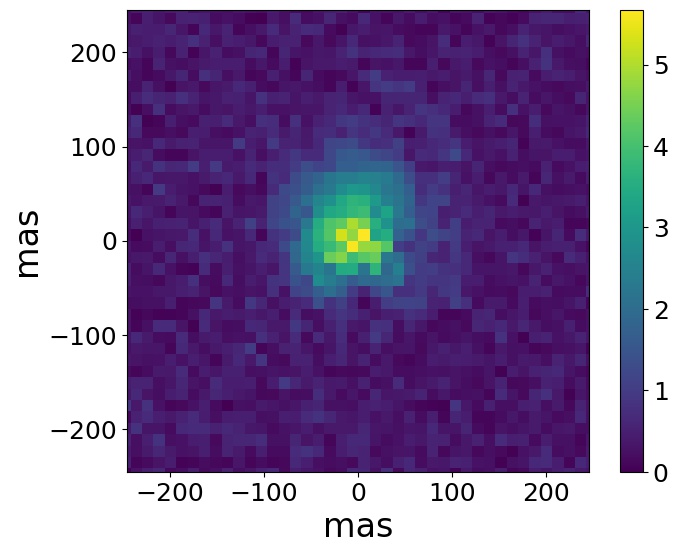}
    \includegraphics[width=0.5\columnwidth]{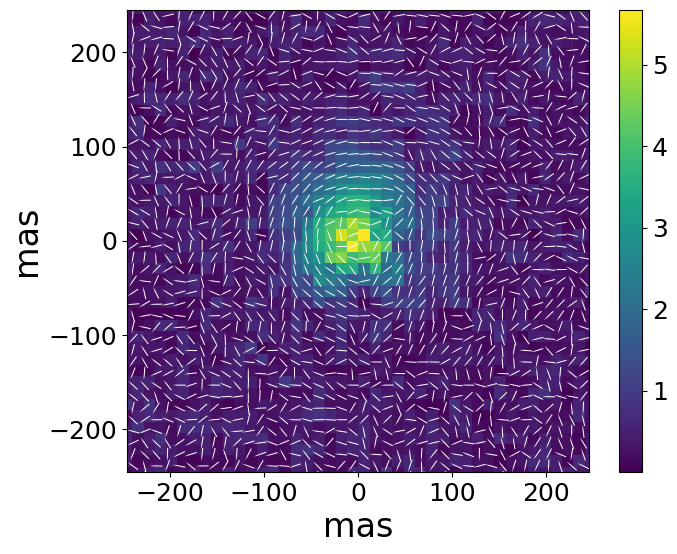}

    \caption{Same as Fig.~\ref{fig:u_mon} for AC\,Her.
    \label{fig:ac_her}}
\end{figure*}

In Fig.~\ref{fig:u_mon},~\ref{fig:iw_car},~\ref{fig:hr_4049},~\ref{fig:iras15469-5311},~\ref{fig:iras17038-4815},~\ref{fig:ru_cen},~\ref{fig:ac_her}, we present polarimetric Q$_\phi$, U$_\phi$ and total intensity images as well as the corresponding angle of linear polarization (AoLP) grouped by individual targets. The top row of each figure represents the data before the correction for the unresolved central polarization, and the bottom row represents the data after this procedure. We provide figures for seven out of eight targets in our sample (except IRAS\,08544-4431). A similar figure for IRAS\,08544-4431 is presented in the main text of the paper (see Fig.~\ref{fig:stellar}). All images are presented on an inverse hyperbolic scale and oriented with North to up. We note that for U\,Mon we use the mean combined data of two individual observations to improve the signal-to-noise ratio.

\section{The deconvolved total polarized intensity}
\label{sec:ap_deconv_totalpol_imag}

\begin{figure*}
    
    \includegraphics[width=0.5\columnwidth]{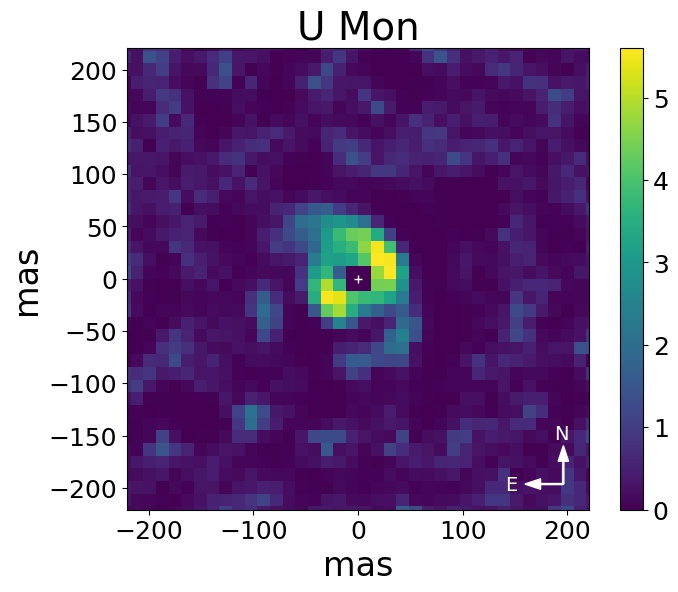}
    \includegraphics[width=0.5\columnwidth]{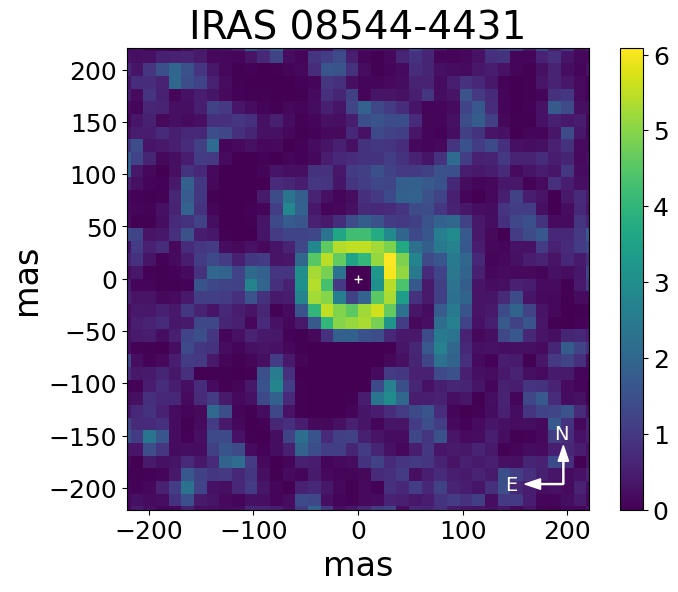}
    \includegraphics[width=0.5\columnwidth]{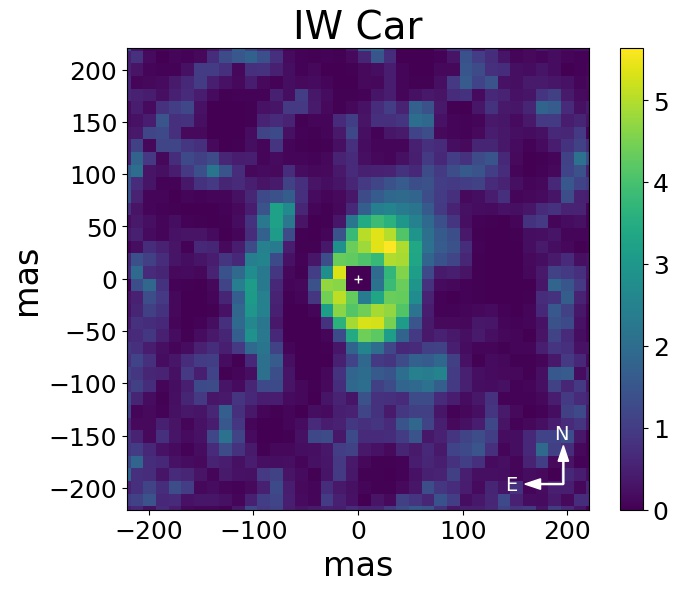}
    \includegraphics[width=0.5\columnwidth]{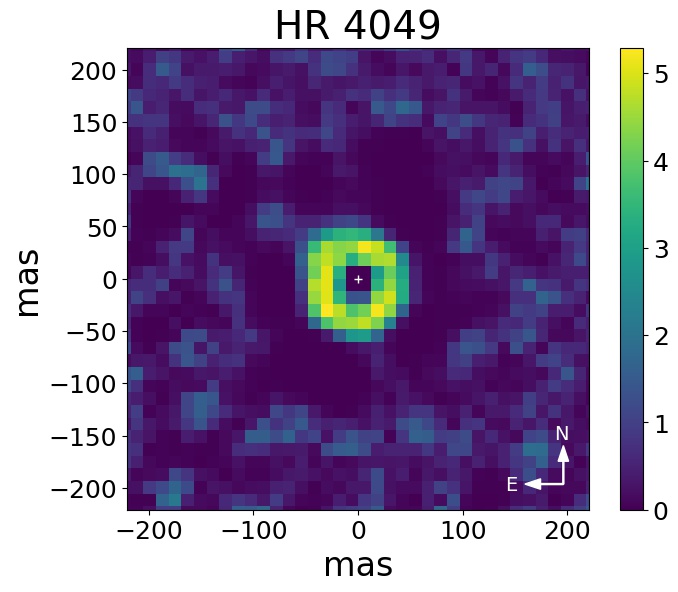}
    \includegraphics[width=0.5\columnwidth]{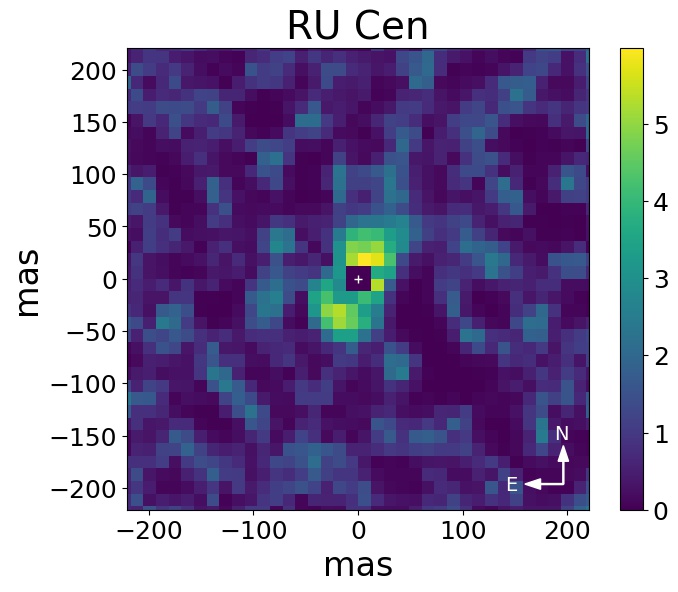}
    \includegraphics[width=0.5\columnwidth]{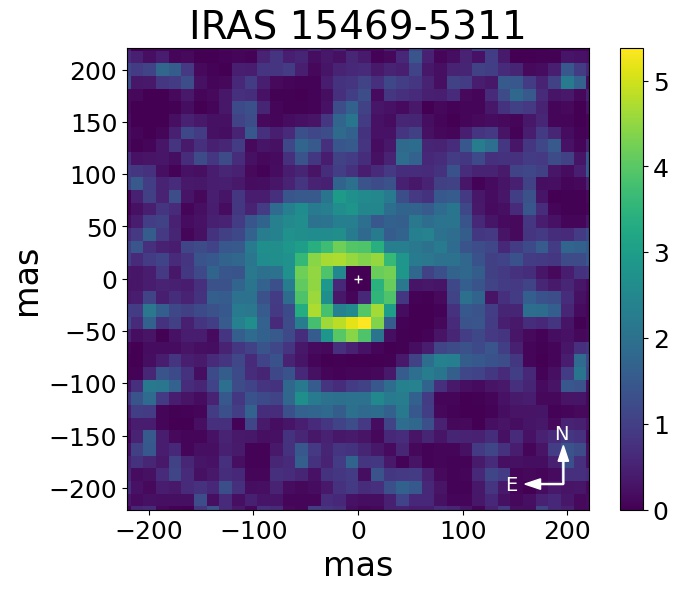}
    \includegraphics[width=0.5\columnwidth]{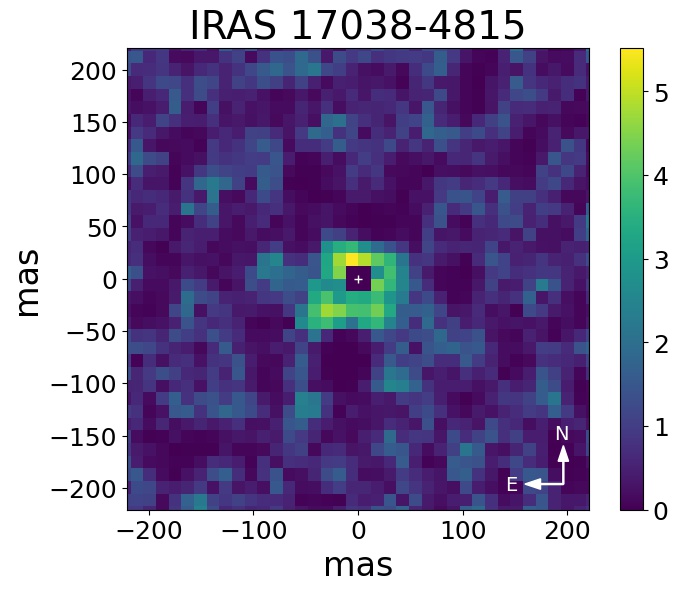}
    \includegraphics[width=0.5\columnwidth]{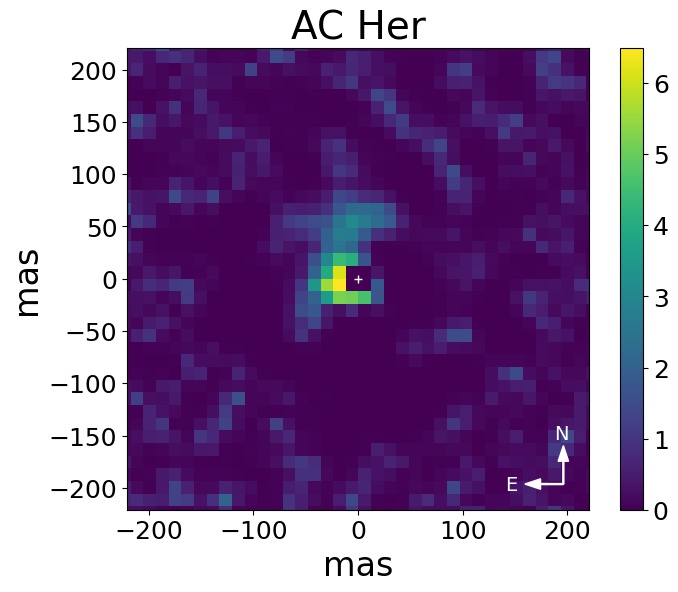}
    
    \caption{Deconvolved total polarized intensity images for all objects in our sample. All images are presented on an inverse hyperbolic scale. The low intensity of the central 4x4 pixel region of each image is a reduction bias caused by over-correction of the unresolved central polarization (see Section~\ref{sec:unresolved}).\label{fig:I_pol}}
    
\end{figure*}

In Fig.~\ref{fig:I_pol}, we present the deconvolved total polarized intensity (I$_{\rm pol}$) for all targets in the sample (see Section~\ref{fig:deconvolution}). 

\section{Estimation of the signal-to-noise ratio (SNR) and the size of the resolution element}
\label{sec:snr}

The signal-to-noise ratio (SNR) for our data was calculated in two ways. Firstly, we adopted the methodology presented in \citet{Ertel2019AJ....157..110E} to derive the SNR based on the peak count of the source. In brief, the authors measured the pixel-to-pixel root mean squared (RMS) in the background of the image and compared it to the source's peak counts. To determine the background signal for our datasets, we used the annulus of the image that did not contain any signal from the target and was also used for the background correction during the IRDAP reduction procedure. The measured peak SNR values for all targets in the sample are tabulated in Table~\ref{tab:snr}. 

\begin{table}
	\centering
	\caption{Peak signal-to-noise ratio (SNR) for all targets in our sample.\label{tab:snr}}
	
	\begin{center}
    	    
    	\begin{tabular}{|c|l|c|} 
    		\hline
    		 \#ID & Name & Peak S/N\\
    		\hline
    		\multicolumn{3}{c}{Full discs}\\
    		\hline
            1 & U\,Mon & 274\\
            2 & IRAS\,08544-4431 & 181\\
            3 & IW\,Car & 152\\
            4& HR 4049 & 139\\
            5&IRAS 15469-5311 & 111\\
            6&IRAS 17038-4815 & 78\\
            \hline
    		\multicolumn{3}{c}{Transition discs}\\
    		\hline
            7 & RU\,Cen  & 231\\
            8 & AC\,Her & 829\\
            \hline
    	\end{tabular}
    \end{center}
	\begin{tablenotes}
        
    \small
\item \textbf{Notes:} To improve the signal-to-noise ratio for the U\,Mon, we use the mean combined reduced polarised images of two independent observations (see Section~\ref{sec:observations}).\\
    \end{tablenotes}
	
\end{table}

Then, we calculated the SNR for each pixel in the azimuthally polarized image to determine the statistically significant region. In this case, the signal was defined as each pixel value instead of only one peak count per image. In further analysis, we rely on the region of the azimuthally polarized image with SNR $\geq5$ (see Fig.\ref{fig:snr}).

Out of interest, we have also calculated the resolution of our observational data to compare it with the expected performance of the SPHERE/IRDIS camera. Following the methodology presented in \citet{Hunziker2021A&A...648A.110H}, we performed the gaussian fitting of the PSF intensity profile and used the full width at half maximum (FWHM) as the size of the resolution element. The resulting resolution values fall in the range of 45-50\,mas for all individual datasets and are comparable with the expected values for the observing conditions listed in Table~\ref{tab:weather}.

\section{Presence of the non-azimuthal polarized signal}
\label{sec:ap_sign_uphi}

\citet{Canovas2015A&A...582L...7C} showed that a significant amount of the non-azimuthal polarized signal (U$\phi$ component) could be produced by dusty circumstellar discs with inclinations $i \gtrsim 40^\circ$. To investigate a scenario where the U$_\phi$ components of our dataset contain polarized signal from the disc, we adopted the methodology proposed by \citet{Avenhaus2018ApJ...863...44A}. In brief, the authors calculated the distribution of mean fluxes in U$_\phi$ image across a number of radially tabulated annuli, divided by their respective $\delta$ (the standard deviation $\sigma_a$ of nondeconvolved U$_\phi$ in the corresponding area, divided by the square root of the number of resolution elements $N$ in it):

\begin{equation}
    \delta =\frac{\sigma_a}{\sqrt{N}}.
	\label{uncert}
\end{equation}

Due to the DR process made in \citet{Avenhaus2018ApJ...863...44A}, the signal in the U$_\phi$ image was set to be zero, on average. Hence, the authors expect the distribution to follow a standardised normal distribution. If this is confirmed, we can conclude that the U$_\phi$ image contains only noise.

\begin{figure*}
    
    \includegraphics[width=0.5\columnwidth]{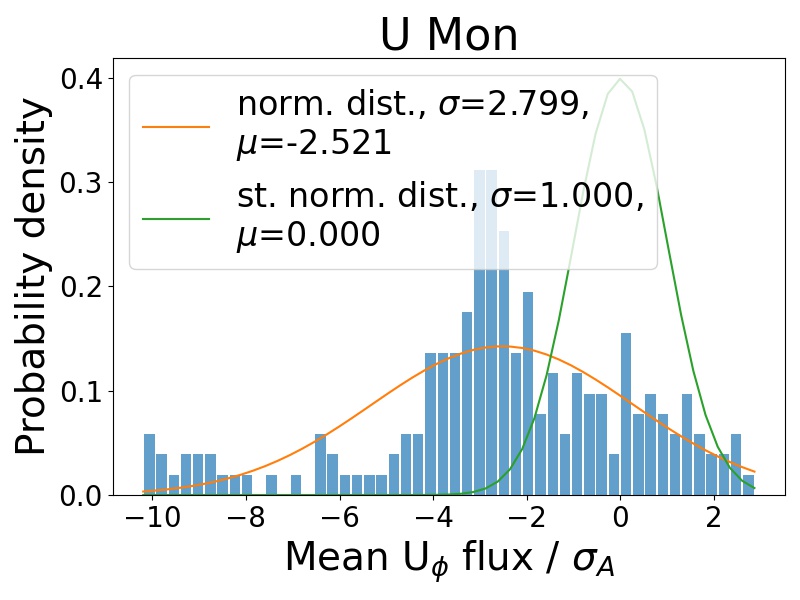}    
    \includegraphics[width=0.5\columnwidth]{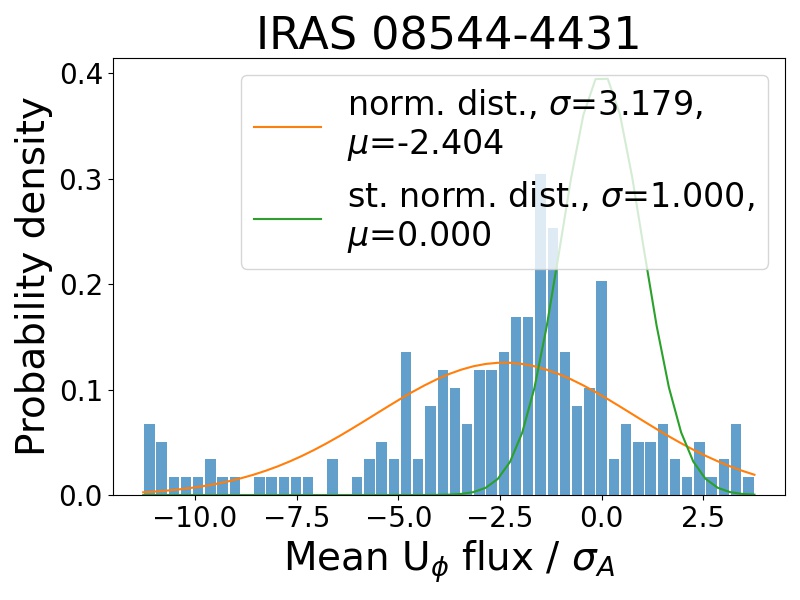}
    \includegraphics[width=0.5\columnwidth]{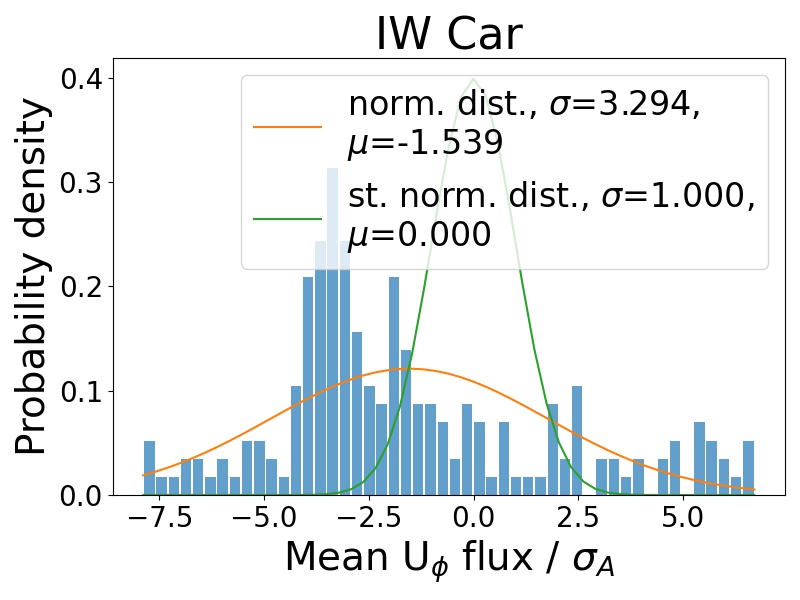}
    \includegraphics[width=0.5\columnwidth]{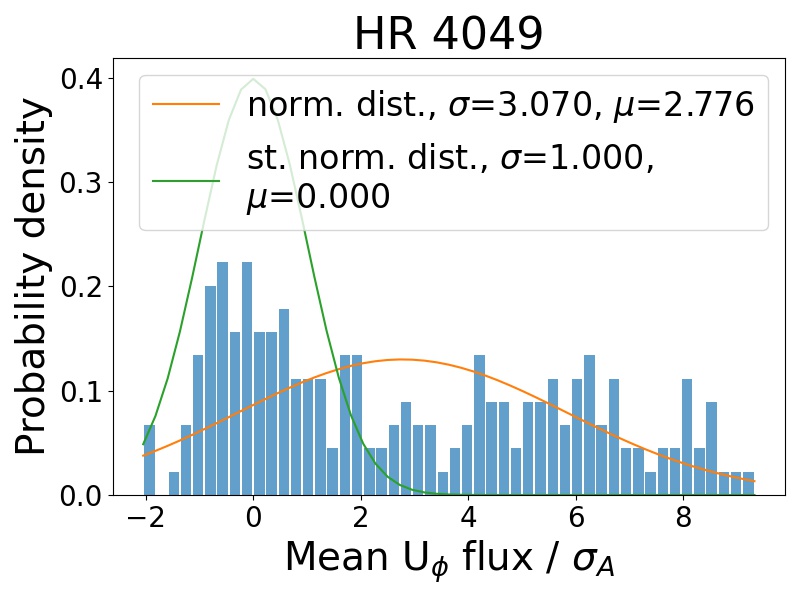}
    
    \includegraphics[width=0.5\columnwidth]{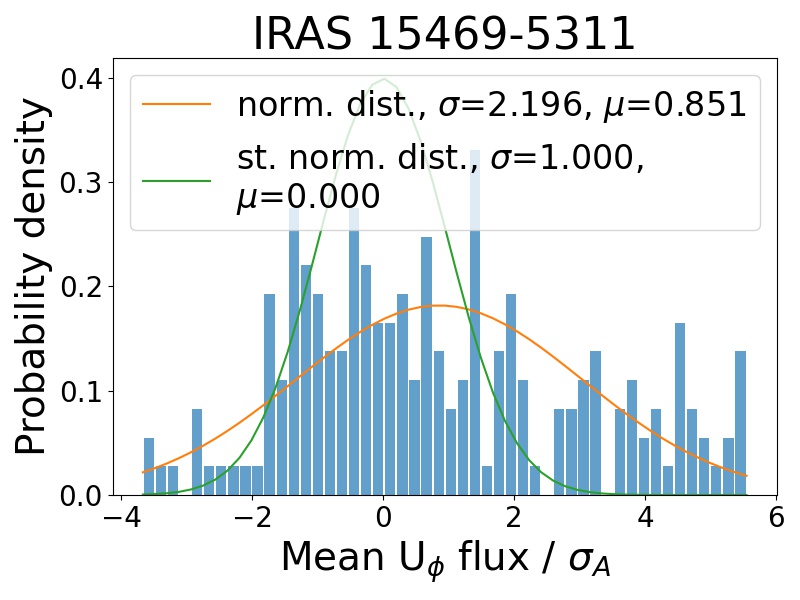}
    \includegraphics[width=0.5\columnwidth]{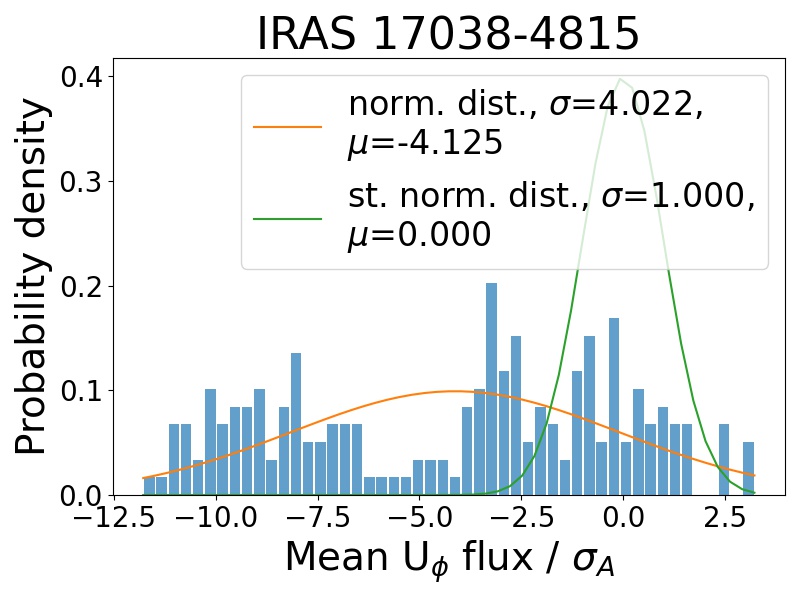}
    \includegraphics[width=0.5\columnwidth]{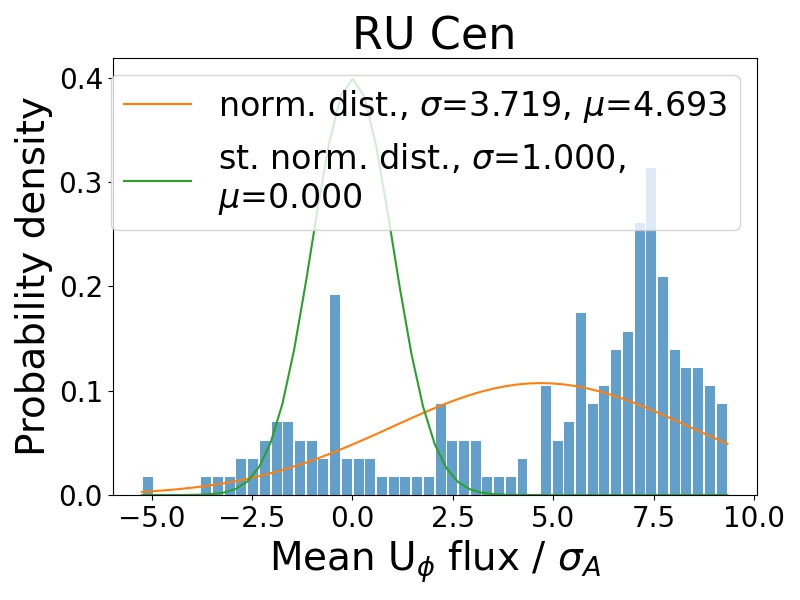}
    \includegraphics[width=0.5\columnwidth]{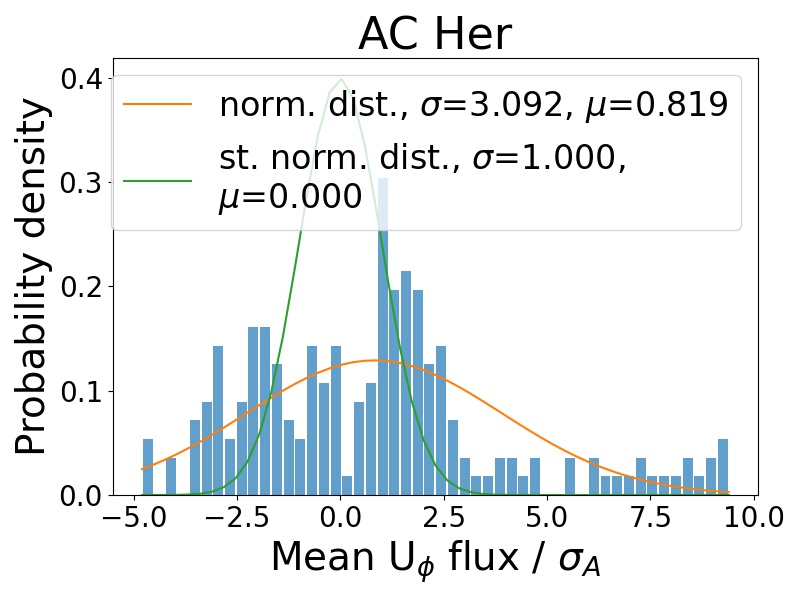}
    
    \caption{The distribution of mean fluxes in U$_\phi$ across a number of radially tabulated annuli for all targets in the sample. The green line represents the standardized normal distribution, while the orange line represents the normal distribution fitted to the data. See Appendix~\ref{sec:ap_sign_uphi} for details.\label{fig:hist}}
    
\end{figure*}

We performed the same technique to test whether we have the non-azimuthal polarized signal from the disc in our data. The resulting histograms are presented in Fig.\ref{fig:hist}. We found that the distribution of mean fluxes in U$_\phi$ image for each individual object does not follow either a standardised normal distribution (marked in green in Fig.\ref{fig:hist}) or a usual normal distribution (marked in orange), which indicates that U$_\phi$ image contains some astrophysical signal in addition to noise. As a result, we can only use the U$_\phi$ image as a rough estimation of the upper noise limit for our observations. This is likely due to the DR method we used, as the IRDAP software uses the modelled values for instrumental polarization, which is a more accurate method than the minimisation of the U$_\phi$ signal.

Despite the confirmation of the presence of the polarized signal from the disc in U$_\phi$ images for all targets in our sample, in the current paper, we analyse only the positive signal of Q$_\phi$ image because, in the case of the non-azimuthal polarized signal, we cannot differentiate between the scattered light from the disc, reduction bias, and noise.

\section{Angle of Linear polarization (AoLP)}
\label{sec:ap_Aolp}

\begin{figure*}
    
    \includegraphics[width=0.5\columnwidth]{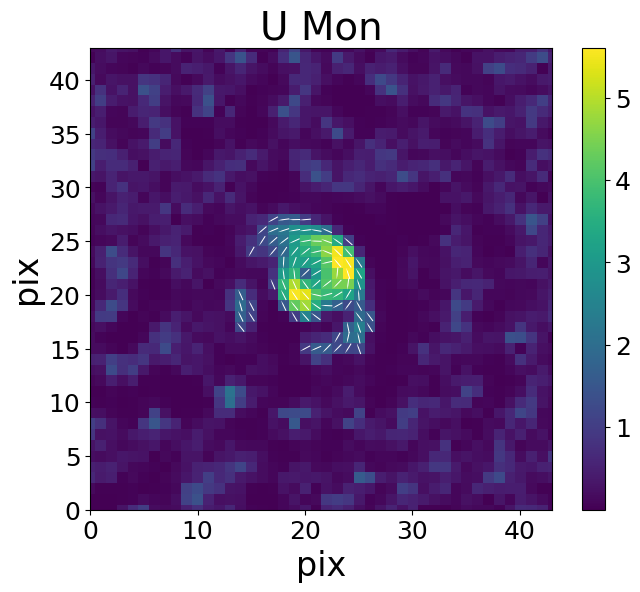}
    \includegraphics[width=0.5\columnwidth]{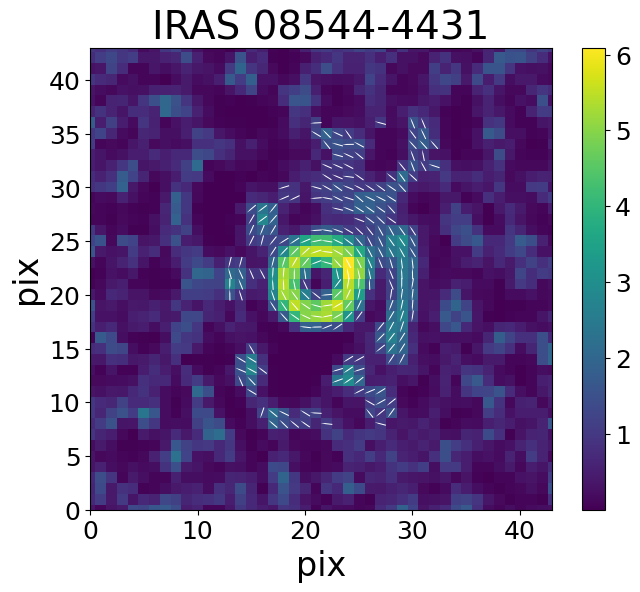}
    \includegraphics[width=0.5\columnwidth]{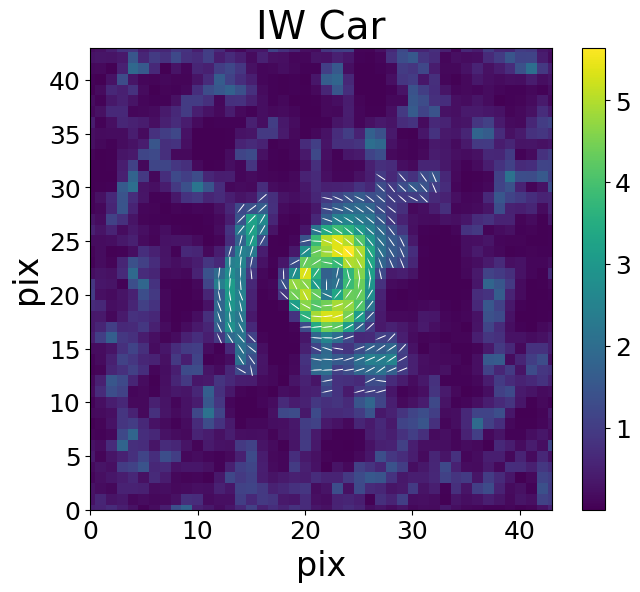}
    \includegraphics[width=0.5\columnwidth]{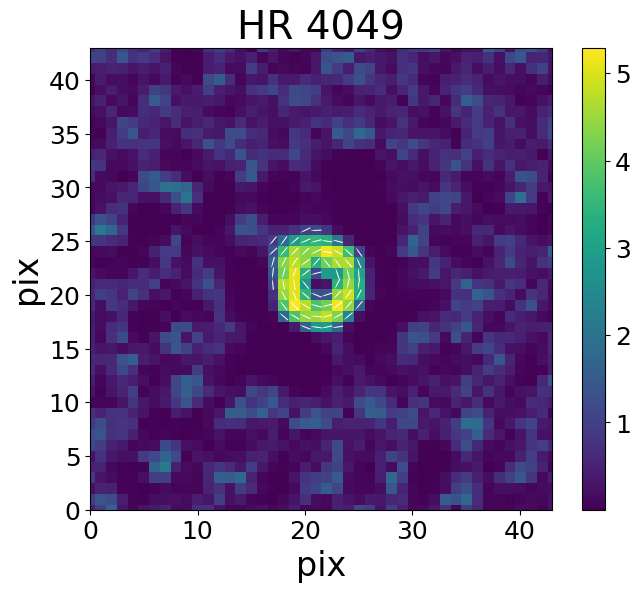}
    
    \includegraphics[width=0.5\columnwidth]{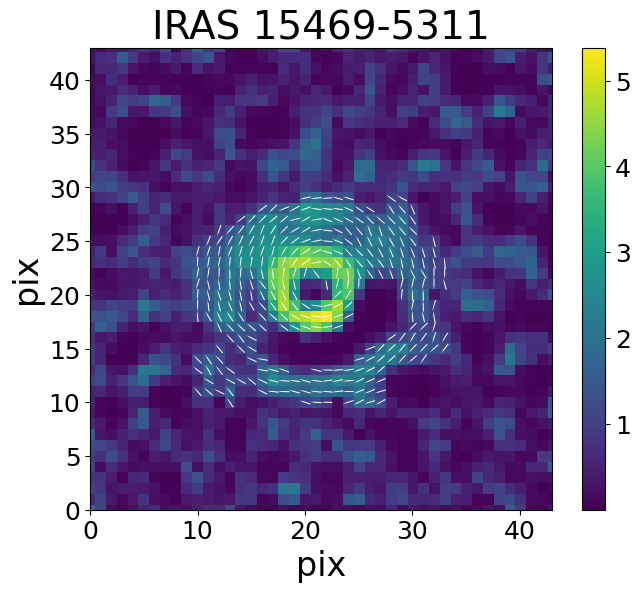}
    \includegraphics[width=0.5\columnwidth]{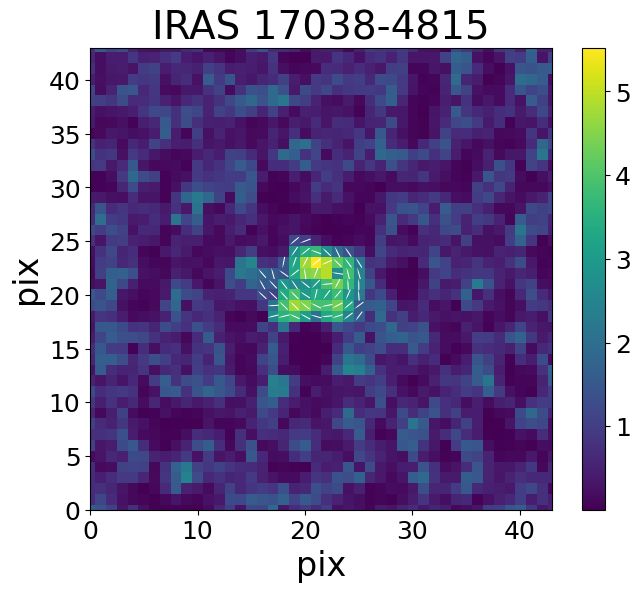}
    \includegraphics[width=0.5\columnwidth]{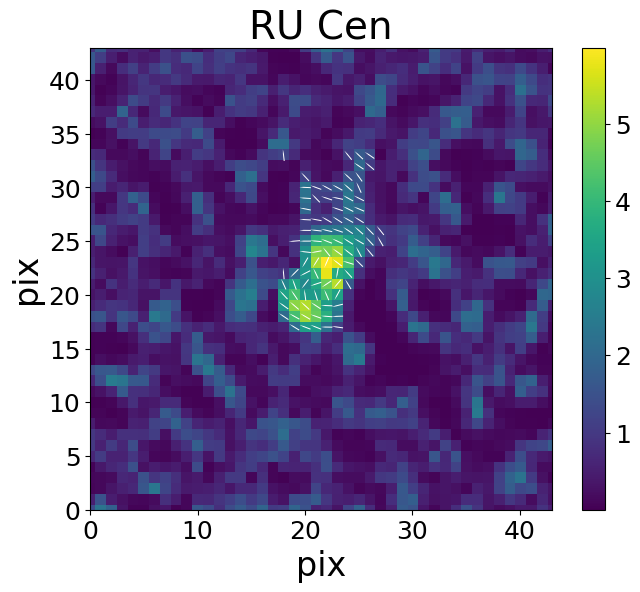}
    \includegraphics[width=0.5\columnwidth]{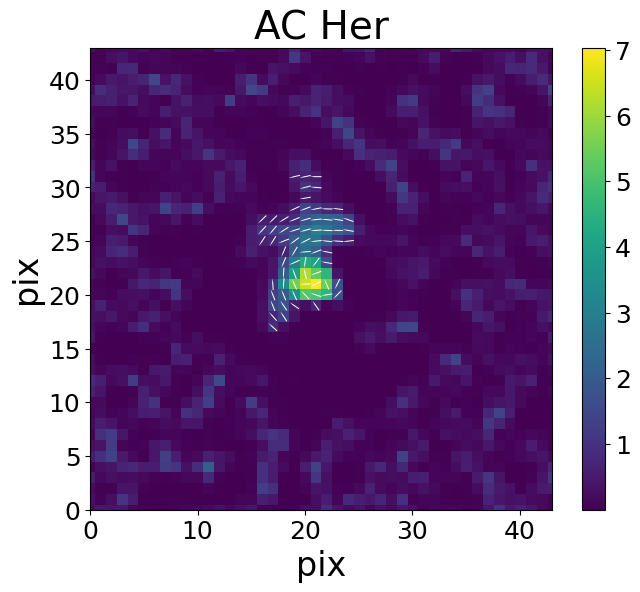}
    \caption{I$_{\rm pol}$ and Angle of Linear Polarization (in white) for confirmed disc structures. I$_{\rm pol}$ is shown on an inverse hyperbolic scale, and the lines indicate the local angle of linear  polarization. See Section~\ref{sec:substructures} for details.\label{fig:Aolp}}
    
\end{figure*}

In Fig.\ref{fig:Aolp}, we present the local Angle of Linear polarization (AoLP) for confirmed disc structures (in white) plotted over the total polarized brightness image (I$_{\rm pol}$) for all targets in our sample (see Section~\ref{sec:substructures}). All images are presented on an inverse hyperbolic scale.

\section{Brightness profiles}
\label{sec:ap_brprof}
\begin{figure*}
    
    \includegraphics[width=\columnwidth]{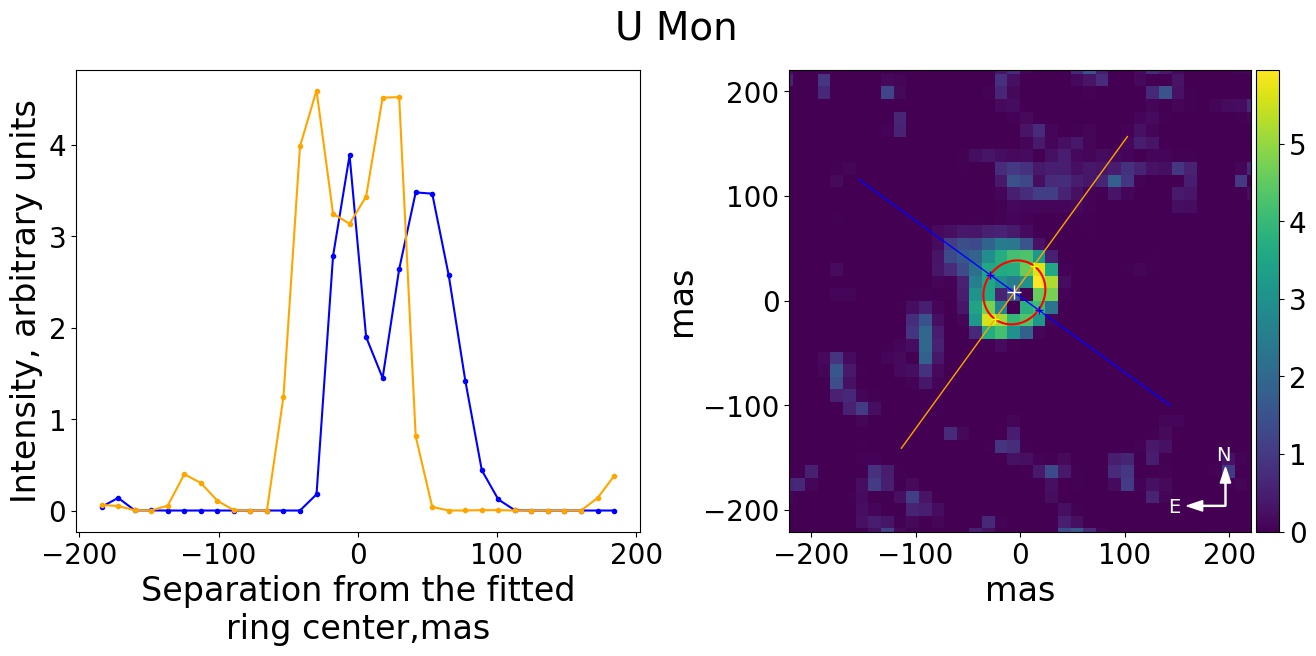}
    \includegraphics[width=\columnwidth]{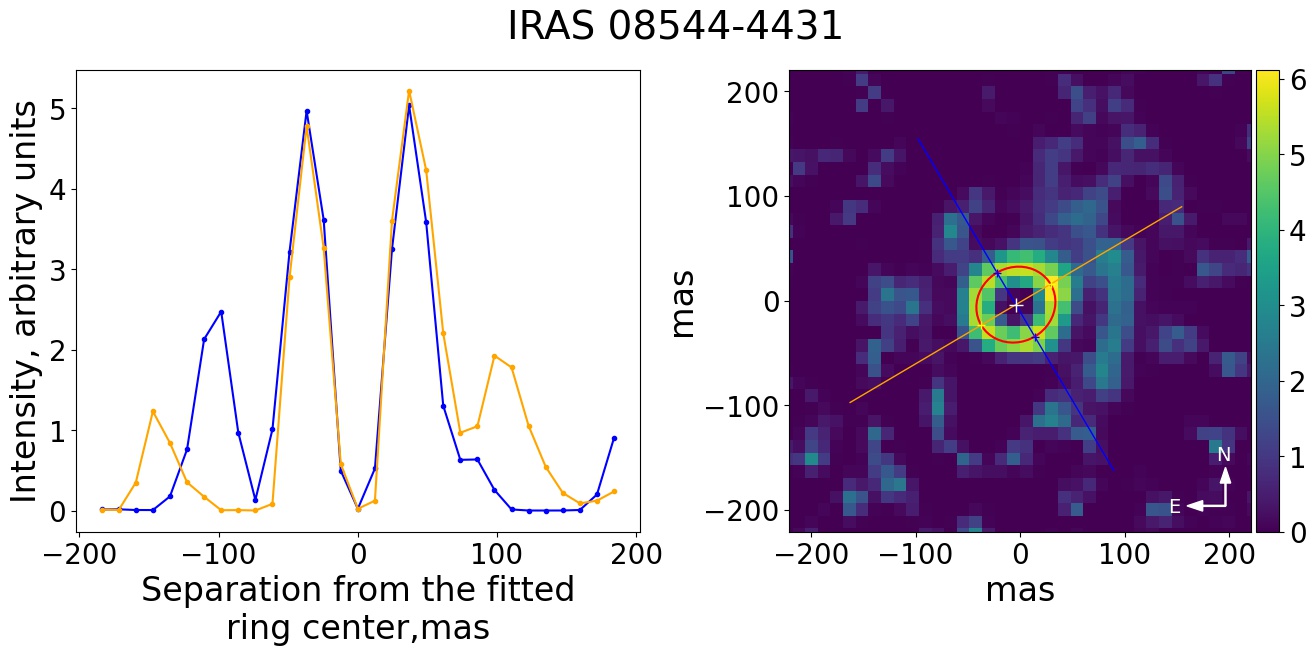}
    \includegraphics[width=\columnwidth]{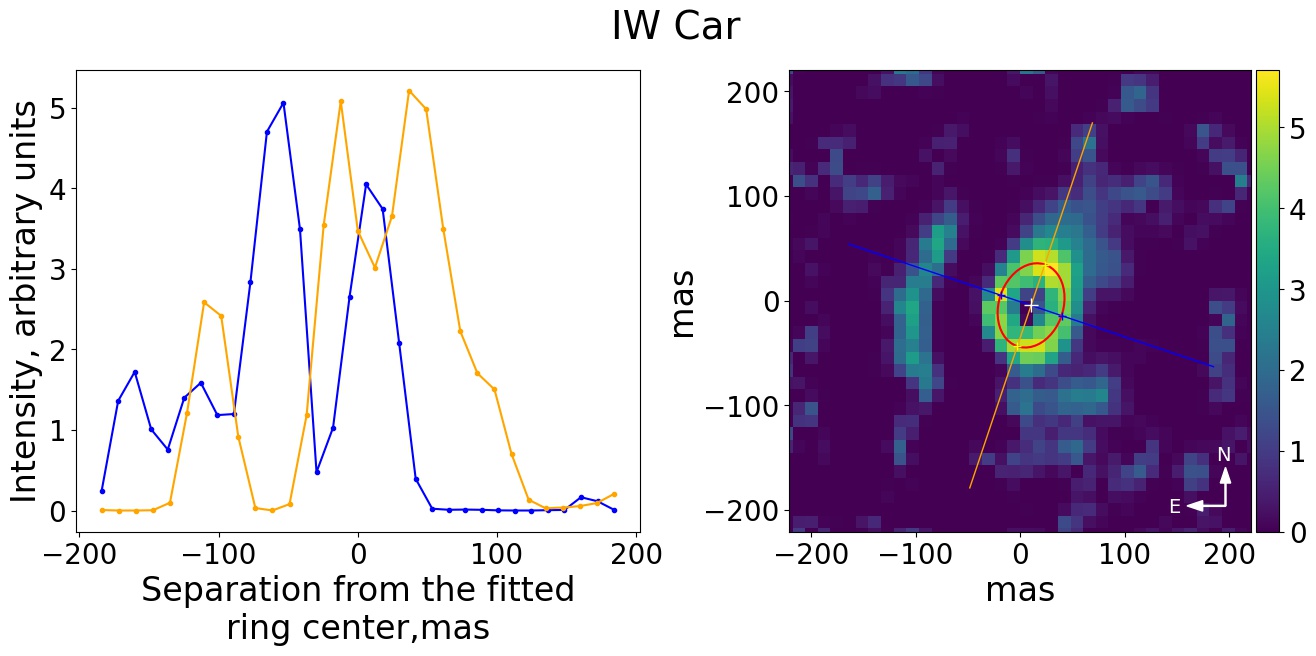}
    \includegraphics[width=\columnwidth]{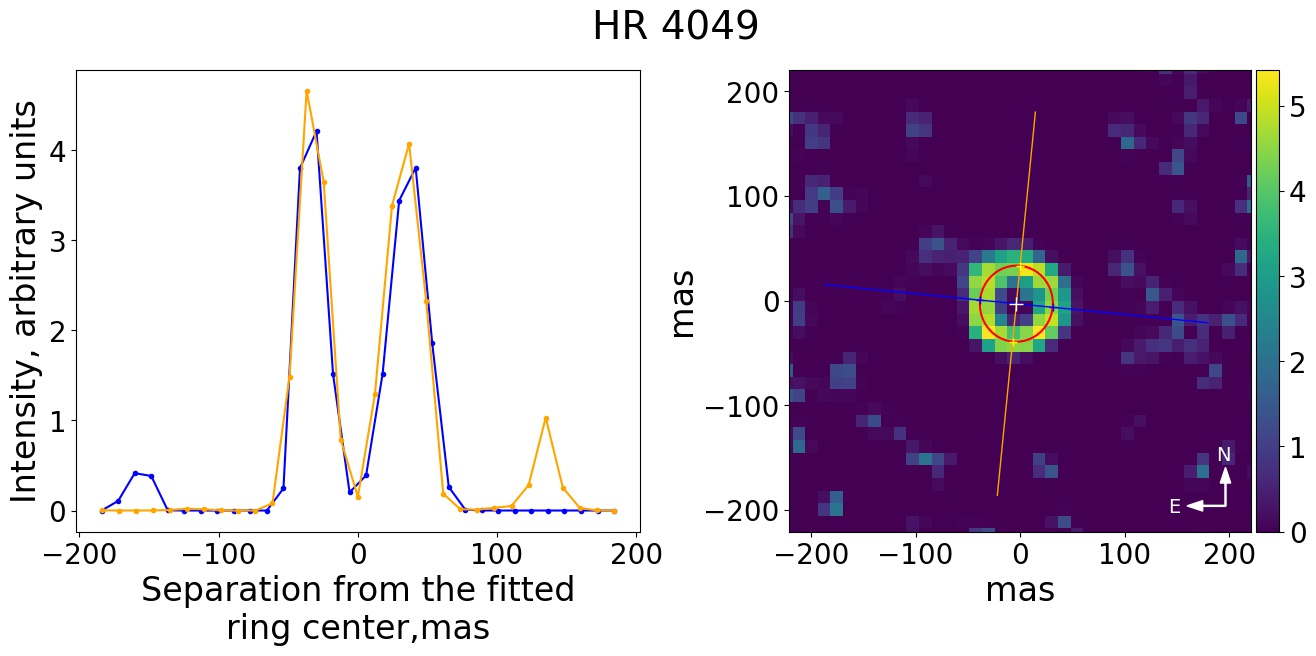}
    \includegraphics[width=\columnwidth]{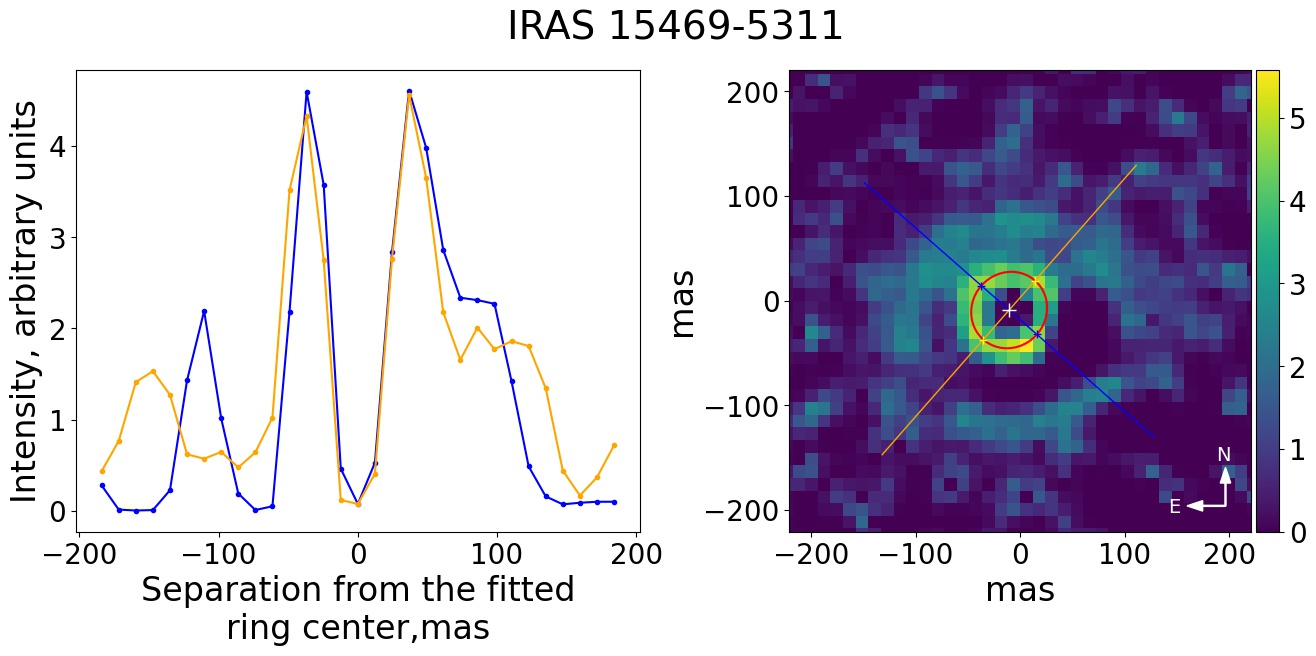}
    \includegraphics[width=\columnwidth]{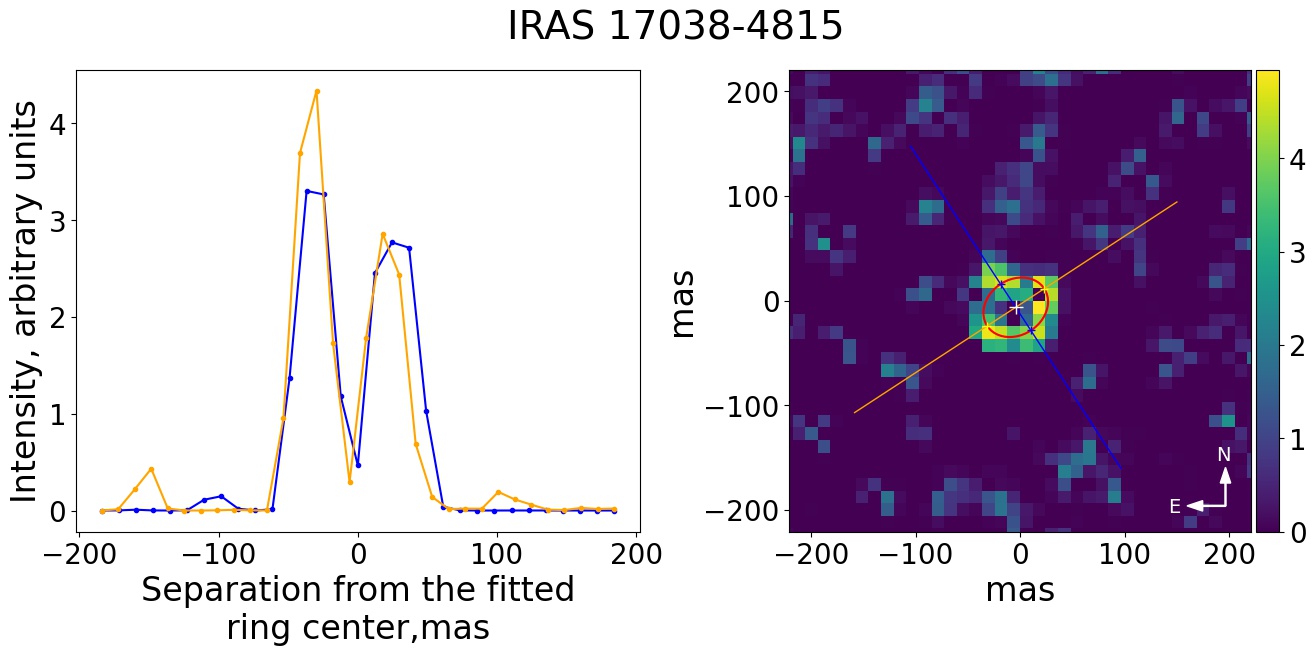}
    \includegraphics[width=\columnwidth]{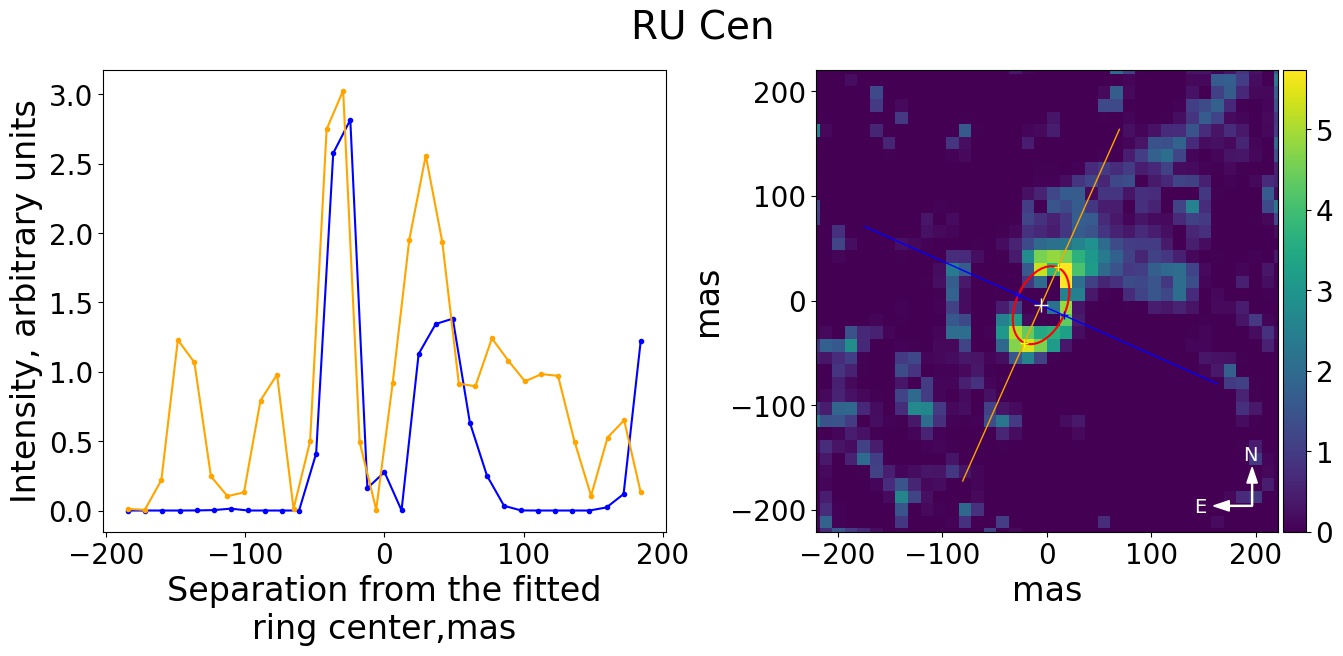}
    \includegraphics[width=\columnwidth]{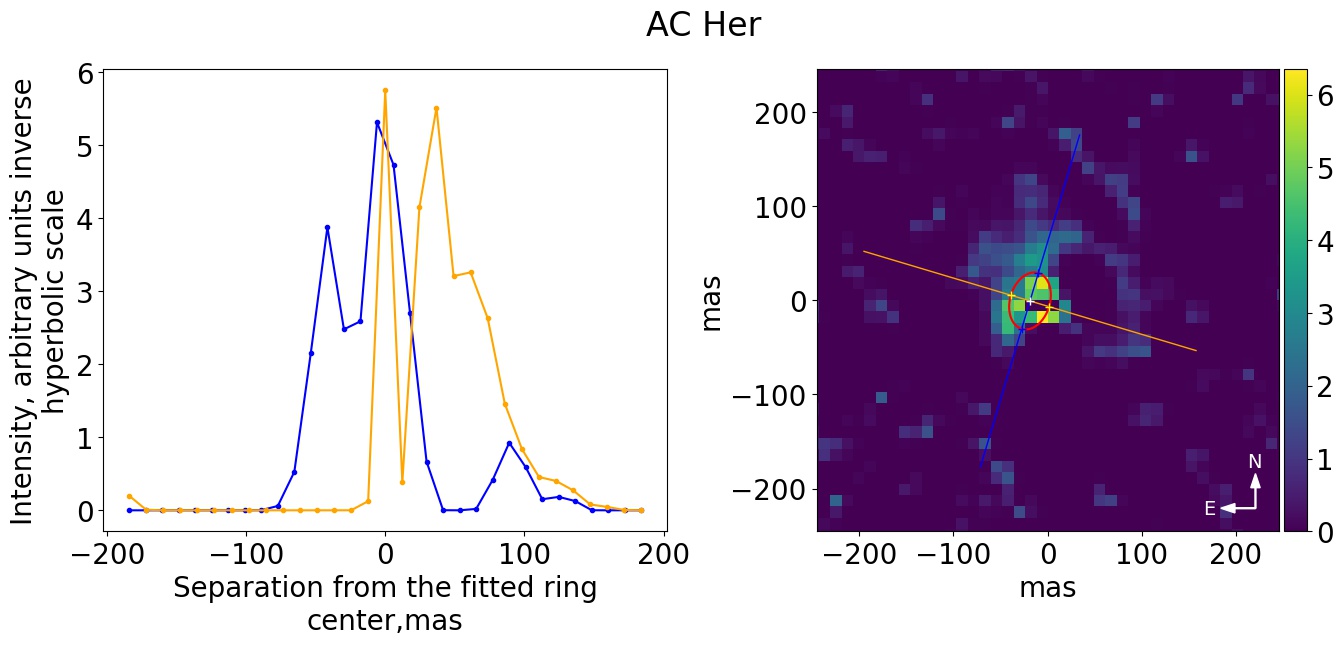}
    \caption{Linear brightness profiles (first and third column) of Q$_\phi$ image (second and fourth column) along the major and minor axes of the fitted ellipse. Profiles are presented on a logarithmic scale. Q$_\phi$ image is presented on an inverse hyperbolic scale with corresponding positions of axes. See Section~\ref{sec:profile} for details. \label{fig:linearprof}}
\end{figure*}

\begin{figure*}

    \includegraphics[width=1\columnwidth]{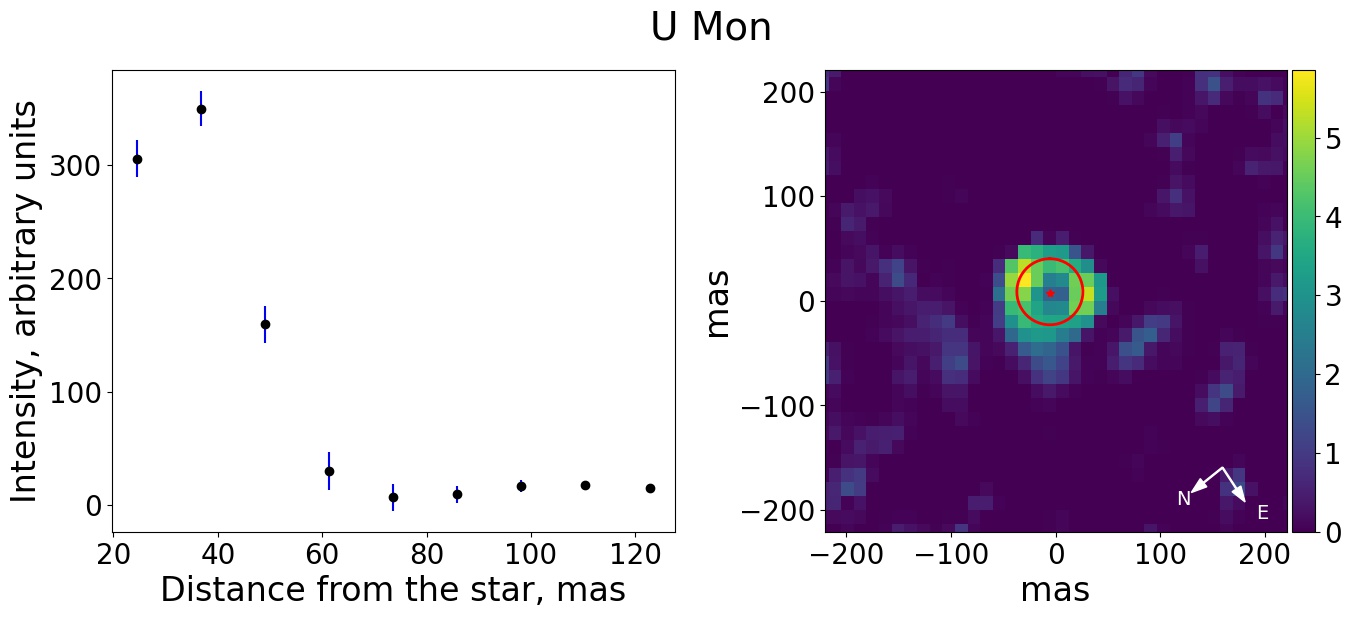}
    \includegraphics[width=1\columnwidth]{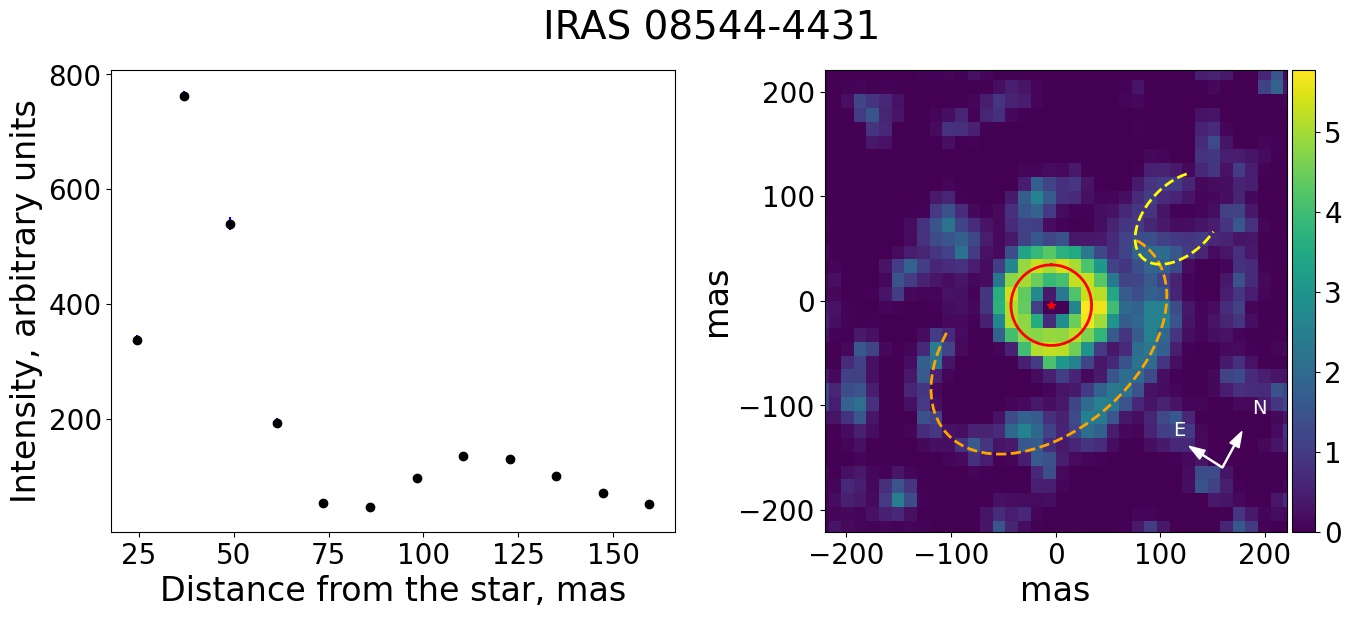}
    \includegraphics[width=1\columnwidth]{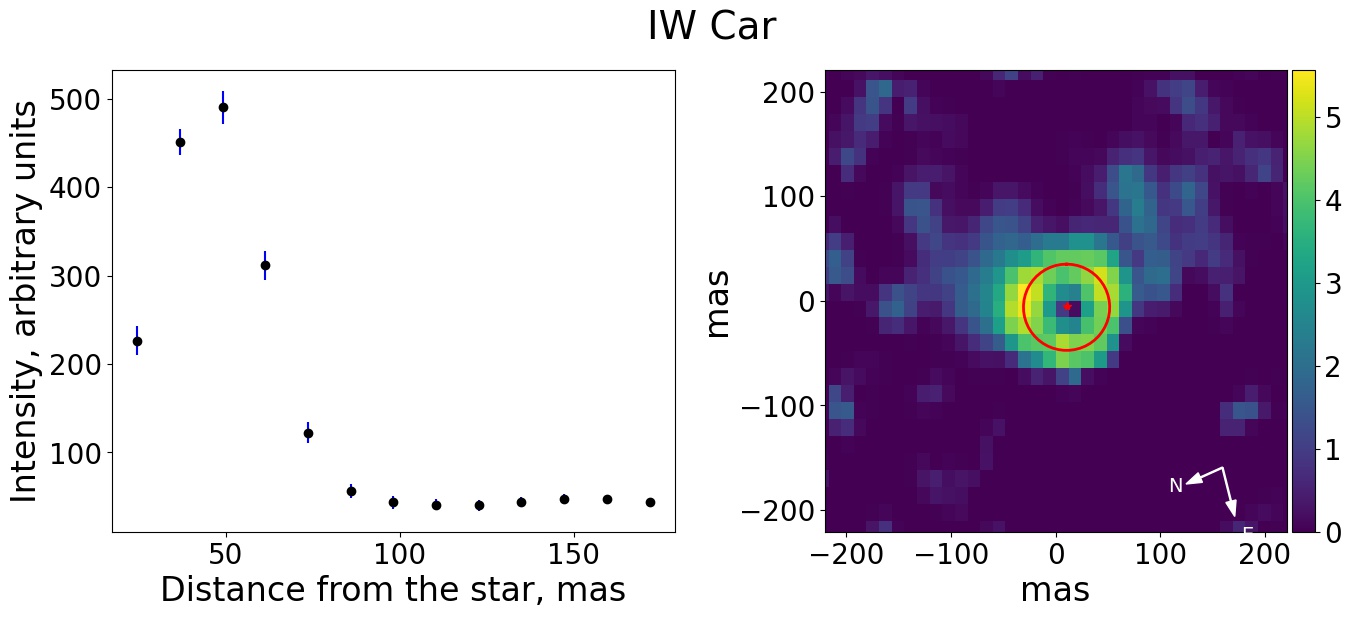}
    \includegraphics[width=1\columnwidth]{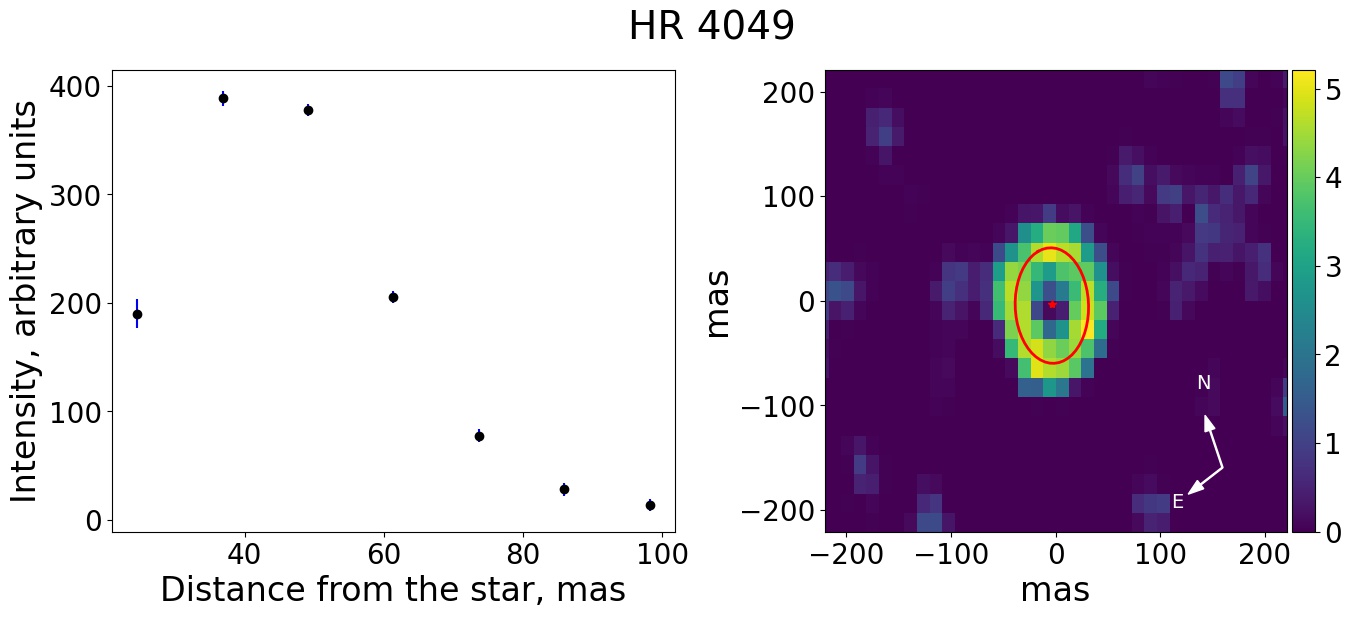}
   
    \includegraphics[width=1\columnwidth]{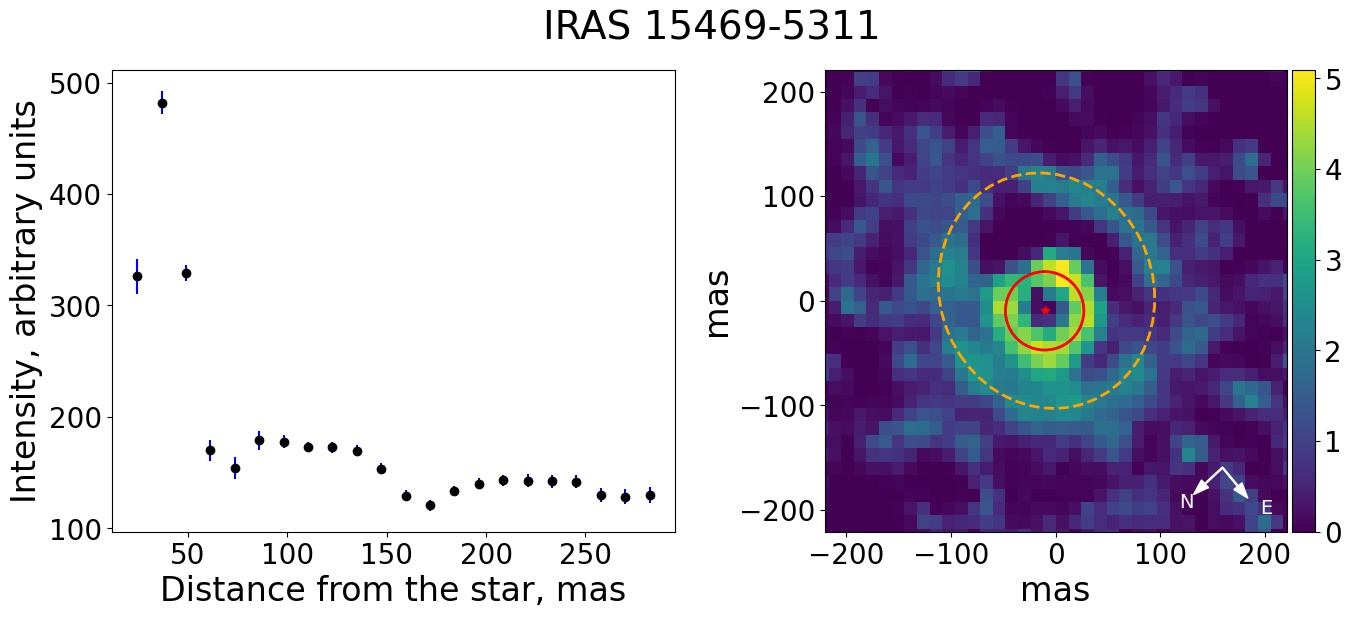}
    \includegraphics[width=1\columnwidth]{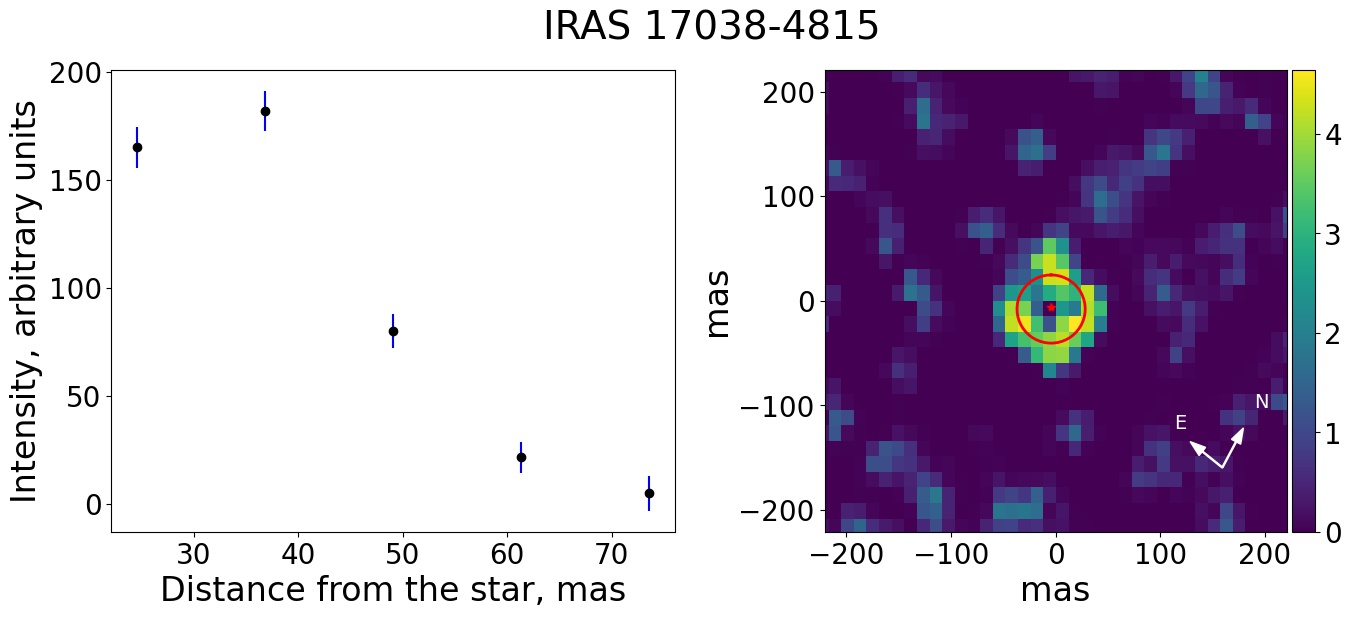}
    \caption{Radial brightness profiles (first and third column) of the deprojected Q$_\phi$ images (second and fourth column) for six out of eight targets in the sample (except for AC\,Her and RU\,Cen). The ellipses illustrate the most plausible orientation of discs, while dashed lines highlight significant substructures (see Section~\ref{sec:orientation} and Section~\ref{sec:substructures}). The orientation of each Q$_\phi$ image is provided in the bottom right corner. The low intensity of the central 4x4 pixel region of each Q$_\phi$ image is a reduction bias caused by over-correction of the unresolved central polarization (Section~\ref{sec:unresolved}). See Section~\ref{sec:profile} for details.\label{fig:radialprof}}
\end{figure*}

\begin{figure*}
   
    \includegraphics[width=1\columnwidth]{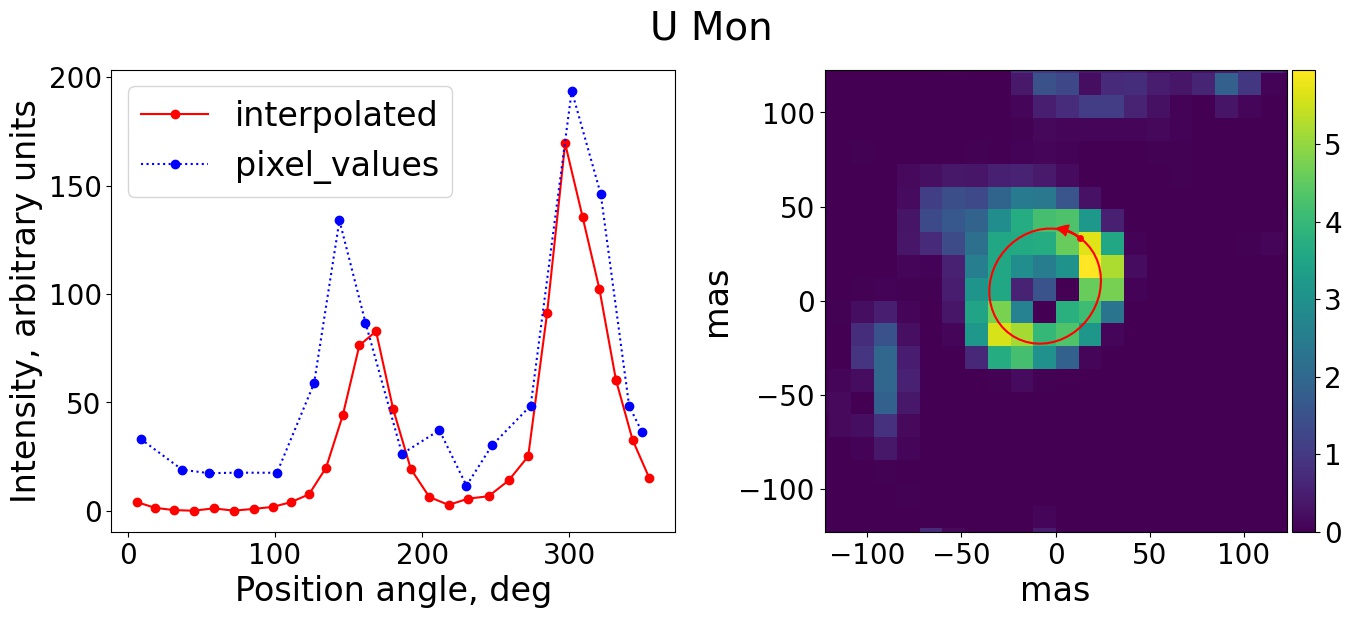}
    \includegraphics[width=1\columnwidth]{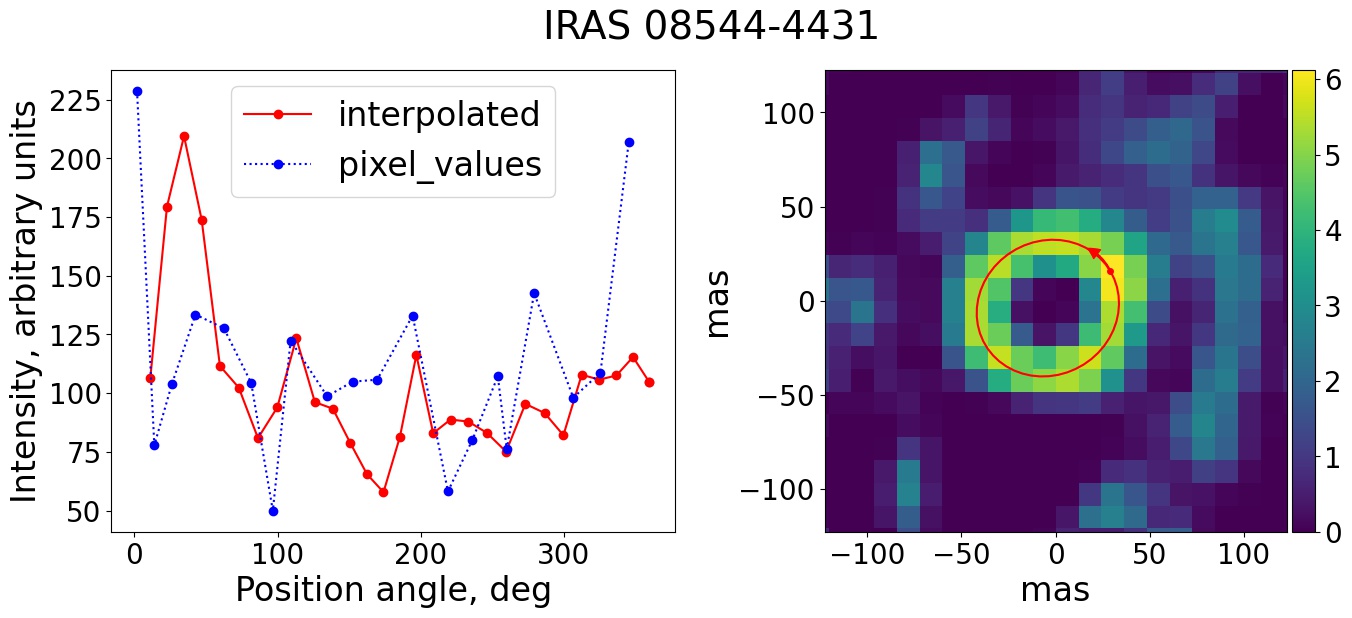}
    \includegraphics[width=1\columnwidth]{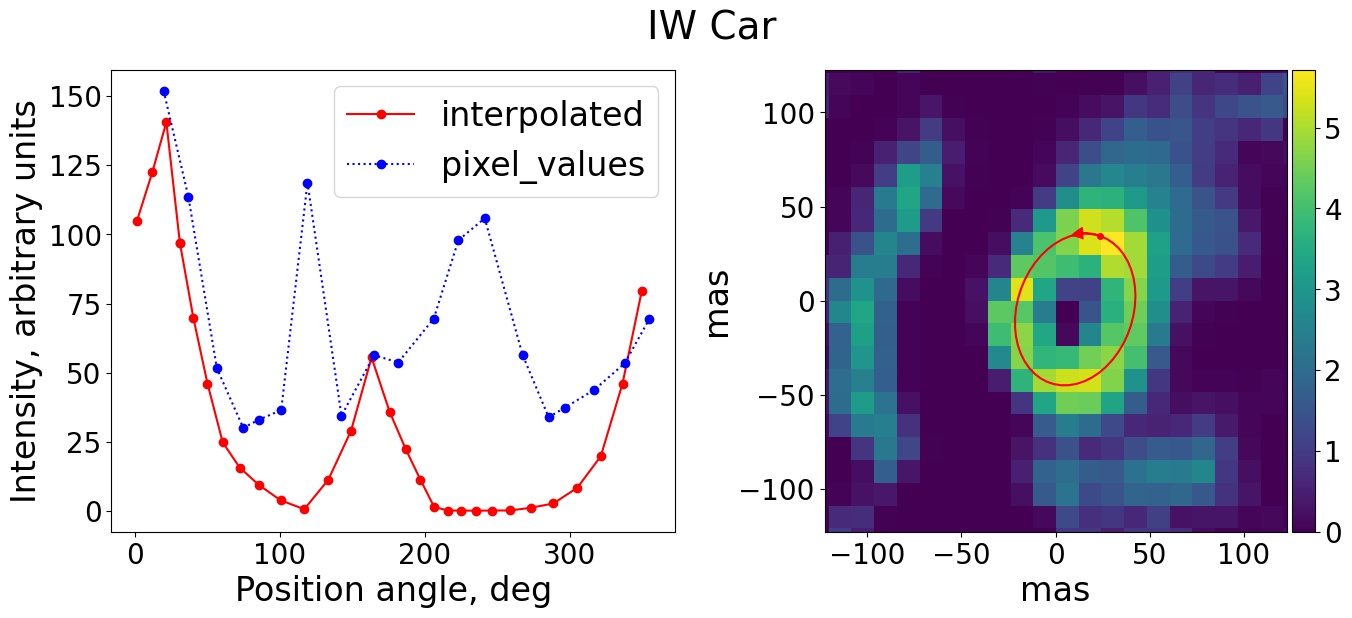}
    \includegraphics[width=1\columnwidth]{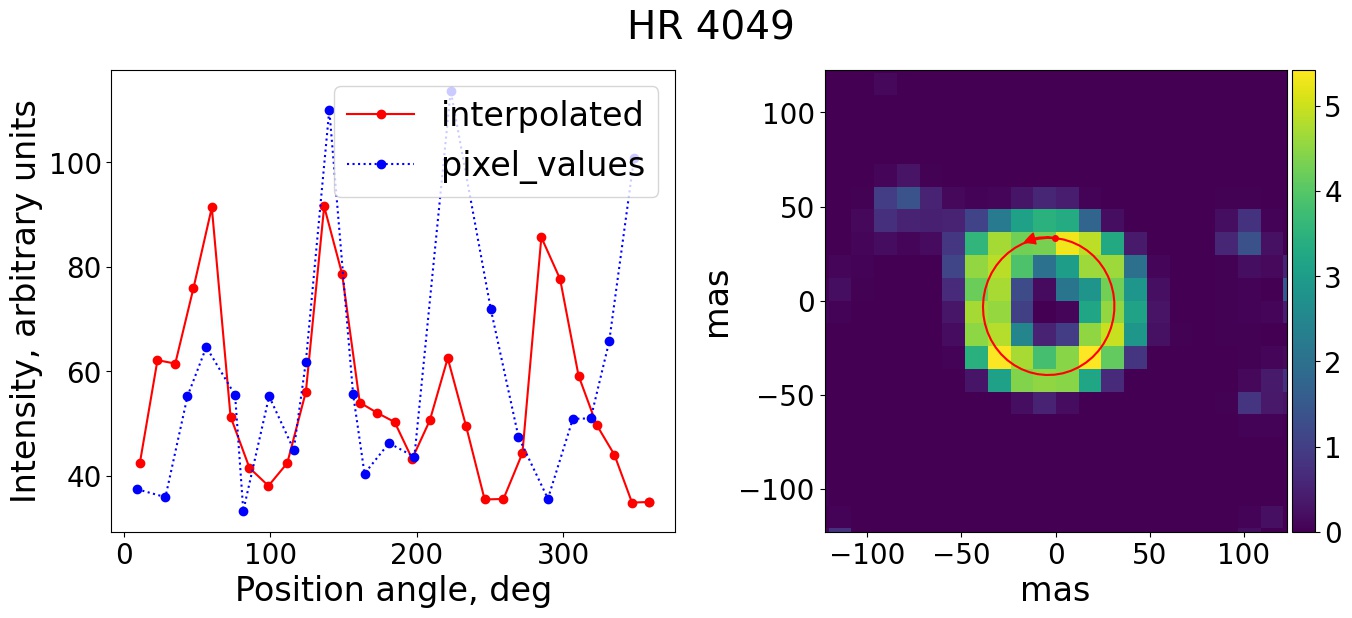}
    \includegraphics[width=1\columnwidth]{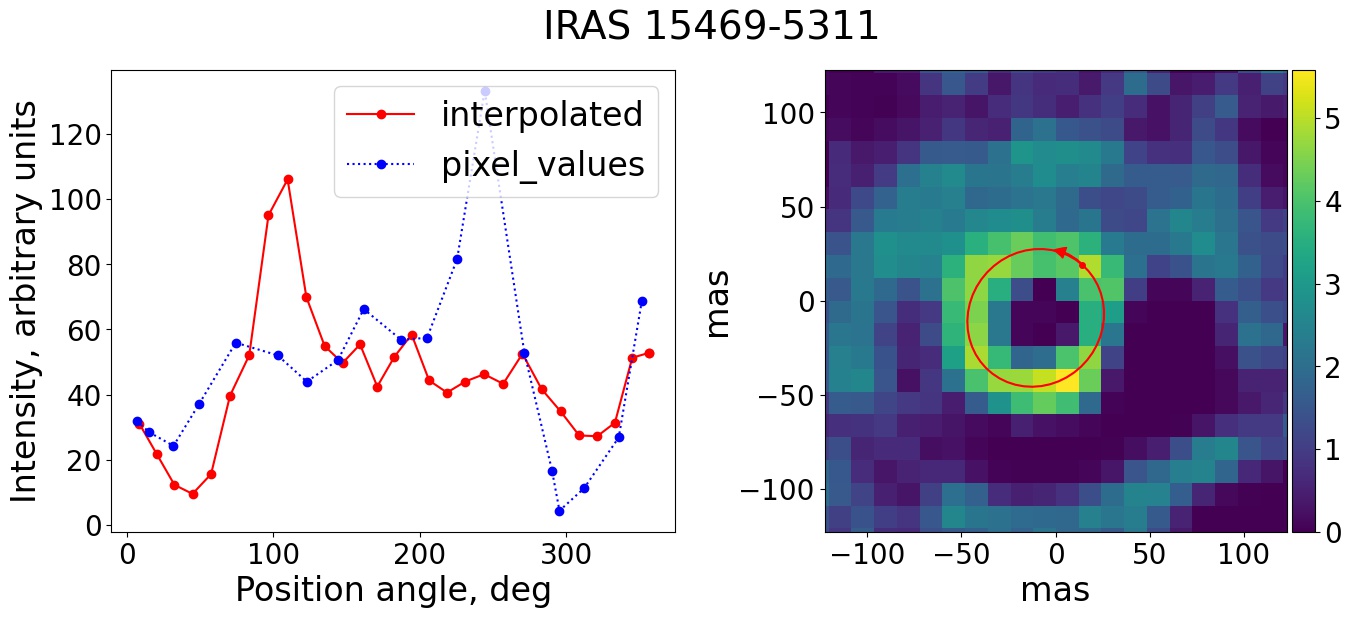}
    \includegraphics[width=1\columnwidth]{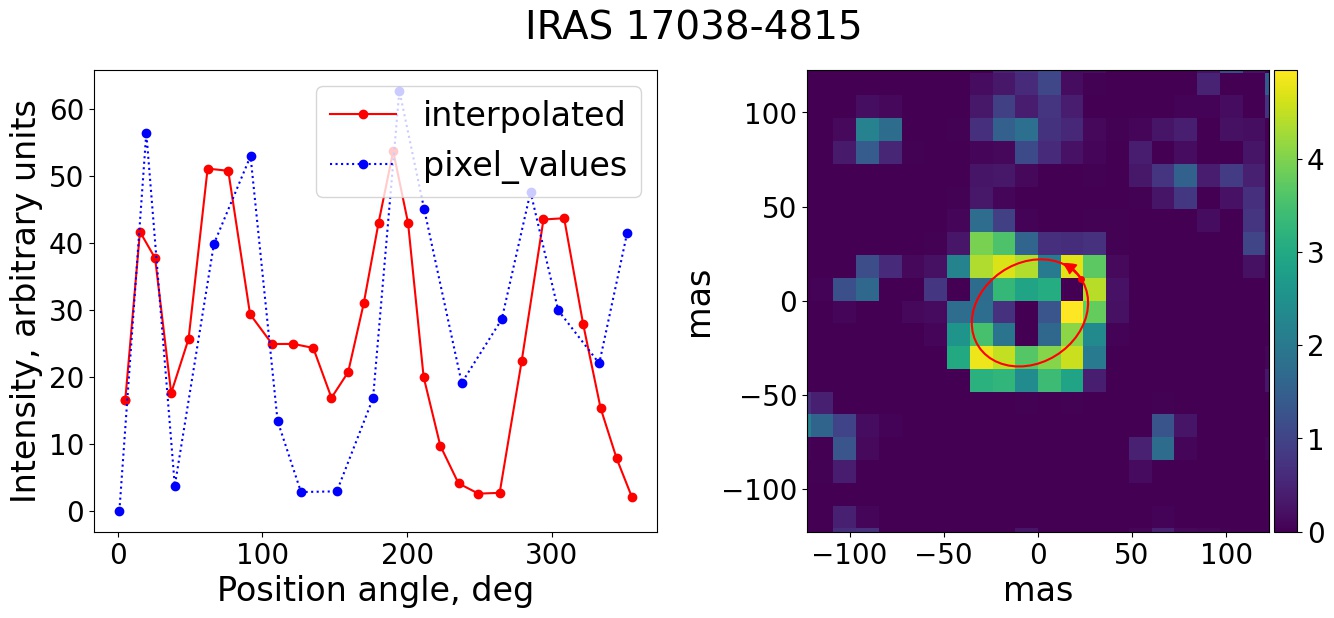}
   
    \caption{Azimuthal brightness profiles (first and third column) and corresponding Q$_\phi$ images (second and fourth column) for six out of eight targets in the sample (except for AC\,Her and RU\,Cen). The blue line represents original pixel values, while the red line represents Bi-linearly interpolated intensities. The red dot and arrow in the Q$_\phi$ image mark the starting point and the direction of the azimuthal brightness profile evaluation. See Section~\ref{sec:profile} for details. \label{fig:azprof}}
\end{figure*}

In Fig.\ref{fig:linearprof}, \ref{fig:radialprof} and \ref{fig:azprof}, we present three types of brightness profiles for each post-AGB binary in the sample: radial profile of the azimuthally polarized  Q$_\phi$ image deprojected onto face-on position, profile along major and minor axes of the visible 'ring' structure, and azimuthal brightness profile along the 'ring' surface (starting from the eastern end of principal axes and going counterclockwise). More details on the methodology are presented in Section~\ref{sec:profile}.

\bsp	
\label{lastpage}
\end{document}